\documentclass[12pt,draftcls,onecolumn]{IEEEtran}
\ifCLASSINFOpdf
  \usepackage[pdftex]{graphicx}
  \graphicspath{{../pdf/}{../jpeg/}}
  \DeclareGraphicsExtensions{.pdf,.jpeg,.png}
\else
  \usepackage[dvips]{graphicx}
  \graphicspath{{../eps/}}
  \DeclareGraphicsExtensions{.eps}
\fi\usepackage{graphicx}

\usepackage{graphics}
\usepackage{epsfig}
\usepackage{epstopdf}
\usepackage{stfloats}
\usepackage[cmex10]{amsmath}
\usepackage{algorithmic}
\usepackage{array}
\usepackage{mdwmath}
\usepackage{mdwtab}
\usepackage{graphicx}
\usepackage{subfigure}
\usepackage{color}
\usepackage{amsfonts,amssymb}
\usepackage{hyperref} 
\usepackage{multirow}
\usepackage{diagbox}
\usepackage{caption}
\usepackage{verbatim}
\usepackage{float}
\usepackage{graphicx}

\usepackage{amsfonts,amsthm,array} 
\makeatletter
\renewcommand{\maketag@@@}[1]{\hbox{\m@th\normalsize\normalfont#1}}
\makeatother

\newtheorem{remark}{Remark}

\begin{document}

\title{On Secure NOMA-CDRT Systems with Physical Layer Network Coding \thanks{Manuscript received.}}

\author{Hongjiang~Lei, 
	Xusheng~She,
	Ki-Hong~Park, 
	Imran Shafique Ansari, \\
	Zheng~Shi, 
	Jing~Jiang, 
	and~Mohamed-Slim~Alouini 
}

\maketitle
\begin{abstract}
	
This paper proposes a new scheme to enhance the secrecy performance of a NOMA-based coordinated direct relay transmission system (NOMA-CDRT) with an untrusted relay.
The physical-layer network coding and the non-orthogonal multiple access scheme are combined to improve the spectrum efficiency.
Furthermore, the inter-user interference and friendly jamming signals are utilized to suppress the eavesdropping ability of the untrusted relay without affecting the acceptance quality of legitimate users. Specifically,
the far user in the first slot and the near user in the second slot act as jammers to generate jamming signals to ensure secure transmissions of the confidential signals.
We investigate the secrecy performance of the proposed scheme in NOMA-CDRT systems and derive the closed-form expression for the ergodic secrecy sum rate.
The asymptotic analysis at high signal-to-noise ratio is performed to obtain more insights.
Finally, simulation results are presented to demonstrate the effectiveness of the proposed scheme and the correctness of the theoretical analysis.

\end{abstract}


\section{Introduction}
\label{sec:introduction}

\subsection{Backgroud and Related Work}

During the past decade, several great technologies have been proposed, such as massive multiple-input multiple-output, millimeter-wave, cognitive radio network, and non-orthogonal multiple access (NOMA), etc., to address a large number of connections with diverse requirements in terms of data rates, latency, and spectral efficiency (SE).
Compared with conventional orthogonal multiple access, NOMA support multiple users to simultaneously access the same wireless resources. It resolves massive connectivity requirement and conserves spectrum resources in Internet of Things (IoT) network, which utilizes superimposed coding and successive interference cancellation (SIC) to offer a significant improvement in reliable performance \cite{DingZ2017JSAC}, \cite{LvL2017TCOM}.

Cooperative NOMA systems have attracted significant attention since they can extend the NOMA users' coverage and enhance the system performance through diversity technology.
There are two architectures for cooperative NOMA systems. A dedicated relay is utilized to forward the signal to far users (FUs) due to the deep fading between the base station and the FUs. The other is the near user (NU) acting as a relay to forward the signals to the FUs based on relaying strategies.
A downlink NOMA-based coordinated direct and relay transmission (NOMA-CDRT) systems was introduced, wherein parallel communications between multiple links were allowed \cite{KimJB2015CL}.
Specifically, the NU does not interfere with the relay forwarding signal in the second slot because the signal for the FU in the first time slot has been decoded with SIC technology. Thus, the relay forwards signals to the FU and the base station simultaneously transmits new signals to the NU.
Thus, the higher ergodic sum rate and SE were achieved.
Liu \emph{et al.} studied the outage performance of a satellite-assisted NOMA-CDRT system and derived the closed-form expression for the exact and asymptotic outage probability (OP) in \cite{LiX2020CL}.
The authors in \cite{PandeyG2021CL} proposed a new signal-space-diversity-based CDRT scheme to improve the achievable ergodic sum rate of the cooperative NOMA system, in which the in-phase or quadrature component of the signals was superimposed and transmitted. The closed-form expressions for the exact and asymptotic OP were derived and compared with non-CDRT scheme.
To enhance SE, Nguyen \emph{et al.} proposed an IoT-based CDRT scheme in which the based station transmits the superimposed signals in the first slot in \cite{TTNguyen2021WCL}. The IoT controller node works as a relay to decode and forward the superimposed signals to the FU and IoT user in the second slot.
The closed-form and approximate expressions for the OP and ergodic sum-rate were derived.

In \cite{NguyenTT2021WCL}, the CDRT scheme was utilized in the underlay cognitive NOMA system, and closed-form and asymptotic expressions for the OP of the primary and cognitive users were derived.
Thai-Hoc \emph{et al.} considered an underlay NOMA-based CDRT system with imperfect SIC, imperfect channel state information {(CSI)}, and co-channel interference in \cite{VuTH2021TCOM}. They proposed multiple relay selection schemes to improve the system throughput. The closed-form expressions for the exact OP of both primary and cognitive users and the system throughput were derived. Moreover, a deep learning framework was designed to predict the performance of the considered system.
To fully utilize the spectrum resources, Zou \emph{et al.} derived the closed-form expression of the ergodic sum-rate for the device-to-device (D2D)-aided NOMA-CDRT systems in which the relay transmits an additional D2D signal to NU while forwarding FU's signals in the second slot in \cite{ZouL2020CL}. 
The authors in \cite{XuY2021TCOM} proposed a new spectrum-efficient scheme for the NOMA-CDRT system in which the uplink and downlink transmissions were held simultaneously via physical-layer network coding (PNC) scheme. The analytical expressions for the ergodic sum-rate, energy efficiency, and Jain's fairness index of the system were derived.
Based on \cite{XuY2021TCOM}, the decoding condition was considered, and an adaptive forward strategy was proposed to implement bidirectional communication in \cite{XuY2021TWC}.
The closed-form expressions for the OP, outage throughput, and ergodic sum-rate were derived. Moreover, the power allocation coefficient was optimized to maximize the ergodic sum-rate.
Yang \emph{et al.} extended NOMA-CDRT with multiple NUs and a best-NU scheduling scheme in which the NU with minimum OP was selected to forward the signals for the FU in \cite{YangM2021CL}.
The closed-form expressions for the exact and asymptotic OP were derived.

Compared with the half-duplex (HD) model, the full-duplex (FD) model overcomes the problem of the limited SE of HD, which can not simultaneously transmit and receive on the same resource block, but at the cost of inevitable self-interference \cite{LiaoY2015CM}.
Si \emph{et al.} studied the performance of the NOMA-CDRT system with an FD relay in \cite{SiQ2020TVT}. Specifically, the signal for the NU was mapped to an $M$-ary modulated symbol, and the signal for the FU was mapped to a spatial modulation symbol. Then the SE was improved and the energy consumption was reduced by making full use of the antenna resources at the relay. The analytical expressions for the ergodic rate and bit error rate of the proposed scheme were derived.
The performance of the NOMA-CDRT system with both FD and HD protocols and multiple FUs was investigated in \cite{PeiX2020TCOM}. Considering residual self-interference, the analytical expressions for the exact OP and the asymptotic ergodic sum-rate were derived. The results showed that the performance with HD outperformed that with FD in the larger-SNR region and self-interference has a more significant impact on the performance of the NU.
The power allocation problem of the NOMA-CDRT system was investigated and the optimal closed-form power allocation policies under the HD and FD protocols were derived in \cite{ChenX2019TWC}. Then, an adaptive relaying scheme was designed to maximize the minimum user achievable rate.

The parallel transmission between multiple links enhances SE and throughput but makes the secrecy performance analysis more complicated and challenging.
Lv \textit{et al.} \cite{LvL2021SCIS} studied the secrecy performance of a NOMA-CDRT system with an untrusted relay in both uplink and downlink scenarios.
Two novel interference-assisted jamming schemes were proposed in which the inter-user interference and the jamming signals were intelligently designed to
suppress the reception quality at the relay. The analytical expressions for the lower bound of the exact and asymptotic ergodic secrecy sum-rate (ESSR) were derived.
New adaptive jamming schemes were proposed to enhance the secrecy performance of downlink and uplink NOMA-CDRT systems in \cite{LvL2020TCOM}.
The problems of maximizing ESSR through the optimization of the jamming power for both downlink and uplink scenarios were studied. The analytical expressions for the lower bound and asymptotic ESSR were derived.
The secrecy performance of a NOMA-CDRT system with multiple NUs was investigated in \cite{LvL2021SCISOMA}  and the best-user scheduling in which the NU with maximum SNR was selected was proposed to enhance the security. The analytical expressions for the lower bound of the exact and asymptotic ergodic secrecy rate (ESR) were derived.

\subsection{Motivation and Contributions}

The PNC scheme improves system throughput while increasing the risk of information eavesdropping.
The CDRT scheme improves system throughput with limited SE.
However, the research on the security performance of the CDRT systems with the PNC scheme is still in its infancy.
This work considers a joint uplink-downlink NOMA-CDRT system with two legitimate users and an untrusted relay, utilizing friendly jammer signals and inter-user interference to provide secure transmission.
The main contributions of this paper are summarized as follows.
\begin{enumerate}
	\item A new scheme, termed as NOMA-CDRT-PNC, is proposed to enhance the secrecy performance of a CDRT with an untrusted relay in which both PNC and NOMA schemes are utilized to improve the spectrum efficiency. The inter-user interference and friendly jamming signals are utilized to suppress the eavesdropping ability of untrusted relay without affecting the SINR of legitimate signals.
	Specifically, the FU in the first slot and the NU in the second slot work as jammers to transmit jamming signals to ensure secure transmission of the confidential signals.
	
    \item We investigate the secrecy performance of the considered CDRT system and derive the closed-form expression for the lower bound of exact and asymptotic ESSR. Simulation results are presented to prove the accuracy of the derived analytical expressions.

    \item Relative to \cite{XuY2021TCOM} and \cite{XuY2021TWC}, in which the PNC scheme with a trusted relay was utilized to enhance SE and outage performance and ergodic sum-rate were investigated, we studied the secrecy performance and the closed-form expression for the lower bound of exact and asymptotic ESSR.
    Relative to \cite{LvL2021SCIS} and \cite{LvL2020TCOM}, in which two interference-assisted jamming schemes were proposed for the downlink and uplink CDRT, we proposed a new jamming scheme combined with PNC for the scenarios with downlink and uplink transmissions to improve both secrecy performance and SE simultaneously.

\end{enumerate}

\subsection{Organization}
The rest of this paper is organized as follows. Section \ref{sec:SystemModel} describes the system model. In Section \ref{sec: ESSR Analyis}, the analytical expressions for the exact ESSR of NOMA-CDRT-PNC system are derived and analyzed. Asymptotic ESSR of NOMA-CDRT-PNC system is derived to gain more insight in Sections \ref{sec: Asy Analyis}.  Section \ref{sec: RESULTS} presents the numerical and simulation results to demonstrate the analysis of the security performance of this system and the paper is concluded in Section \ref{sec: Conclusion}.

\section{System Model}
\label{sec:SystemModel}

\begin{figure*}[!t]
	\centering
	\includegraphics[width = 6 in]{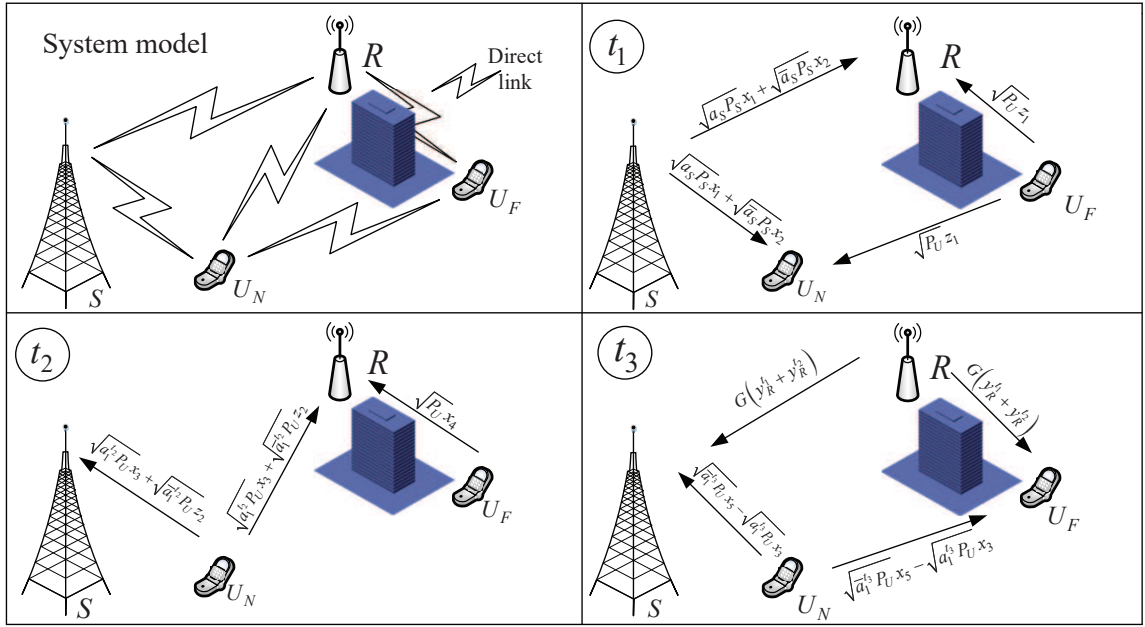}
	\caption{System model consisting of a base station ($S$), two users (${U_N}$ and ${U_F}$), and an untrusted relay (${R}$).}
	\label{figmodel}
\end{figure*}


Fig. \ref{figmodel} illustrates the system model consisting of a base station denoted by $S$, a NU denote by ${U_N}$, and a FU denote by ${U_F}$.
There is no direct link between $S$ and $U_F$ due to deep fading and shadowing, then communication link between $S$ and $U_F$ must be deployed via an intermediate relay ${R}$, which is trusted at the service level while untrusted at the data level \cite{LvL2020TCOM, LvL2019TIFS}.
In other words, $R$ is a potential eavesdropping node to eavesdrop on the confidential information for  $U_i$ $\left( {i \in \left\{ {N,F} \right\}} \right)$.	
The average channel gains and channel coefficient between source $i$ and destination $j$ are denoted by ${\lambda_{ij}}$ and ${h_{ij}}$ for $i,j \in \left\{ {S,{U_N},{U_F},R} \right\}$ $\left( {i \ne j} \right)$.
All the wireless links are assumed to experience quasi-static independent Rayleigh fading and reciprocal.
The transmit power at $S$ is denoted as ${P_S}$ and the transmit power at $U_N$, $U_F$, $R$ are denoted as ${P_U}$.
The data transmission in each fading block has three consecutive and equal phases, as elaborated below.

In the first time slot $\left( {{t_1}} \right)$, according to downlink NOMA scheme, ${S}$ broadcasts a superimposed signal
of $x_1$ and $x_2$ to $U_N$ and $U_F$.
At the same time, ${U_F}$ sends a  jamming signal ${z_1}$\footnote{In this work, it is assumed that the jamming signals is able to be a Gaussian pseudo-random sequence, or utilizes deterministic waveforms similar to the structure of the desired signal as \cite{LvL2019TIFS, LiB2019TC, ZhengTX2020TWC}. Then
all the jamming signals are known for the legitimate receivers and the transmitter, which means that the jamming signals do not affect the receiving quality of the legitimate users.} to effectively disrupt the eavesdropping quality of $R$. Then the received signals at the destination $d$ is expressed as
\begin{equation}
	y_d^{{t_1}} = {h_{Sd}}\left( {\sqrt {{a  _S}{P_S}} {x_1}+ \sqrt {{\bar a  _S}{P_S}} {x_2}} \right) + {h_{{U_F}d}}\sqrt {P_U} {z_1} + n_d^{{t_1}},
	\label{yRt1}
\end{equation}
where ${d \in \left\{ {{U_N},R} \right\}}$,
${a_S}$ denotes the power allocation coefficient for $x_1$, ${{{\bar a  }_S} = 1 - {a  _S}}$,
and
${n_d^{{t_1}}}$ signifies the additive Gaussian white noise (AWGN)  in slot ${{t_1}}$.
After deleting $z_1$ and utilizing perfect SIC detection following the decoding order of ${{x_2} \to {x_1}}$ base on transmit power \cite{KimJB2015CL}, the SNR of $x_1$ at $U_N$ is expressed as
\begin{equation}
    {\gamma ^{{x_1}} = {a  _S}{\rho _S}{\left| {{h_{S{U_N}}}} \right|^2}},
    \label{snrx1}
\end{equation}
where ${\rho _S} = \frac{P_S}{\sigma ^2}$ and $\sigma ^2$ signifies the noise power.

\begin{remark}
It must be noted there is a premise for the CDRT system that ${U_N}$ firstly decode the signal for ${U_F}$. 
which denotes that the power allocation coefficient for ${x_2}$ should be greater than 0.5 $\left( {{a_S} < 0.5} \right)$.
\textcolor[rgb]{1.00,0.00,0.00}{It is assumed ${{\lambda _{SR}} < {\lambda _{S{U_N}}}}$ to ensure that $U_N$ firstly decodes $U_F$ signal and then cancel it, which is a premise for the CDRT system.}
\end{remark}

It is assumed that
${R}$ acts as a potential eavesdropper to decode confidential information based on parallel interference cancellation (PIC) technology \cite{LvL2020TCOM}.
The SINRs of $x_1$ and $x_2$ at $R$ is obtained as
\begin{equation}
     \gamma _R^{{x_1}} = \frac{{{a_S}{\rho _S}{{\left| {{h_{SR}}} \right|}^2}}}{{{{\bar a}_S}{\rho _S}{{\left| {{h_{SR}}} \right|}^2} + {\rho _U}{{\left| {{h_{{U_F}R}}} \right|}^2} + 1}},
	\label{snrRx1}
\end{equation}
and
\begin{equation}
    \gamma _R^{{x_2}} = \frac{{{{\bar a}_S}{\rho _S}{{\left| {{h_{SR}}} \right|}^2}}}{{{a_S}{\rho _S}{{\left| {{h_{SR}}} \right|}^2} + {\rho _U}{{\left| {{h_{{U_F}R}}} \right|}^2} + 1}},
	\label{snrRx2}
\end{equation}
respectively, where ${\rho _U} = \frac{P_U}{\sigma ^2}$.


In the second time slot $\left( {{t_2}} \right)$, ${U_N}$ broadcasts a superimposed signal of the desired signal $x_3$ and a jamming signal $z_2$ to $S$ and $R$, 
${U_F}$ synchronously sends its signal $x_4$ to $S$.
Then the received signals at ${S}$ and $R$ are expressed as
\begin{equation}
	y_S^{{t_2}} = {h_{S{U_N}}}\left( {\sqrt {a  _1^{{t_2}}{P_U}} {x_3} + \sqrt {\bar a  _1^{{t_2}}{P_U}} {z_2}} \right) + n_S^{{t_2}},
	\label{ySt2}
\end{equation}
\begin{equation}
	y_R^{{t_2}} = {h_{{U_N}R}}\left( {\sqrt {a  _1^{{t_2}}{P_U}} {x_3} + \sqrt {\bar a  _1^{{t_2}}{P_U}} {z_2}} \right) + {h_{{U_F}R}}\sqrt {{P_U}} {x_4} + n_R^{{t_2}},	
	\label{yRt2}
\end{equation}
respectively, where $a_1^{{t_2}}$ denotes the power allocation for $x_3$, $\bar a_1^{{t_2}} = 1 - a_1^{{t_2}}$.
After canceling $z_2$, the SNR of $x_3$ at $S$ is written as
\begin{equation}
      {\gamma ^{{x_3}}} = a  _1^{{t_2}}{\rho _U}{\left| {{h_{S{U_N}}}} \right|^2}.
      \label{snrx3}
\end{equation}
With PIC method, the SINRs of $x_3$ and $x_4$ at $R$ are expressed as
\begin{equation}
     \gamma _R^{{x_3}} = \frac{{{\rho _U}a_1^{{t_2}}{{\left| {{h_{{U_N}R}}} \right|}^2}}}{{{\rho _U}\bar a_1^{{t_2}}{{\left| {{h_{{U_N}R}}} \right|}^2} + {\rho _U}{{\left| {{h_{{U_F}R}}} \right|}^2} + 1}},
	\label{snrRx3}
\end{equation}
and
\begin{equation}
\gamma _R^{{x_4}} = \frac{{{\rho _U}{{\left| {{h_{{U_F}R}}} \right|}^2}}}{{{\rho _U}{{\left| {{h_{{U_N}R}}} \right|}^2} + 1}},
	\label{snrRx4}
\end{equation}
respectively.

In the third time slot $\left( {{t_3}} \right)$, $R$ amplifies the received signals and then broadcasts them with power $P_U$.
Simultaneously, ${U_N}$ transmits superimposed signals of a new signal ${x_5}$ and ${x_3}$ to $S$ and ${R}$.
It should be noted that ${x_5}$ can not be wiretapped by ${R}$ since ${R}$ is transmitting then achieve prefect secrecy transmission.
Transmitting ${x_3}$ aims to linearly eliminate the interference of the forwarded signal from $R$ at ${U_F}$.
The received signals at ${S}$ is expressed as
\begin{equation}
	\begin{aligned}
		y_S^{{t_3}} &= {h_{S{U_N}}}\left( {\sqrt {\bar a  _1^{{t_{\rm{3}}}}{P_U}} {x_5} - \sqrt {a  _1^{{t_3}}{P_U}} {x_3}} \right)+ G{h_{SR}}\left( {y_R^{{t_1}} + y_R^{{t_2}}} \right)+ n_S^{{t_3}}\\
		& = G{h_{SR}}{h_{SR}}\left( {\sqrt {{a  _S}{P_S}} {x_1} + \sqrt {{\bar a  _S}{P_S}} {x_2}} \right) + \left( {G{h_{SR}}{h_{{U_N}R}}\sqrt {a  _1^{{t_{\rm{2}}}}{P_U}}  - {h_{S{U_N}}}\sqrt {a  _1^{{t_3}}{P_U}} } \right){x_3}\\
		& + G{h_{SR}}{h_{{U_F}R}}\sqrt {{P_U}} {x_4} + {h_{S{U_N}}}\sqrt {\bar a  _1^{{t_{\rm{3}}}}{P_U}} {x_5}+ G{h_{SR}}{h_{{U_F}R}}\sqrt {{P_U}} {z_1} \\
		&+ G{h_{SR}}{h_{{U_N}R}}\sqrt {{\bar a  _1^{{t_2}}}{P_U}} {z_2}+ G{h_{SR}}n_R^{{t_1}}  + G{h_{SR}}n_R^{{t_2}} + n_S^{{t_3}}\\
		& \mathop  = \limits^{\left( a \right)} G{h_{SR}}{h_{{U_F}R}}\sqrt {{P_U}} {x_4} + {h_{S{U_N}}}\sqrt {\bar a  _1^{{t_{\rm{3}}}}{P_U}} {x_5} + {n_S},
		\label{ySt3}
	\end{aligned}
\end{equation}
where
${{G^2} = \frac{{{\rho _U}}}{{{\rho _S}{\lambda _{SR}} + 2{\rho _U}{\lambda _{R{U_F}}} + {\rho _U}{\lambda _{R{U_N}}} + 2}}}$ denotes amplifying coefficient in fixed-gain relay scheme \cite{LvL2021SCIS}, \cite{LvL2020TCOM}, \cite{LvL2019TIFS},
${{n_S} = G{h_{SR}}n_R^{{t_1}} + G{h_{SR}}n_R^{{t_2}} + n_S^{{t_3}}}$,
${a  _1^{{t_3}}}$ denotes the power allocation coefficient for $x_3$, ${\bar a  _1^{{t_3}} = 1 - a  _1^{{t_3}}}$,
and step $\left( a \right)$ is obtained since $x_1$, $x_2$, $x_3$, $z_1$, and $z_2$ are known at $S$.
Then ${S}$ uses SIC technology following the decoding order of ${{x_5} \to {x_4}}$ and
\textcolor[rgb]{1.00,0.00,0.00}{based on average channel gain of ${\lambda _{SR}} < {\lambda _{S{U_N}}}$ \cite{LvL2020TCOM},}
the SINRs of  $x_5$ and $x_4$  at $S$ are obtained as
\begin{equation}
	{{\gamma ^{{x_5}}} = \frac{{\bar a  _1^{{t_{\rm{3}}}}{\rho _U}{{\left| {{h_{S{U_N}}}} \right|}^2}}}{{{G^2}{{\left| {{h_{SR}}} \right|}^2}\left( {{\rho _U}{{\left| {{h_{{U_F}R}}} \right|}^2} + 2} \right) + 1}}},
	\label{snrx5}
\end{equation}
and
\begin{equation}
	{{\gamma ^{{x_4}}} = \frac{{{G^2}{\rho _U}{{\left| {{h_{SR}}} \right|}^2}{{\left| {{h_{{U_F}R}}} \right|}^2}}}{{2{G^2}{{\left| {{h_{SR}}} \right|}^2} + 1}}},
	\label{snrx4}
\end{equation}
respectively.

Similarly, the received signals at ${U_F}$ is expressed as
\begin{equation}
	\begin{aligned}
		y_{{U_F}}^{{t_3}} &= {h_{{U_N}{U_F}}}\left( {\sqrt {\bar a  _1^{{t_{\rm{3}}}}{P_U}} {x_5} - \sqrt {a  _1^{{t_3}}{P_U}} {x_3}} \right) + G{h_{R{U_F}}}\left( {y_R^{{t_1}} + y_R^{{t_2}}} \right) + n_{{U_F}}^{{t_3}}\\
		& \mathop  = \limits^{\left( b \right)} G{h_{SR}}{h_{R{U_F}}}\left( {\sqrt {{a_S}{P_S}} {x_1} + \sqrt {{{\bar a}_S}{P_S}} {x_2}} \right)+ {\omega _0}{x_3} + {h_{{U_N}{U_F}}}\sqrt {\bar a_1^{{t_3}}{P_U}} {x_5} + {n_2},
		\label{yU2t3}
	\end{aligned}
\end{equation}
where
${\omega _0} = {G\left| {{h_{R{U_F}}}} \right|\left| {{h_{R{U_N}}}} \right|\sqrt {a_1^{{t_2}}{\rho _U}}  - \left| {{h_{{U_N}{U_F}}}} \right|\sqrt {a_1^{{t_3}}{\rho _U}}}$,
${{n_F} = G{h_{R{U_F}}}n_R^{{t_1}} + G{h_{R{U_F}}}n_R^{{t_2}} + n_{{U_F}}^{{t_3}}}$,
step $\left( b \right)$ is obtained since $x_4$, $z_1$, and $z_2$ are known at $U_F$.
To remove $x_3$ at ${U_F}$, ${a  _1^{{t_3}}}$ must satisfy ${\omega _0} = 0$
Then, we have
\begin{equation}
     a  _1^{{t_3}} = \frac{{{G^2}{{\left| {{h_{R{U_F}}}} \right|}^2}{{\left| {{h_{{U_N}R}}} \right|}^2} a  _1^{{t_2}}}}{{{{\left| {{h_{{U_N}{U_F}}}} \right|}^2}}}.
    \label{pa1t3}
\end{equation}
Taking the expectation operation for $a  _1^{{t_3}}$  \cite{LvL2021SCISOMA}, we have
\footnote{\textcolor[rgb]{1.00,0.00,0.00}{In this work, it is assumed ${{\lambda _{{U_N}R}} \le {\lambda _{{U_N}{U_F}}}}$ to ensure ${\mathbb{E}}\left[ a  _1^{{t_3}} \right] < 1$}.}
\begin{equation}
	\begin{aligned}
		{\mathbb{E}}\left[ a  _1^{{t_3}} \right] &= \frac{{{G^2}a  _1^{{t_2}}{\lambda _{{U_N}R}}{\lambda _{R{U_F}}}}}{{{\lambda _{{U_N}{U_F}}}}}\\
		&= \frac{{{\rho _U}a  _1^{{t_2}}{\lambda _{{U_N}R}}{\lambda _{R{U_F}}}}}{{{\lambda _{{U_N}{U_F}}}\left( {{\rho _S}{\lambda _{SR}} + 2{\rho _U}{\lambda _{R{U_F}}} + {\rho _U}{\lambda _{R{U_N}}} + 2} \right)}}\\
		&< \frac{{a  _1^{{t_2}}}}{{{\lambda _{{U_N}{U_F}}}}}\frac{{{\lambda _{{U_N}R}}{\lambda _{R{U_F}}}}}{{{\lambda _{R{U_F}}} + {\lambda _{R{U_N}}}}}\\
		&  < \frac{{{\lambda _{{U_N}R}}a_1^{{t_2}}}}{{{\lambda _{{U_N}{U_F}}}}}\\
		& \le a_1^{{t_2}} < 1.
		\label{a1t3}
	\end{aligned}
\end{equation}

It must be noted that ${\mathbb{E}}\left[ a  _1^{{t_3}} \right] < 1$ may not guarantee $a  _1^{{t_3}} < 1$.
To meet the causality constraint on power coefficient, we set $a  _1^{{t_3}} = \frac{{{G^2}a  _1^{{t_2}}{\lambda _{{U_N}R}}{\lambda _{R{U_F}}}}}{{{\lambda _{{U_N}{U_F}}}}}$.
Then ${U_F}$ uses SIC detection following the decoding order of ${{x_5} \to {x_2}}$,
the SINR of ${x_2}$ is obtained as
\begin{equation}
	\begin{aligned}
		{\gamma ^{{x_2}}} &= \frac{{{G^2}{{\bar a}_S}{\rho _S}{{\left| {{h_{SR}}} \right|}^2}{{\left| {{h_{R{U_F}}}} \right|}^2}}}{{{G^2}{{\left| {{h_{R{U_F}}}} \right|}^2}\left( {{a_S}{\rho _S}{{\left| {{h_{SR}}} \right|}^2} + 2} \right) + {\omega _0^2} + 1}}\\
		&\approx \frac{{{G^2}{{\bar a  }_S}{\rho _S}{{\left| {{h_{SR}}} \right|}^2}{{\left| {{h_{R{U_F}}}} \right|}^2}}}{{{G^2}{{\left| {{h_{R{U_F}}}} \right|}^2}\left( {{a  _S}{\rho _S}{{\left| {{h_{SR}}} \right|}^2} + 2} \right) + 1}}.
		\label{snrx2}
	\end{aligned}
\end{equation}

The instantaneous secrecy rate of $x_j$ is expressed as \cite{BlochM2008TiT}
\begin{equation}
    C_s^{{x_j}} = {\left[ {\ln \left( {1 + {\gamma^{{x_j}}}} \right) - \ln \left( {1 + \gamma_R^{{x_j}}} \right)} \right]^ + },
    \label{secrecyrate}
\end{equation}
where $j = 1, \cdots ,5$ and ${\left[ x \right]^ + } = \max \left\{ {x,0} \right\}$.

\section{Ergodic Secrecy Sum Rate Analysis}
\label{sec: ESSR Analyis}

In this section, we analyze the ESSR of the proposed NOMA-CDRT-PNC scheme, which is expressed as
\begin{equation}
	\begin{aligned}
	{{\bar C}_{{\rm{ESSR}}}} &= \sum\limits_{j = 1}^5 {\bar C_{{\rm{ESR}}}^{{x_j}}} \\
	&= \frac{1}{3} \sum\limits_{j = 1}^4 { \mathbb{E}{{\left[ {{C^{{x_j}}} - C_R^{{x_j}}} \right]}^ + }} + \frac{1}{3} \mathbb{E}\left[ {{C^{{x_5}}}} \right],
	\label{ESSR}
    \end{aligned}
\end{equation}
where
${\bar C_{{\rm{ESR}}}^{{x_j}}}$ denotes the ergodic secrecy rate of $x_j$,
${C^{{x_j}}} = \ln \left( {1 + {\gamma ^{{x_j}}}} \right)$,
$C_R^{{x_j}} = \ln \left( {1 + \gamma _R^{{x_j}}} \right)$,
$\mathbb{E}\left[  \cdot  \right]$ denotes expectation operator.
By utilizing Jensen's inequality, the lower bound of the ESSR is expressed as
\begin{equation}
	\begin{aligned}
		\bar C_{{\rm{ESSR}}}^{\rm{L}} &= \sum\limits_{j = 1}^5 {\bar C_{{\rm{ESR}}}^{{x_j},{\rm{L}}}} \\
		&= \frac{1}{3}\sum\limits_{i = 1}^4 {{{\left[ {{{\bar C}^{{x_j}}} - \bar C_R^{{x_j}}} \right]}^ + }}  + \frac{1}{3}{{\bar C}^{{x_5}}},
		\label{ESSRlower}
	\end{aligned}
\end{equation}
where
${{\bar C}^{{x_j}}} = \mathbb{E}\left[ { {C^{{x_j}}}} \right]$,
$\bar C_R^{{x_j}} = \mathbb{E}\left[ {C_R^{{x_j}}} \right]$,
and
the superscript `${\rm{L}}$' denotes lower bound.
Based on the probability theory, ergodic capacity is expressed as
\begin{equation}
  \begin{aligned}
      \bar C &= \mathbb{E}\left[ {\ln \left( {1 + \gamma } \right)} \right]\\
      & = \int_0^\infty  {\ln \left( {1 + x} \right){f_\gamma }\left( x \right)dx} \\
      & = \int_0^\infty  {\frac{{1 - {F_\gamma }\left( x \right)}}{{1 + x}}dx},
      \label{ec}
  \end{aligned}
\end{equation}
where ${{f_\gamma }\left( x \right)}$ and ${{F_\gamma }\left( x \right)}$ are probability density function (PDF) and cumulative distribution function (CDF) of $\gamma$, respectively.

The analytical expressions for ergodic rate of all the signals in Section II are derived as follows.

Substituting (\ref{snrx1}) into (\ref{ec}) and utilizing \cite[(4.337.2)]{Gradshteyn2007Book}, ${{\bar C}^{{x_1}}}$ is obtained as
\begin{equation}
   \begin{aligned}
	{{\bar C}^{{x_1}}} &= \mathbb{E}\left[ {\ln \left( {1 + {a  _S}{\rho _S}{{\left| {{h_{S{U_N}}}} \right|}^2}} \right)} \right]\\
	&= \frac{1}{{{\lambda _{S{U_N}}}}}\int_0^\infty  {\ln \left( {1 + {a_S}{\rho _S}x} \right){e^{ - \frac{x}{{{\lambda _{S{U_N}}}}}}}dx} \\
	&= {\phi _1}\left( {{a_S}{\rho _S}{\lambda _{S{U_N}}}} \right),
	\label{ecx1}
   \end{aligned}
\end{equation}
where
${\phi _1}\left( x \right) =  - \exp \left( {\frac{1}{x}} \right){\rm{Ei}}\left( { - \frac{1}{x}} \right)$
and
${\rm{Ei}}\left( x \right) =  - \int_{ - x}^\infty  {{t^{ - 1}}\exp \left( { - t} \right)dt}$ is exponential integral function, defined by \cite[(8.211.1)]{Gradshteyn2007Book}.

Based on (\ref{snrRx1}), we obtain the CDF of ${\gamma _R^{{x_1}}}$ as
\begin{equation}
	\begin{aligned}
		{F_{\gamma _R^{{x_1}}}}\left( x \right) &= \Pr \left\{ {\frac{{{a_S}{\rho _S}{{\left| {{h_{SR}}} \right|}^2}}}{{{{\bar a}_S}{\rho _S}{{\left| {{h_{SR}}} \right|}^2} + {\rho _U}{{\left| {{h_{{U_F}R}}} \right|}^2} + 1}} < x} \right\}\\
		&= \left\{ {\begin{array}{*{20}{c}}
				{\Pr \left\{ {{{\left| {{h_{SR}}} \right|}^2} < \frac{{{\rho _U}x{{\left| {{h_{{U_F}R}}} \right|}^2} + x}}{{\left( {{a_S} - {{\bar a}_S}x} \right){\rho _S}}}} \right\},}&{x < \frac{{{a_S}}}{{{{\bar a}_S}}}},\\
				{1,}&{x > \frac{{{a_S}}}{{{{\bar a}_S}}}},
		\end{array}} \right.\\
		&= \left\{ {\begin{array}{*{20}{c}}
				{\int_0^\infty  {{F_{{{\left| {{h_{SR}}} \right|}^2}}}\left( {\frac{{{\rho _U}xy + x}}{{\left( {{a_S} - {{\bar a}_S}x} \right){\rho _S}}}} \right){f_{{{\left| {{h_{R{U_F}}}} \right|}^2}}}\left( y \right)dy} ,}&{x < \frac{{{a_S}}}{{{{\bar a}_S}}}},\\
				{1,}&{x > \frac{{{a_S}}}{{{{\bar a}_S}}}},
		\end{array}} \right.\\
		&= \left\{ {\begin{array}{*{20}{c}}
				{1 - \frac{{{a_S} - {{\bar a}_S}x}}{{\left( {{\omega _1} - {{\bar a}_S}} \right)x + {a_S}}}{e^{ - \frac{x}{{{\rho _S}{\lambda _{SR}}\left( {{a_S} - {{\bar a}_S}x} \right)}}}},}&{x < \frac{{{a_S}}}{{{{\bar a}_S}}}},\\
				{1,}&{x > \frac{{{a_S}}}{{{{\bar a}_S}}}},
		\end{array}} \right.
		\label{CDFRX1}
	\end{aligned}
\end{equation}
where ${\omega _1} = \frac{{{\rho _U}{\lambda _{R{U_F}}}}}{{{\rho _S}{\lambda _{SR}}}}$.
Substituting (\ref{CDFRX1}) into  (\ref{ec}) and utilizing \cite[(3.352.4)]{Gradshteyn2007Book}, $\bar C_R^{{x_1}}$ is derived as
\setcounter{equation}{22} 
\begin{equation}
	\begin{aligned}
		\bar C_R^{{x_1}} &= \int_0^\infty  {\frac{1}{{x + 1}}\left( {1 - {F_{\gamma _R^{{x_1}}}}\left( x \right)} \right)dx} \\
	    &= \frac{{{\phi _2}\left( {{\rho _S}{\lambda _{SR}},{{\bar a}_S}{\rho _S}{\lambda _{SR}}} \right)}}{{1 - {\omega _1}}} - \frac{{{\omega _1}{a_S}{\phi _2}\left( {{\omega _1}{\rho _S}{\lambda _{SR}},{{\bar a}_S}{\rho _S}{\lambda _{SR}}} \right)}}{{\left( {1 - {\omega _1}} \right)\left( {{\omega _1} - {{\bar a}_S}} \right)}},
		\label{ecRx1}
	\end{aligned}
\end{equation}
where ${{\phi _2}\left( {a,b} \right) = {\phi _1}\left( a \right) - {\phi _1}\left( b \right)}$.
Substituting (\ref{ecx1}) and (\ref{ecRx1}) into (\ref{ESSRlower}), the closed-form expression for lower bound of ${\bar C_{{\rm{ESR}}}^{{x_1}}}$ is obtained.

\begin{remark}
	Based on (\ref{snrx1}) and (\ref{snrRx1}), one can find that both $\gamma ^{{x_1}}$ and $\gamma _R ^{{x_1}}$ increase as increasing of $a_S$. However, increasing of $\gamma ^{{x_1}}$ is faster than the increasing of $\gamma _R^{{x_1}}$ because $\gamma _R^{{x_1}} < \frac{{{a_S}{\rho _S}{{\left| {{h_{SR}}} \right|}^2}}}{{{\rho _U}{{\left| {{h_{{U_F}R}}} \right|}^2} + 1}}$ and ${\gamma ^{{x_1}} = {a  _S}{\rho _S}{\left| {{h_{S{U_N}}}} \right|^2}}$ is proportional to ${a_S}$. Thus, The ESR of $x_1$ increases as increasing $a_S$.
\end{remark}
	
The CDF of ${x_2}$ is obtained as
\begin{equation}
	\begin{aligned}
		{F_{{\gamma ^{{x_{_2}}}}}}\left( x \right) &= \Pr \left\{ {\frac{{{G^2}{{\bar a}_S}{\rho _S}{{\left| {{h_{R{U_F}}}} \right|}^2}{{\left| {{h_{SR}}} \right|}^2}}}{{{G^2}{{\left| {{h_{R{U_F}}}} \right|}^2}\left( {{a_S}{\rho _S}{{\left| {{h_{SR}}} \right|}^2} + 2} \right) + 1}} < x} \right\}\\
		&= \left\{ {\begin{array}{*{20}{c}}
				{\Pr \left\{ {{{\left| {{h_{SR}}} \right|}^2} < \frac{{\left( {2{G^2}{{\left| {{h_{R{U_F}}}} \right|}^2} + 1} \right)x}}{{\left( {{{\bar a}_S} - {a_S}x} \right){G^2}{\rho _S}{{\left| {{h_{R{U_F}}}} \right|}^2}}}} \right\},}&{x < \frac{{{{\bar a}_S}}}{{{a_S}}}},\\
				{1,}&{x > \frac{{{{\bar a}_S}}}{{{a_S}}}},
		\end{array}} \right.\\
		&= \left\{ {\begin{array}{*{20}{c}}
				{\int_0^\infty  {{F_{{{\left| {{h_{SR}}} \right|}^2}}}\left( {\frac{{2xy + x/{G^2}}}{{{\rho _S}\left( {{{\bar a}_S} - {a_S}x} \right)y}}} \right){f_{{{\left| {{h_{R{U_F}}}} \right|}^2}}}\left( y \right)dy} ,}&{x < \frac{{{{\bar a}_S}}}{{{a_S}}}},\\
				{1,}&{x > \frac{{{{\bar a}_S}}}{{{a_S}}}},
		\end{array}} \right.\\
		&= \left\{ {\begin{array}{*{20}{c}}
				{1 - {\phi _3}\left( {\frac{x}{{{\beta _S}\left( {{{\bar a}_S} - {a_S}x} \right)}}} \right){e^{ - \frac{{2x}}{{{\rho _S}{\lambda _{SR}}\left( {{{\bar a}_S} - {a_S}x} \right)}}}},}&{x < \frac{{{{\bar a}_S}}}{{{a_S}}}},\\
				{1,}&{x > \frac{{{{\bar a}_S}}}{{{a_S}}}},
		\end{array}} \right.
		\label{cdfx2}
	\end{aligned}
\end{equation}
where ${\beta _S}  = \frac{{{G^2}{\rho _S}{\lambda _{SR}}{\lambda _{R{U_F}}}}}{4}$, ${\phi _3}\left( x \right) = \sqrt x {K_1}\left( {\sqrt x } \right)$,
and
${{K_v}\left(  \cdot  \right)}$ is the $v$-order modified Bessel function of the second kind, defined in \cite[(8.432.1)]{Gradshteyn2007Book}.
Substituting (\ref{cdfx2}) into (${\ref{ec}}$) and utilizing \cite[(3.324.1)]{Gradshteyn2007Book}, we obtain
\begin{equation}
	\begin{aligned}
		{{\bar C}^{{x_2}}} & = \int_0^\infty  {\frac{{{\beta _S}}}{{{\beta _S}x + 1}}{e^{ - \frac{{{G^2}{\lambda _{R{U_F}}}x}}{2}}}\sqrt x {K_1}\left( {\sqrt x } \right)dx} \\
		&- \int_0^\infty  {\frac{{{a_S}{\beta _S}}}{{{a_S}{\beta _S}x + 1}}{e^{ - \frac{{{G^2}{\lambda _{R{U_F}}}x}}{2}}}\sqrt x {K_1}\left( {\sqrt x } \right)dx} \\
		&= {\beta _S} {\phi _4}\left( {{\beta _S} ,\frac{{{G^2}{\lambda _{R{U_F}}}}}{2}} \right) - {a_S}{\beta _S} {\phi _4}\left( {{a_S}{\beta _S} ,\frac{{{G^2}{\lambda _{R{U_F}}}}}{2}} \right),	
		\label{ecx2}
	\end{aligned}
\end{equation}
where 
${\phi _4}\left( {a,b} \right) = \int_0^\infty  {\frac{1}{{ax + 1}}{e^{ - bx}}\sqrt x {K_1}\left( {\sqrt x } \right)dt}$.
Utilizing \cite[(9.34.3)]{Gradshteyn2007Book}, (10) and (11) of \cite{Adamchik1990}, and \cite[(1.2)]{MathaiAM2010}, ${{\phi _4}\left( {a,b} \right)}$ is denoted as
\begin{equation}
	\begin{aligned}
		{\phi _4}\left( {a,b} \right) &= \int_0^\infty  {\frac{1}{{ax + 1}}{e^{ - bx}}\sqrt x {K_1}\left( {\sqrt x } \right)dt} \\
		&= \int_0^\infty  {G_{1,1}^{1,1}\left[ {ax\left| {_0^0} \right.} \right]G_{0,1}^{1,0}\left[ {bx\left| {_0^ - } \right.} \right]G_{0,2}^{2,0}\left[ {\frac{x}{4}\left| {_{1,0}^ - } \right.} \right]dx} \\
		&= \frac{1}{b}H_{0,0:0,0:0,0}^{0,1:1,1:0,2}\left[ {\left. {_ - ^1} \right|\left. {_0^0} \right|\left. {_{1,0}^ - } \right|\frac{a}{b},\frac{1}{{4b}}} \right],
		\label{gx}
	\end{aligned}
\end{equation}
where ${G_{c,d}^{a,b}\left[  \cdot  \right]}$ is the Meijer's $G$-function defined by \cite[(9.301)]{Gradshteyn2007Book}
and
${H_{c,d:p,r:a  ,\beta }^{0,b:m,n:\gamma ,\varepsilon }\left[  \cdot  \right]}$ is the Extended Generalized Bivariate Fox's H-function (EGBFHF) defined by \cite[(2.57)]{MathaiAM2010}.

With the same method as (\ref{ecRx1}), $\bar C_R^{{x_2}} $ is obtained as
\begin{equation}
	\begin{aligned}
	\bar C_R^{{x_2}} &= \frac{{{\phi _2}\left( {{\rho _S}{\lambda _{SR}},{a_S}{\rho _S}{\lambda _{SR}}} \right)}}{{1 - {\omega _1}}}- \frac{{{\omega _1}{{\bar a}_S}{\phi _2}\left( {{\omega _1}{\rho _S}{\lambda _{SR}},{a_S}{\rho _S}{\lambda _{SR}}} \right)}}{{\left( {1 - {\omega _1}} \right)\left( {{\omega _1} - {a_S}} \right)}}.
	\label{ecRx2}
	\end{aligned}
\end{equation}
Substituting (\ref{ecx2}) and (\ref{ecRx2}) into (\ref{ec}), the closed-form for lower bound of ${\bar C_{{\rm{ESR}}}^{{x_2}}}$ is obtained.

\begin{remark}
	Based on (\ref{snrRx2}) and (\ref{snrx2}), one can find that both $\gamma ^{{x_2}}$ and $\gamma _R ^{{x_2}}$ tend to a constant independent to $\rho$ at the high-SNR region.
	Moreover, one can find that the effect of $a_S$ on the ESR of $x_2$ is the opposite of the effect of $a_S$ on the ESR of $x_1$ because increasing $a_S$ signifies decreasing of the transmit power for $x_2$, which lead to the deterioration of the ESR of $x_2$, testified in \cite{LeiH2017CL}.	
\end{remark}
	
With the same method as (\ref{ecx1}) and (\ref{ecRx1}), we obtain ${{\bar C}^{{x_3}}}$ and $\bar C_R^{{x_3}}$ as
\begin{equation}
	{{\bar C}^{{x_3}}} = {\phi _1}\left( {a_1^{{t_2}}{\rho _U}{\lambda _{S{U_N}}}} \right),
	\label{ecx3}
\end{equation}
and
\begin{equation}
  \begin{aligned}
	\bar C_R^{{x_3}} &=\frac{{\bar a_1^{{t_2}}{\phi _2}\left( {\bar a_1^{{t_2}}{\rho _U}{\lambda _{R{U_N}}},{\rho _U}{\lambda _{R{U_F}}}} \right)}}{{{\omega _2} - \bar a_1^{{t_2}} }} - \frac{{{\phi _2}\left( {{\rho _U}{\lambda _{R{U_N}}},{\rho _U}{\lambda _{R{U_F}}}} \right)}}{{ {\omega _2} - 1}},
	\label{ecRx3}
  \end{aligned}
\end{equation}
respectively, where
${\omega _2} = \frac{{{\lambda _{R{U_F}}}}}{{{\lambda _{R{U_N}}}}}$.
Substituting (\ref{ecx3}) and (\ref{ecRx3}) into (\ref{ec}), the closed-form for lower bound of ${\bar C_{{\rm{ESR}}}^{{x_3}}}$ is obtained.

\begin{remark}
	Although the equation of the ESR of $x_3$ is similar to that of the ESR of $x_1$, it must be noted that the effect of ${a _1^{{t_2}}}$ on the ESR of $x_3$ is different from the effect of $a_S$ on the ESR of ${x_1}$ since there is $0 \le {a _1^{{t_2}}} \le 1$ while $a_S < 0.5$. 	
	Based on (\ref{snrRx3}), one can find that increasing means that ${a _1^{{t_2}}}$ denotes more power is allocated to $x_3$ and less power is allocated to $z_2$, which leads to increasing both $\gamma _R^{{x_3}}$ and $\gamma ^{{x_3}}$. However,  $\gamma _R^{{x_3}}$ at $0.5 < {a _1^{{t_2}}} < 1$ region increases faster than that at $0 < {a _1^{{t_2}}} < 0.5$ because the power allocated to $z_2$ tends to zero. Thus the ESR of $x_3$ increases first and then decreases as ${a _1^{{t_2}}}$ increases. Thus, there exists an optimal ${a_1^{{t_2}}}$ to maximize the ESR of $x_3$.
\end{remark}

With the same method as (\ref{cdfx2}), we obtain
\begin{equation}
	\begin{aligned}
		{F_{{\gamma ^{{x_4}}}}}\left( x \right) = 1 - {e^{ - \frac{{2x}}{{{\rho _U}{\lambda _{R{U_F}}}}}}}{\phi _3}\left( {\frac{x}{{{\beta _U}}}} \right),
		\label{cdfx4}
	\end{aligned}
\end{equation}
where ${\beta _U} = \frac{{{G^2}{\rho _U}{\lambda _{SR}}{\lambda _{R{U_F}}}}}{4}$.
Similar as (\ref{ecx2}), we obtain
\begin{equation}
	{{\bar C}^{{x_4}}} = {\beta _U}{\phi _4}\left( {{\beta _U},\frac{{{G^2}{\lambda _{SR}}}}{2}} \right).
	\label{ecx4}
\end{equation}
The CDF of ${\gamma _R^{{x_4}}}$ is expressed as
\begin{equation}
  \begin{aligned}
	{F_{\gamma _R^{{x_4}}}}\left( x \right) &= \Pr \left\{ {{{\left| {{h_{{U_F}R}}} \right|}^2} < \frac{{{\rho _U}x{{\left| {{h_{{U_N}R}}} \right|}^2} + x}}{{{\rho _U}}}} \right\}\\
	&= \int_0^\infty  {{F_{{{\left| {{h_{R{U_F}}}} \right|}^2}}}\left( {\frac{{{\rho _U}xy + x}}{{{\rho _U}}}} \right){f_{{{\left| {{h_{R{U_N}}}} \right|}^2}}}\left( y \right)dy} \\
	&= 1 - \frac{{{\lambda _{R{U_F}}}}}{{{\lambda _{R{U_N}}}x + {\lambda _{R{U_F}}}}}{e^{ - \frac{x}{{{\rho _U}{\lambda _{R{U_F}}}}}}}.
	\label{CDFRX4}
  \end{aligned}
\end{equation}
Substituting (\ref{CDFRX4}) into (${\ref{ec}}$), utilizing \cite[(3.352.4)]{Gradshteyn2007Book}, we obtain
\begin{equation}
	\begin{aligned}
	\bar C_R^{{x_4}} &= \int_0^\infty  {\frac{1}{{x + 1}}\left( {1 - {F_{\gamma _R^{{x_4}}}}\left( x \right)} \right)dx} \\
	&= \int_0^\infty  {\frac{1}{{x + 1}}\frac{{{\lambda _{R{U_F}}}}}{{{\lambda _{R{U_N}}}x + {\lambda _{R{U_F}}}}}{e^{ - \frac{x}{{{\rho _U}{\lambda _{R{U_F}}}}}}}dx} \\
	&=\frac{{{\omega _2}{\phi _2}\left( {{\rho _U}{\lambda _{R{U_F}}},{\rho _U}{\lambda _{R{U_N}}}} \right)}}{{{\omega _2} - 1}}.
	\label{ecRx4}
	\end{aligned}
\end{equation}
Substituting (\ref{ecx4}) and (\ref{ecRx4}) into (\ref{ec}), the closed-form for lower bound of ${\bar C_{{\rm{ESR}}}^{{x_4}}}$ is obtained.

Based on (\ref{snrx5}), the CDF of ${x_5}$ is obtained as
\begin{equation}
	\begin{aligned}
		{F_{{\gamma ^{{x_5}}}}}\left( x \right) &= \Pr \left\{ {{{\left| {{h_{S{U_N}}}} \right|}^2} < \frac{{{G^2}{{\left| {{h_{SR}}} \right|}^2}x\left( {{\rho _U}{{\left| {{h_{R{U_F}}}} \right|}^2} + 2} \right) + x}}{{\bar a_1^{{t_3}}{\rho _U}}}} \right\}\\
		&= \int_0^\infty  {\int_0^\infty  {{F_{{{\left| {{h_{S{U_N}}}} \right|}^2}}}\left( {\frac{{{G^2}xy\left( {{\rho _U}z + 2} \right) + x}}{{\bar a_1^{{t_3}}{\rho _U}}}} \right)} }{f_{{{\left| {{h_{SR}}} \right|}^2}}}\left( y \right){f_{{{\left| {{h_{R{U_F}}}} \right|}^2}}}\left( z \right)dydz\\
		&= 1 - \frac{{\bar a_1^{{t_3}}{\lambda _{S{U_N}}}}}{{{G^2}{\lambda _{SR}}{\lambda _{R{U_F}}}x}}{e^{ - \frac{x}{{\bar a_1^{{t_3}}{\rho _U}{\lambda _{S{U_N}}}}}}}{\phi _1}\left( {\frac{2}{{{\rho _U}{\lambda _{R{U_F}}}}} + \frac{{\bar a_1^{{t_3}}{\lambda _{S{U_N}}}}}{{{G^2}{\lambda _{SR}}{\lambda _{R{U_F}}}x}}} \right).
		\label{cdfx5}
	\end{aligned}
\end{equation}

Then the ergodic rate of ${x_5}$ is expressed as
\begin{equation}
\begin{aligned}
	{{\bar C}^{{x_5}}} &= \frac{{\bar a_1^{{t_3}}{\lambda _{S{U_N}}}}}{{{G^2}{\lambda _{SR}}{\lambda _{R{U_F}}}}}\int_0^\infty  {\frac{1}{{x\left( {x + 1} \right)}}{e^{ - \frac{x}{{\bar a_1^{{t_3}}{\rho _U}{\lambda _{S{U_N}}}}}}}} {\phi _1}\left( {\frac{2}{{{\rho _U}{\lambda _{R{U_F}}}}} + \frac{{\bar a_1^{{t_3}}{\lambda _{S{U_N}}}}}{{{G^2}{\lambda _{SR}}{\lambda _{R{U_F}}}x}}} \right)dx.
    \label{NoBoundecx5}
\end{aligned}
\end{equation}
The closed-form expression for (\ref{NoBoundecx5}) is difficult to obtain since there is a complicated integral of exponential integral function, then its lower bound is derived based on the method proposed in \cite{LvL2021SCISOMA}.
We define ${X = \bar a  _1^{{t_3}}{\rho _U}{\left| {{h_{S{U_N}}}} \right|^2}}$ and ${Y={{G^2}{{\left| {{h_{SR}}} \right|}^2}\left( {{\rho _U}{{\left| {{h_{R{U_F}}}} \right|}^2} + 2} \right)}}$, and using Jensen’s inequality, we obtain
\begin{equation}
	\begin{aligned}
    {C^{{x_5}}} &= \mathbb{E}\left[ {\ln \left( {1 + {\gamma ^{{x_5}}}} \right)} \right]\\
     &= \mathbb{E}\left[ {\ln \left( {1 + \frac{X}{{1 + Y}}} \right)} \right]\\
     &= \mathbb{E}\left[ {\ln \left( {1 + {e^{\ln \left( {\frac{X}{{1 + Y}}} \right)}}} \right)} \right]\\
     &\ge {C^{{x_5},{\rm{L}}}} = \ln \left( {1 + {e^{\mathbb{E}\left[ {\ln \left( {\frac{X}{{1 + Y}}} \right)} \right]}}} \right)\\
     &= \ln \left( {1 + {{\mathop{\rm e}\nolimits} ^\Phi }} \right),
		\label{ecx5}
	\end{aligned}
\end{equation}
where
$\Phi  = {\mathbb{E}}\left[ {\ln \left( X \right)} \right] - {\mathbb{E}}\left[ {\ln \left( {1 + Y} \right)} \right]$.
Utilizing \cite[(4.311.1)]{Gradshteyn2007Book}, we  have
\begin{equation}
	\begin{aligned}
		\mathbb{E}\left[ {\ln \left( X \right)} \right] &= \int_0^\infty  {\ln \left( {\bar a_1^{{t_3}}{\rho _U}x} \right){f_{{{\left| {{h_{S{U_N}}}} \right|}^2}}}\left( x \right)dx} \\
		&= \ln \left( {\bar a_1^{{t_3}}{\rho _U}{\lambda _{S{U_N}}}} \right) - {\rm{C}},
		\label{Xec}
	\end{aligned}
\end{equation}
where ${\rm{C}}=0.577215$ denotes Euler’s constant \cite[(8.367.1)]{Gradshteyn2007Book}.
By utilizing
\begin{equation}
	\int_{{x_0}}^{{x_1}} {e^ {{ - ax - \frac{b}{x}}}dx}  \approx \left( {{e^{ - a{x_0}}} - {e^{ - a{x_1}}}} \right)\sqrt {\frac{{4b}}{a}} {K_1}\left( {\sqrt {4ab} } \right),
	\label{H232}
\end{equation}
which is verified in \cite{YanM2012WCL}, the CDF of $Y$ is obtained as
\begin{equation}
	\begin{aligned}
         {F_Y}\left( y \right) &= \Pr \left\{ {{{\left| {{h_{SR}}} \right|}^2} < \frac{y}{{{G^2}\left( {{\rho _U}{{\left| {{h_{R{U_F}}}} \right|}^2} + 2} \right)}}} \right\}\\
         &= \int_0^\infty  {{F_{{{\left| {{h_{SR}}} \right|}^2}}}\left( {\frac{y}{{{G^2}\left( {{\rho _U}x + 2} \right)}}} \right){f_{{{\left| {{h_{R{U_F}}}} \right|}^2}}}\left( x \right)dx} \\
	     &= 1 - \frac{1}{{{\lambda _{R{U_F}}}}}\int_0^\infty  {{e^{ - \frac{y}{{{G^2}{\lambda _{SR}}\left( {{\rho _U}x + 2} \right)}}}}} {e^{ - \frac{x}{{{\lambda _{R{U_F}}}}}}}dx\\
	     &= 1 - \frac{1}{{{\rho _U}{\lambda _{R{U_F}}}}}{e^{\frac{2}{{{\rho _U}{\lambda _{R{U_F}}}}}}}\int_2^\infty  {{e^{ - \frac{y}{{{G^2}{\lambda _{SR}}t}}}}{e^{ - \frac{t}{{{\rho _U}{\lambda _{R{U_F}}}}}}}dt} \\
	     &\approx 1 - {\phi _3}\left( {\frac{y}{{{\beta _U}}}} \right).
		\label{YCDF}
	\end{aligned}
\end{equation}

Substituting (\ref{YCDF}) into (\ref{ec}), utilizing \cite[(9.31.2)]{Gradshteyn2007Book} and \cite[(21)]{Adamchik1990}, we obtain
\begin{equation}
	\begin{aligned}
      &\mathbb{E}\left[ {\ln \left( {1 + Y} \right)} \right] \\
      &= \int_0^\infty  {\frac{{1 - {F_Y}\left( x \right)}}{{1 + x}}dx} \\
       &= {\beta _U}\int_0^\infty  {G_{1,1}^{1,1}\left[ {{\beta _U}t\left| {_0^0} \right.} \right]G_{0,2}^{2,0}\left[ {\frac{t}{4}\left| {_{1,0}^ - } \right.} \right]dt} \\
       &= G_{1,3}^{3,1}\left[ {\frac{1}{{4{\beta _U}}}\left| {_{1,0,0}^0} \right.} \right].
	\label{Yec}
	\end{aligned}
\end{equation}
Then, we have
\begin{equation}
     \Phi  = \ln \left( {\bar a_1^{{t_3}}{\rho _U}{\lambda _{S{U_N}}}} \right) - G_{1,3}^{3,1}\left[ {\frac{1}{{4{\beta _U}}}\left| {_{1,0,0}^0} \right.} \right] - {\rm{C}}.
	 \label{Phi}
\end{equation}
Substituting (\ref{Phi}) into $ {C^{{x_5},{\rm{L}}}}$, the closed-form for lower bound of ${\bar C_{{\rm{ESR}}}^{{x_5}}}$ is obtained.

\begin{remark}
	Based on $a  _1^{{t_3}} = \frac{{{G^2}a  _1^{{t_2}}{\lambda _{{U_N}R}}{\lambda _{R{U_F}}}}}{{{\lambda _{{U_N}{U_F}}}}}$, one can find that increasing $a  _1^{{t_2}}$ signifies decreasing of $\bar a_1^{{t_3}}$ and ${\bar C_{{\rm{ESR}}}^{{x_5}}}$ since the transmit power for $x_5$, which lead to the deterioration of the ESR of $x_5$.	
\end{remark}

Substituting (\ref{ecx1}), (\ref{ecRx1}), (\ref{ecx2}), (\ref{ecRx2}), (\ref{ecx3}), (\ref{ecRx3}), (\ref{ecx4}), (\ref{ecRx4}), and (\ref{ecx5}) into (\ref{ESSRlower}), the analytical expressions for the lower bound of the ESSR is obtained as
\begin{equation}
	\begin{aligned}
		\bar C_{{\rm{ESSR}}}^{\rm{L}} &= \frac{1}{3}\left[ {{\phi _1}\left( {{a_S}{\rho _S}{\lambda _{S{U_N}}}} \right) - \frac{{{\phi _2}\left( {{\rho _S}{\lambda _{SR}},{{\bar a}_S}{\rho _S}{\lambda _{SR}}} \right)}}{{1 - {\omega _1}}}} \right.{\left. { + \frac{{{\omega _1}{a_S}{\phi _2}\left( {{\omega _1}{\rho _S}{\lambda _{SR}},{{\bar a}_S}{\rho _S}{\lambda _{SR}}} \right)}}{{\left( {1 - {\omega _1}} \right)\left( {{\omega _1} - {{\bar a}_S}} \right)}}} \right]^ + }\\
		&+ \frac{1}{3}\left[ {{\beta _S}{\phi _4}\left( {{\beta _S},\frac{{{G^2}{\lambda _{R{U_F}}}}}{2}} \right) - {a_S}{\beta _S}{\phi _4}\left( {{a_S}{\beta _S},\frac{{{G^2}{\lambda _{R{U_F}}}}}{2}} \right)} \right. - \frac{{{\phi _2}\left( {{\rho _S}{\lambda _{SR}},{a_S}{\rho _S}{\lambda _{SR}}} \right)}}{{1 - {\omega _1}}}\\
		&\;\;\;\;\;\;\;\;\; {\left. { + \frac{{{\omega _1}{{\bar a}_S}{\phi _2}\left( {{\omega _1}{\rho _S}{\lambda _{SR}},{a_S}{\rho _S}{\lambda _{SR}}} \right)}}{{\left( {1 - {\omega _1}} \right)\left( {{\omega _1} - {a_S}} \right)}}} \right]^ + }\\
		&+ \frac{1}{3}\left[ {{\phi _1}\left( {a_1^{{t_2}}{\rho _U}{\lambda _{S{U_N}}}} \right) + \frac{{{\phi _2}\left( {{\rho _U}{\lambda _{R{U_N}}},{\rho _U}{\lambda _{R{U_F}}}} \right)}}{{{\omega _2} - 1}}} \right.{\left. { - \frac{{\bar a_1^{{t_2}}{\phi _2}\left( {\bar a_1^{{t_2}}{\rho _U}{\lambda _{R{U_N}}},{\rho _U}{\lambda _{R{U_F}}}} \right)}}{{{\omega _2} - \bar a_1^{{t_2}}}}} \right]^ + }\\
		&+ \frac{1}{3}\left[ {{\beta _U}{\phi _4}\left( {{\beta _U},\frac{{{G^2}{\lambda _{SR}}}}{2}} \right)} \right.{\left. { - \frac{{{\omega _2}{\phi _2}\left( {{\rho _U}{\lambda _{R{U_F}}},{\rho _U}{\lambda _{R{U_N}}}} \right)}}{{{\omega _2} - 1}}} \right]^ + }+ \frac{1}{3}\ln \left( {1 + e^\Phi} \right).
		\label{ESSR1}
	\end{aligned}
\end{equation}

The analytical expressions is complicated because many factors affect the ESSR of the considered system, specifically, the power coefficients $\left( {{a_S},a_1^{{t_2}}} \right)$, the transmit powers $\left( {{P_S},{P_U}} \right)$, and the average channel gains $\left( {{\lambda _{S{U_N}}},{\lambda _{SR}},{\lambda _{{U_N}R}},{\lambda _{R{U_F}}},{\lambda _{{U_N}{U_F}}}} \right)$.
To obtain more insights, we derive asymptotic expressions of the ESSR in the high transmit power regime in the next section.

\section{Asymptotic Analysis for Ergodic Secrecy Sum Rate}
\label{sec: Asy Analyis}
To gain more insights about the proposed scheme, we analyze asymptotic ESSR expression in high-SNR region in this section. It assumed ${\rho _S} = \nu {\rho _U}$ and  ${{\rho _U} = \rho  \to \infty }$.

Based on (\ref{ecx1}), utilizing ${{\rm{Ei}}\left( x \right)\mathop  \sim \limits^{x \to {0^ - }} {\rm{C}} + \ln \left( { - x} \right) }$ and ${e^x \mathop  \sim \limits^{x \to 0} 1 + x}$\cite{ZhangC2021WCL}, we have ${\phi _1}\left( x \right)\mathop  \approx \limits^{x \to 0} \ln \left( x \right) - {\rm{C}}$.
Then the asymptotic ESR of $x_1$ is obtained as
\begin{equation}
	\begin{aligned}
	  \bar C_{\rm ESR}^{{x_1},\rho  \to \infty } &= {{\bar C}^{{x_1},\rho  \to \infty }} - \bar C_R^{{x_1},\rho  \to \infty }\\
      &= \ln \left( {{a _S}\nu \rho {\lambda _{S{U_N}}}} \right) + \frac{{{\omega _1}{a_S}\ln \left( {{\omega _1}} \right)}}{{\left( {1 - {\omega _1}} \right)\left( {{\omega _1} - {{\bar a}_S}} \right)}} - \frac{{{{\bar a}_S}\ln \left( {{{\bar a}_S}} \right)}}{{{\omega _1} - {{\bar a}_S}}} - {\rm{C}}\\
      &\approx \ln \left( {{a _S}\nu \rho {\lambda _{S{U_N}}}} \right).
	  \label{esrx1asy}
	\end{aligned}
\end{equation}
Similarly, the asymptotic ESR of $x_3$ is obtained as
\begin{equation}
	\begin{aligned}
		\bar C_{{\rm{ESR}}}^{{x_3},\rho  \to \infty } &= {{\bar C}^{{x_3},\rho  \to \infty }} - \bar C_R^{{x_3},\rho  \to \infty }\\
		&= \ln \left( {a_1^{{t_2}}\rho {\lambda _{S{U_N}}}} \right) + \frac{{{\omega _2}a_1^{{t_2}}\ln \left( {{\omega _2}} \right)}}{{\left( {1 - {\omega _2}} \right)\left( {{\omega _2} - \bar a_1^{{t_2}}} \right)}} - \frac{{\bar a_1^{{t_2}}\ln \left( {\bar a_1^{{t_2}}} \right)}}{{{\omega _2} - \bar a_1^{{t_2}}}} - {\rm{C}}\\
		&\approx \ln \left( {a_1^{{t_2}}\rho {\lambda _{S{U_N}}}} \right).
		\label{esrx3asy}
	\end{aligned}
\end{equation}

\begin{remark}
	Based on (\ref{esrx1asy}) and (\ref{esrx3asy}), one can observer that the lower bounds of ESR of $x_1$ and $x_3$ scale as $\ln \left( \rho  \right)$ at high-SNR region.
	This is because SIC deletes the inter-user interference on $U_N$ and the jamming signal does not affect $\gamma^{x_1}$ and $\gamma^{x_3}$, which tends to infinity as  $\rho  \to \infty $.
	However, the inter-user interference and the jamming signal on $R$ can not be deleted by PIC, then $\gamma_R^{x_1}$  and $\gamma_R^{x_3}$ tend to be a constant, which is independent of $\rho$.
	Moreover, the asymptotic ESR of ${x_1}$ is proportional to ${a_S}$, which is same as statement in Remark 2.
\end{remark}


Utilizing $ {{K_1}\left( x \right)\mathop  \approx \limits^{x \to 0} \frac{1}{x} + \frac{x}{2}\ln \left( {\frac{x}{2}} \right)}$ \cite[(17)]{YueX2021TWC}, we have $ {x{K_1}\left( x \right)\mathop  \approx \limits^{x \to 0} 1 + \frac{{{x^2}}}{2}\ln \left( {\frac{x}{2}} \right) \approx 1}$, then ${{\bar C_R}^{{x_4}}}$ is approximated as
\begin{equation}
	\begin{aligned}
		{\bar C^{{x_4},\rho  \to \infty }} = {e^{\frac{2}{{\rho {\lambda _{R{U_F}}}}}}}\left( {\ln \left( {\frac{{\rho {\lambda _{R{U_F}}}}}{2}} \right) - {\rm{C}}} \right).
		\label{ecx4asy}
	\end{aligned}
\end{equation}
With the same method as (\ref{esrx1asy}), we have
\begin{equation}
	\begin{aligned}
		\bar C_R^{{x_4},\rho  \to \infty } = \frac{{{\omega _2}}}{{{\omega _2} - 1}}\ln \left( {{\omega _2}} \right),
		\label{ecRx4asy}
	\end{aligned}
\end{equation}
Then the asymptotic ESR of ${x_4}$ is obtained as
\begin{equation}
	\begin{aligned}
		\bar C_{{\rm{ESR}}}^{{x_4},\rho  \to \infty } &= {e^{\frac{2}{{\rho {\lambda _{R{U_F}}}}}}}\left( {\ln \left( {\frac{{\rho {\lambda _{R{U_F}}}}}{2}} \right) - {\rm{C}}} \right) - \frac{{{\omega _2}}}{{{\omega _2} - 1}}\ln \left( {{\omega _2}} \right)\\
		&\approx \ln \left( {\frac{{\rho {\lambda _{R{U_F}}}}}{2}} \right).
		\label{esrx4asy}
	\end{aligned}
\end{equation}

\begin{remark}
	Based on (\ref{esrx4asy}), one can observe that the lower bound of ESR of $x_4$ scales as $\ln \left( \rho  \right)$ at the high-SNR region.
\end{remark}


Based on (\ref{ecx2}), utilizing $ {{K_1}\left( x \right)\mathop  \approx \limits^{x \to 0} \frac{1}{x} + \frac{x}{2}\ln \left( {\frac{x}{2}} \right)}$ \cite[(17)]{YueX2021TWC},
${{\bar C}^{{x_2}}}$ is approximated as
\begin{equation}
	\begin{aligned}
		{{\bar C}^{{x_2},\rho  \to \infty }}
		&=\underbrace {\int_0^\infty  {\frac{1}{{x + \frac{1}{{{\beta _S}}}}}{e^{ - \frac{{{G^2}{\lambda _{R{U_F}}}x}}{2}}}\left( {1 + \frac{x}{4}\ln \left( {\frac{x}{4}} \right)} \right)dx} }_{ \buildrel \Delta \over = {I_1}} \\
		&- \underbrace {\int_0^\infty  {\frac{1}{{x + \frac{1}{{{a_S}{\beta _S}}}}}{e^{ - \frac{{{G^2}{\lambda _{R{U_F}}}x}}{2}}}\left( {1 + \frac{x}{4}\ln \left( {\frac{x}{4}} \right)} \right)dx} }_{ \buildrel \Delta \over = {I_2}}.
		\label{ecx2asy1}
	\end{aligned}
\end{equation}
Utilizing \cite[(3.352.4), (4.331.1)]{Gradshteyn2007Book}, we obtain $I_1$ as
\begin{equation}
	\begin{aligned}
		{I_1} 
		&= \int_0^\infty  {\frac{1}{{x + \frac{1}{{{\beta _S}}}}}{e^{ - \frac{{{G^2}{\lambda _{R{U_F}}}x}}{2}}}dx} + \frac{1}{4}\int_0^\infty  {\left( {1 - \frac{{\frac{1}{{{\beta _S}}}}}{{x + \frac{1}{{{\beta _S}}}}}} \right){e^{ - \frac{{{G^2}{\lambda _{R{U_F}}}x}}{2}}}\ln \left( {\frac{x}{4}} \right)dx} \\
		&= {\phi _1}\left( {\frac{{2{\beta _S}}}{{{G^2}{\lambda _{R{U_F}}}}}} \right) + \int_0^\infty  {{e^ { - 2{G^2}{\lambda _{R{U_F}}}y}}\ln \left( y \right)dy}- \int_0^\infty  {\frac{1}{{4{\beta _S}y + 1}}{e^ { - 2{G^2}{\lambda _{R{U_F}}}y}}\ln \left( y \right)dy} \\
		&= {\phi _1}\left( {\frac{{\nu \rho {\lambda _{SR}}}}{2}} \right) - \frac{{{\rm{C}} + \ln \left( {2{G^2}{\lambda _{R{U_F}}}} \right)}}{{2{G^2}{\lambda _{R{U_F}}}}} - {\phi _5}\left( {4{\beta _S},2{G^2}{\lambda _{R{U_F}}}} \right),
		\label{I1asy}
	\end{aligned}
\end{equation}
where
${{\phi _5}\left( {a,b} \right)}$ is denoted as
\begin{equation}
	\begin{aligned}
     {\phi _5}\left( {a,b} \right) &= \int_0^\infty  {\frac{1}{{ay + 1}}\exp \left( { - by} \right)\ln \left( y \right)dy}\\
     &= \frac{1}{{2a}}\ln \left( {\frac{b}{a}} \right)\ln \left( {ab} \right)- \frac{{{\psi ^{\left( 1 \right)}}\left( 1 \right)}}{{2a}} + \frac{{{\rm{C}}\ln \left( b \right)}}{a} + \frac{{{{\rm{C}}^2}}}{{2a}},
	\label{psi5}
	\end{aligned}
\end{equation}
where
${\psi ^{\left( k \right)}}\left( x \right)$ is the polygamma function, defined by \cite[(8.363.8)]{Gradshteyn2007Book}.
The detailed derivation of ${{\phi _5}\left( {a,b} \right)}$ is given in Appendix \ref{appendicesA}.
Similarly, we obtain
\begin{equation}
	\begin{aligned}
	{I_2} &= {\phi _1}\left( {\frac{{{a_S}\nu \rho {\lambda _{SR}}}}{2}} \right) - \frac{{{\rm{C}} + \ln \left( {2{G^2}{\lambda _{R{U_F}}}} \right)}}{{2{G^2}{\lambda _{R{U_F}}}}}- {\phi _5}\left( {4{a_S}{\beta _S},2{G^2}{\lambda _{R{U_F}}}} \right).
	\label{I2asy}
	\end{aligned}
\end{equation}
Due to
${\beta _S} = \frac{{{G^2}{\rho _S}{\lambda _{SR}}{\lambda _{R{U_F}}}}}{4} \to \infty $, ${\phi _5}\left( {a,b} \right)\mathop  \approx \limits^{a \to \infty } 0$,
${\phi _1}\left( x \right)\mathop  \approx \limits^{x \to 0} \ln \left( x \right) - {\rm{C}}$,
and
${\phi _2}\left( {a,b} \right)\mathop  \approx \limits^{a,b \to 0} \ln \left( {\frac{a}{b}} \right)$.
Substituting (\ref{I1asy}) and (\ref{I2asy}) into (\ref{ecx2asy1}), we obtain ${{\bar C}^{{x_2},\rho  \to \infty }}$ as
\begin{equation}
		\begin{aligned}
			{{\bar C}^{{x_2},\rho  \to \infty }} &= {\phi _2}\left( {\frac{{\nu \rho {\lambda _{SR}}}}{2},\frac{{{a_S}\nu \rho {\lambda _{SR}}}}{2}} \right)+ {\phi _5}\left( {4{a_S}{\beta _S},2{G^2}{\lambda _{R{U_F}}}} \right) - {\phi _5}\left( {4{\beta _S},2{G^2}{\lambda _{R{U_F}}}} \right)\\
			&\approx \ln \left( {\frac{1}{{{a_S}}}} \right).
			\label{ecx2asy2}
		\end{aligned}
\end{equation}
Based on (\ref{ecRx2}), utilizing ${\phi _1}\left( x \right)\mathop  \approx \limits^{x \to 0} \ln \left( x \right) - {\rm{C}}$, $\bar C_R^{{x_2} }$ is approximated as
\begin{equation}
	\bar C_R^{{x_2},\rho  \to \infty } = \frac{{{a_S}\ln \left( {{a_S}} \right)}}{{{\omega _1} - {a_S}}} - \frac{{{\omega _1}{{\bar a}_S}\ln \left( {{\omega _1}} \right)}}{{\left( {1 - {\omega _1}} \right)\left( {{\omega _1} - {a_S}} \right)}}.
	\label{ecRx2asy}
\end{equation}
Then the asymptotic ESR of ${x_2}$ is obtained as
\begin{equation}
	\begin{aligned}
		\bar C_{{\rm{ESR}}}^{{x_2},\rho  \to \infty } &\approx \frac{{{\omega _1}}}{{{\omega _1} - {a_S}}}\left( { \ln \left( {\frac{1}{{{a_S}}}} \right) + \frac{{{{\bar a}_S}\ln \left( {{\omega _1}} \right)}}{{1 - {\omega _1}}}} \right).
		\label{esrx2asy}
	\end{aligned}
\end{equation}

\begin{remark}
	Based on (\ref{esrx2asy}), one can observe that the asymptotic ESR of $x_2$ tends to be a constant independent of $\rho$ at the high-SNR region.
\end{remark}

Similarly, based on (\ref{snrx5}), one can observe that the ergodic rate of $x_5$ tends to a constant independent on $\rho$ since ${\gamma ^{{x_5}}}$ tends to a constant independent on $\rho$ when ${\rho  \to \infty }$.
Moreover, increasing $a_1^{t_2}$ results in decreasing of ${\gamma ^{{x_5}}}$ since less power is allocated for $x_5$.

Based on (\ref{esrx1asy}), (\ref{esrx3asy}), and (\ref{esrx4asy}), the asymptotic expressions for the lower bound of the ESSR is obtained as
\begin{equation}
	\begin{aligned}
     \bar C_{{\rm{ESSR}}}^{{\rm{L,}}\rho  \to \infty } &\approx  \ln \left( {{a_S}\nu \rho {\lambda _{S{U_N}}}} \right) + \ln \left( {a_1^{{t_2}}\rho {\lambda _{S{U_N}}}} \right) + \ln \left( {\frac{{\rho {\lambda _{R{U_F}}}}}{2}} \right).
		\label{ESSRasy}
	\end{aligned}
\end{equation}

\begin{figure}[t]
	\centering
	\subfigure[]{
		\label{fig21}
		\includegraphics[width = 0.319 \textwidth]{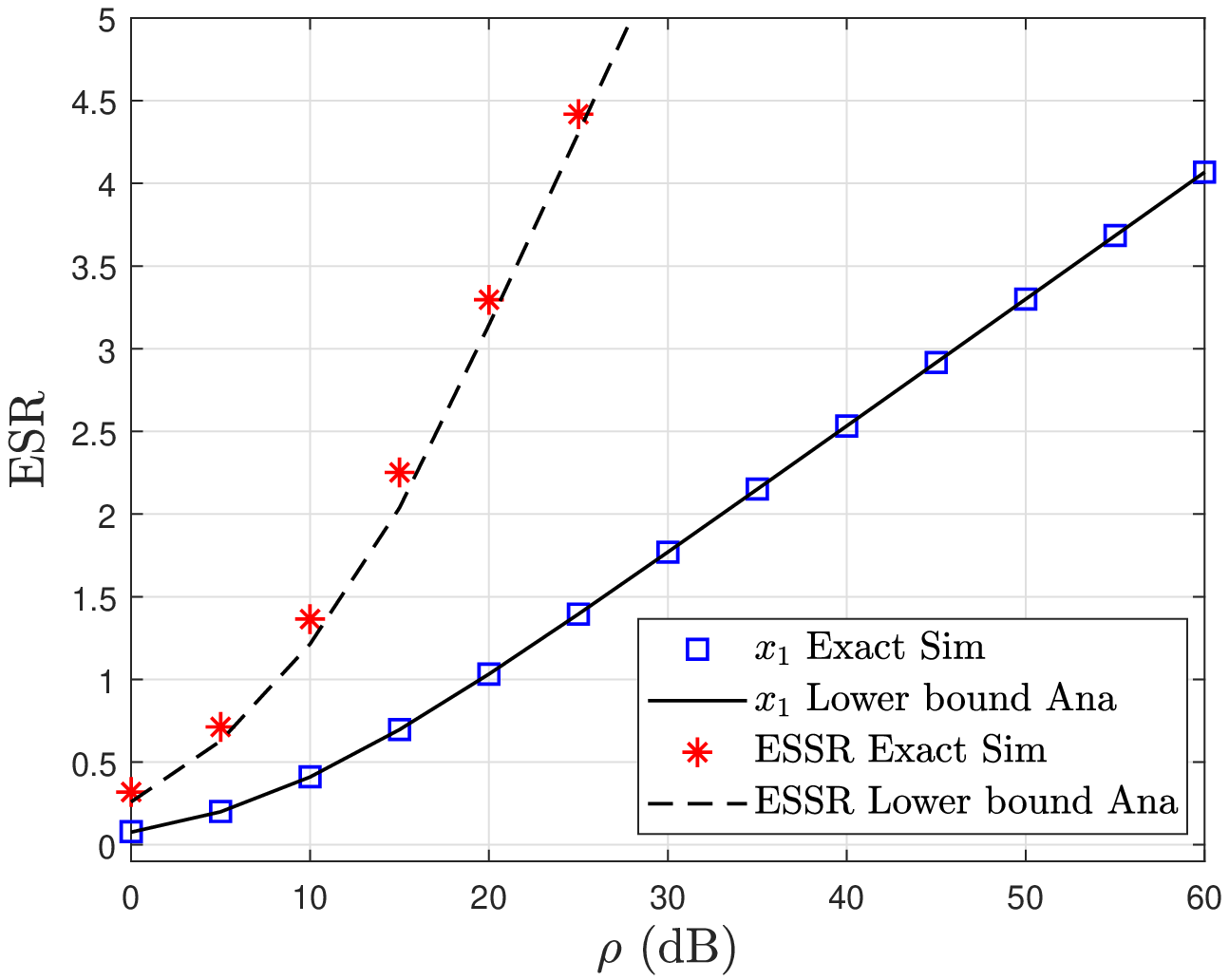}}
	\subfigure[]{
		\label{fig22}
		\includegraphics[width = 0.319 \textwidth]{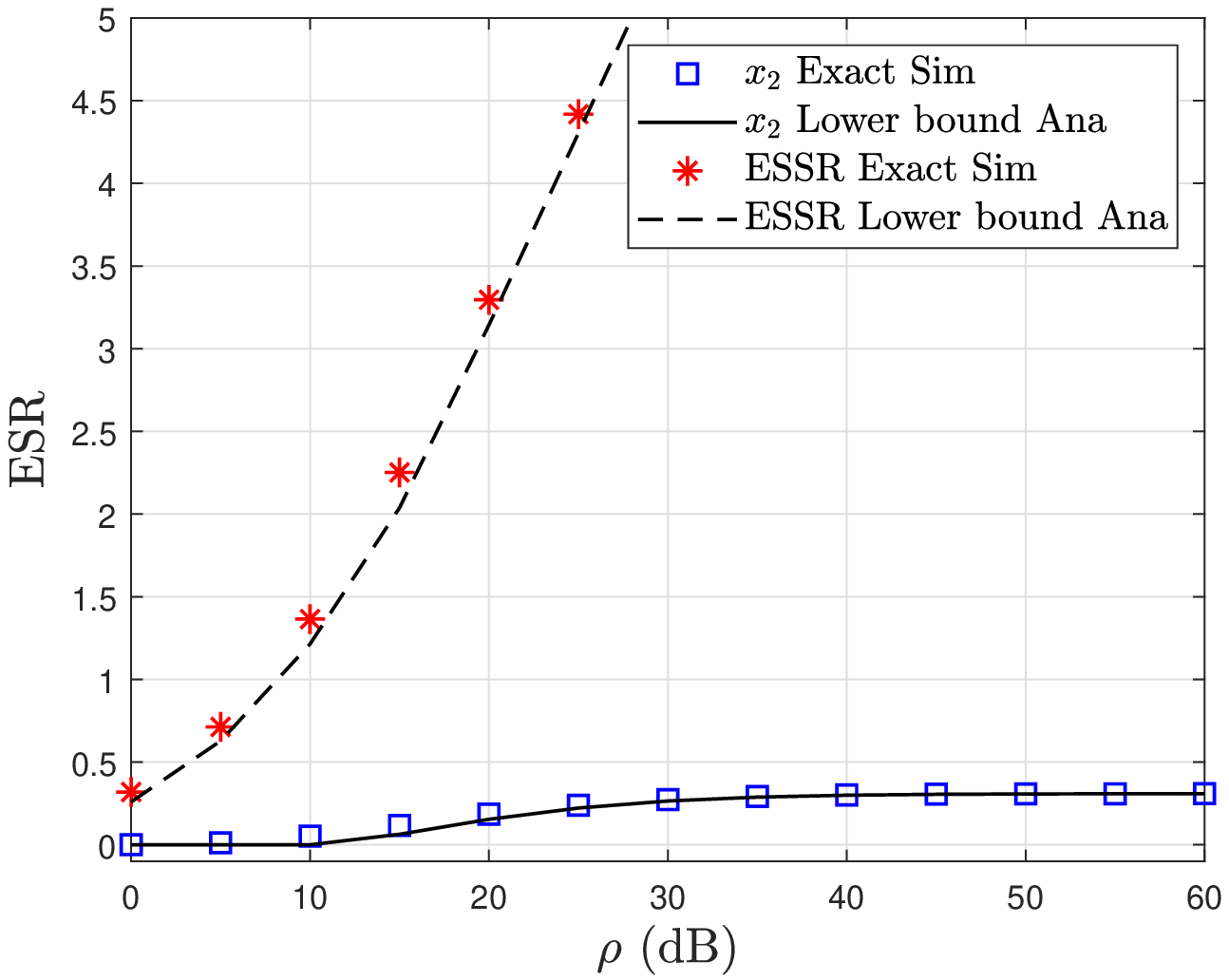}}
	\subfigure[]{
		\label{fig23}
		\includegraphics[width = 0.319 \textwidth]{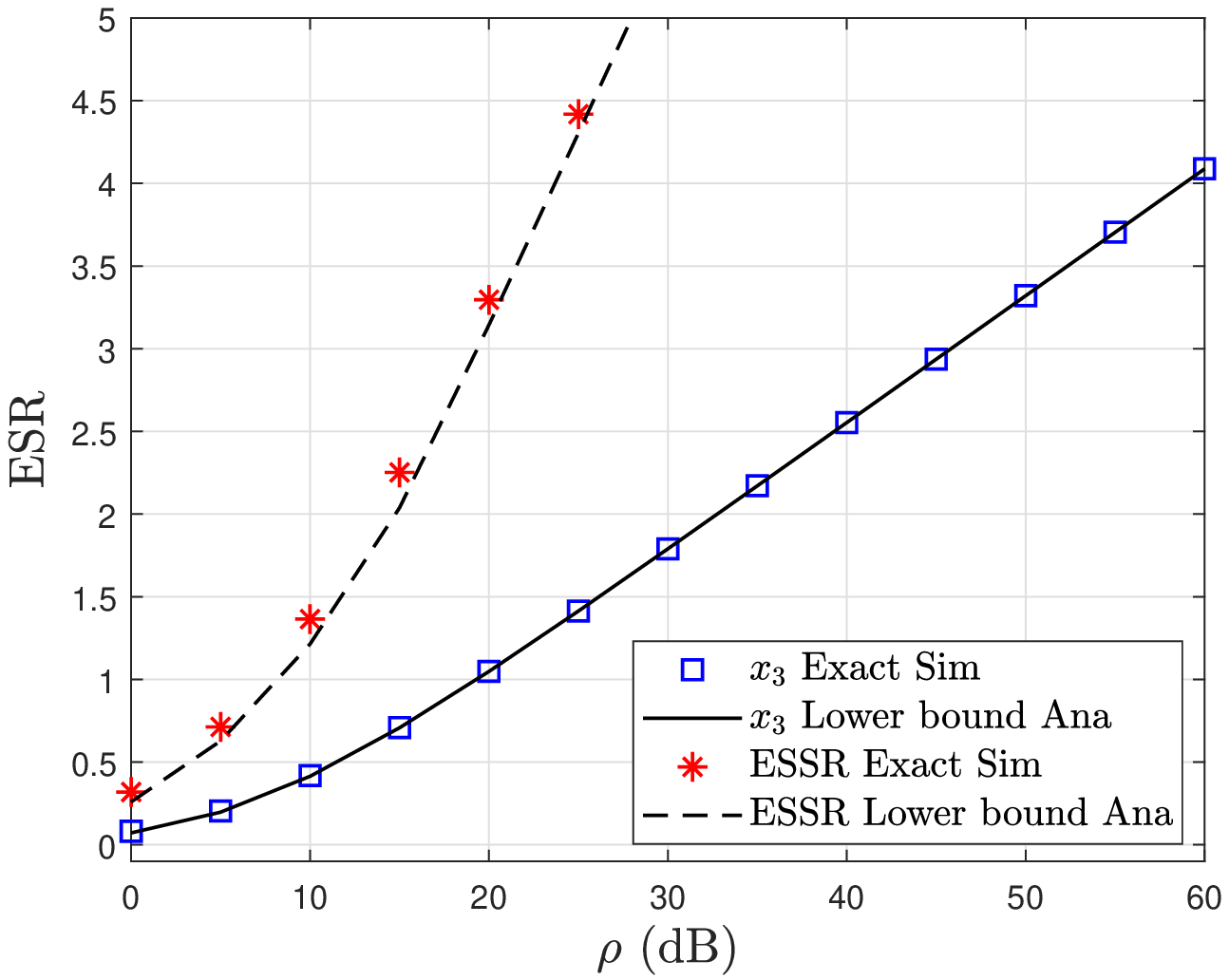}}
	\subfigure[]{
		\label{fig24}
		\includegraphics[width = 0.319 \textwidth]{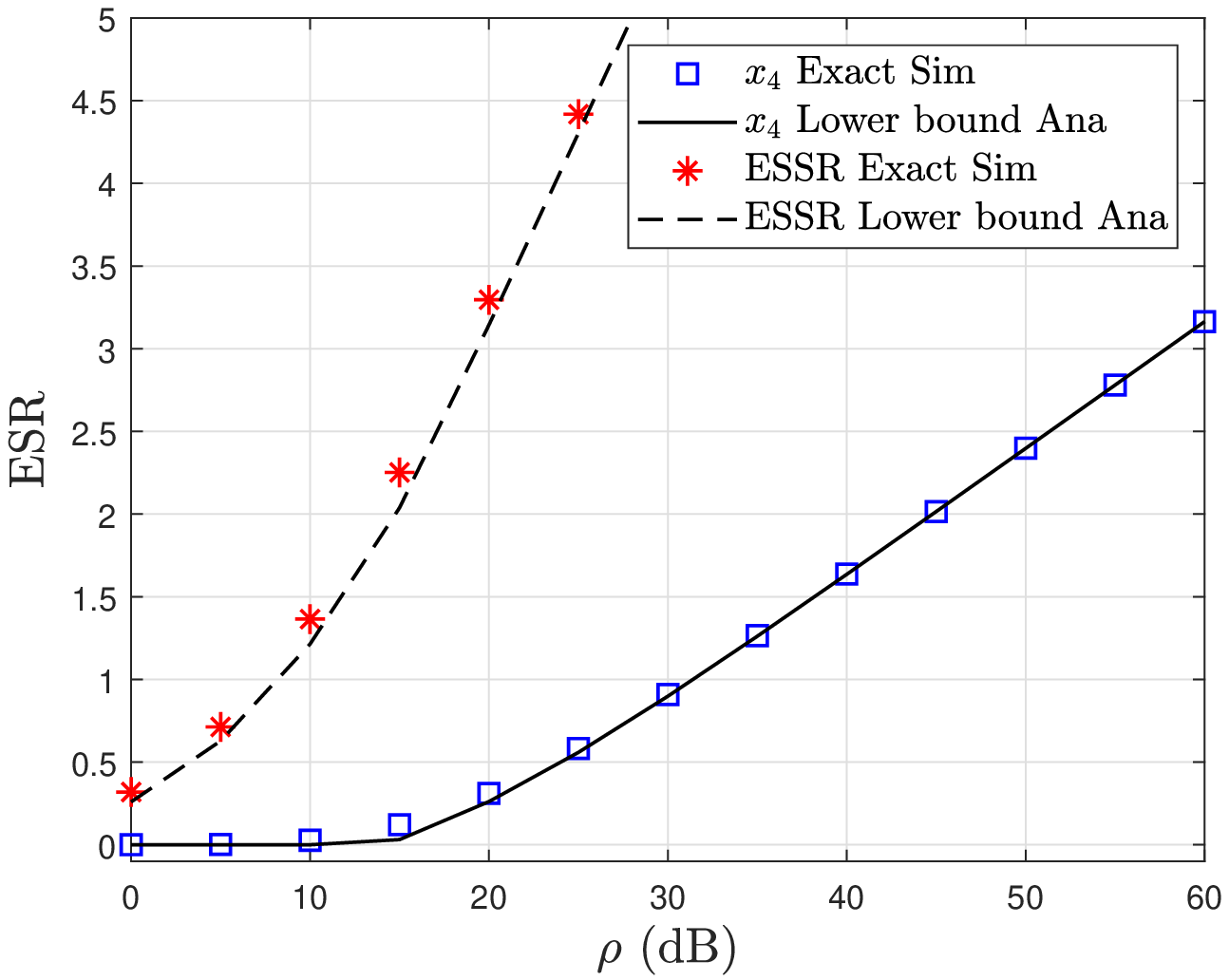}}
	\subfigure[]{
		\label{fig25}
		\includegraphics[width = 0.319 \textwidth]{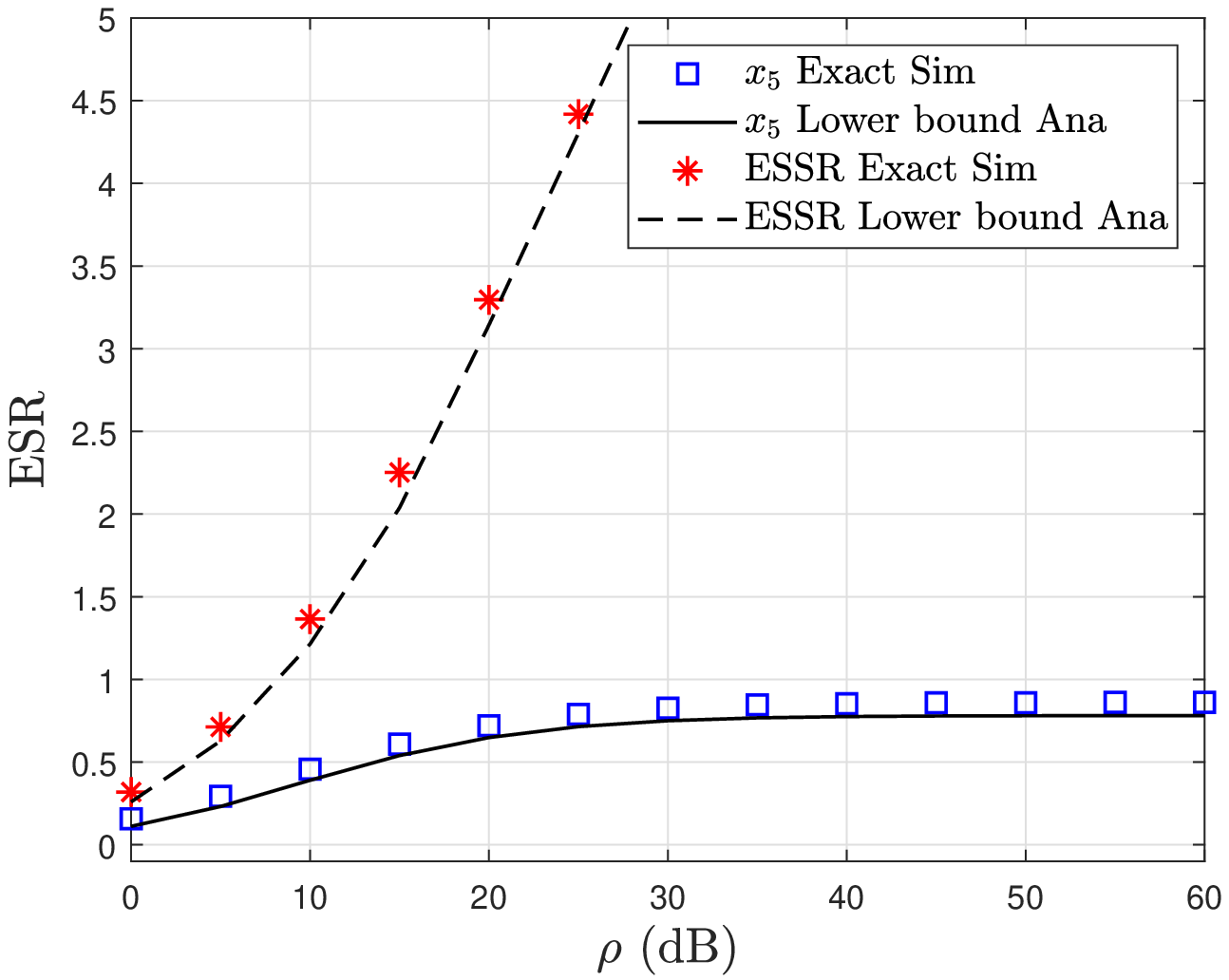}}
	\subfigure[]{
		\label{fig26}
		\includegraphics[width = 0.319 \textwidth]{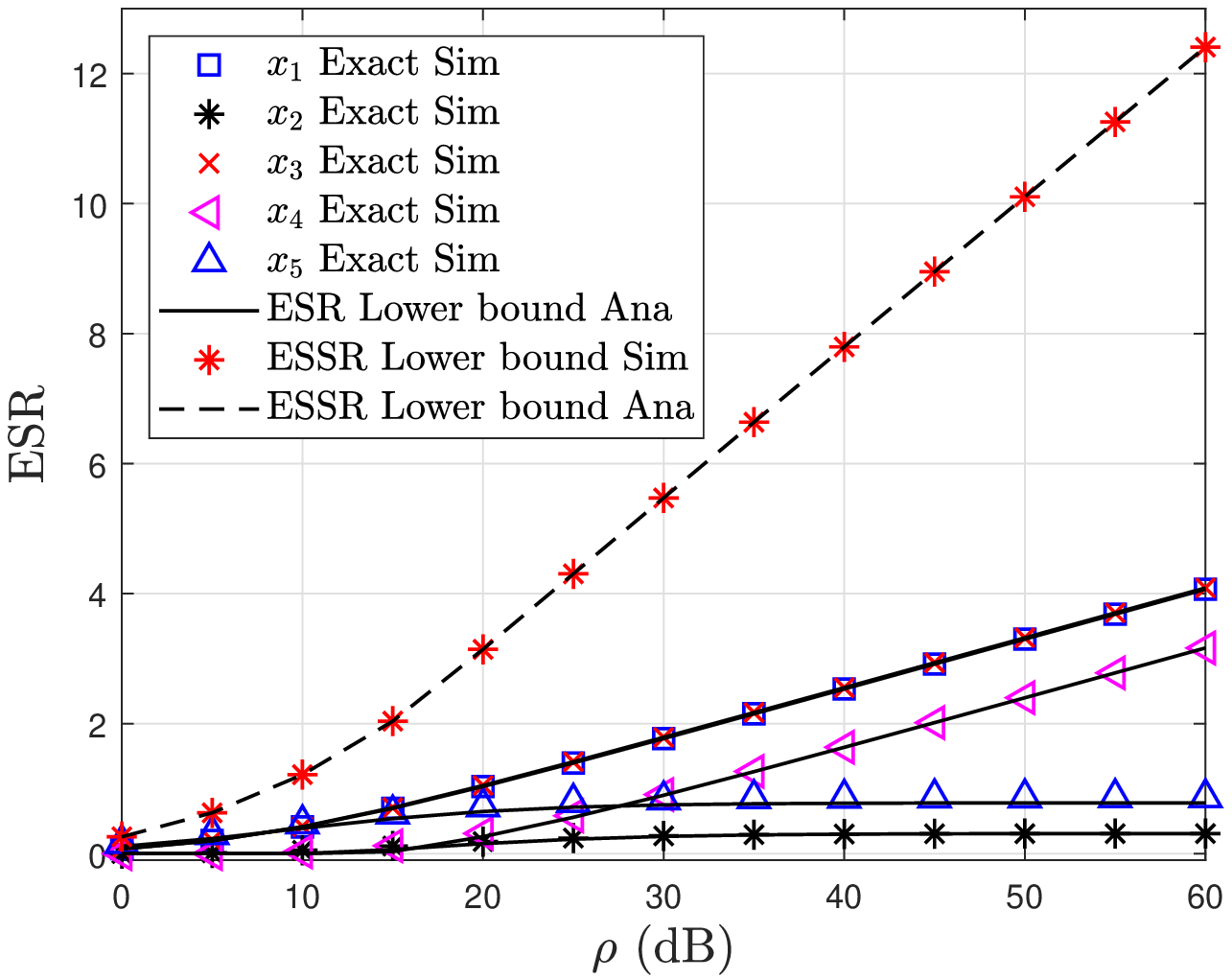}}
	\caption{ESRs and ESSR versus $\rho$.}
	\label{fig2}
\end{figure}

\section{Simulation Results and Discussions}
\label{sec: RESULTS}

In this section, simulation and analysis results are presented to validate the effectiveness of the proposed secure scheme.
The effects of system parameters on secure performance of the considered system, such as transmit SNR, average channel gains, power ratio coefficient, and power allocation coefficients, are investigated.
Without loss of generality, the parameters in the simulation and analysis are set as ${{\lambda _{S{U_N}}} = {\lambda _{R{U_F}}} = 1}$, ${{\lambda _{SR}} = 0.7}$, ${{\lambda _{R{U_N}}} = {\lambda _{{U_N}{U_F}}} = 0.8}$,  ${\nu   = 2}$, ${a_S = 0.2}$, and ${a_1^{{t_2}} = 0.5}$.
`Sim', `Ana', and `Asy' in all the figures denote the simulation, numerical, and asymptotic results.
Since there are five ESRs in the expression of ESSR, we present the effect of each system parameter on the ESRs and ESSR.
To prove the superiority of the proposed scheme, two conventional schemes are considered as benchmarks:

\begin{enumerate}
\item A CDRT system based on NOMA without PNC (`Ben1'):
$S$ broadcasts downlink superimposed signals ${\sqrt {{a_S}{P_S}} {x_1} + \sqrt {{{\bar a}_S}{P_S}} {x_2}}$ and $U_F$ emits jamming ${\sqrt {{P_U}} {z_1}}$ in the first slot $\left( {{t_1}} \right)$, $R$ amplifies received signals and forwards to $U_F$ in the second slot $\left( {{t_2}} \right)$, $U_N$ and $U_F$ broadcast uplink NOMA signals ${\sqrt {a_1^{{t_2}}{P_U}} {x_3} + \sqrt {\bar a_1^{{t_2}}{P_U}} {z_2}}$ and ${\sqrt {{P_U}} {x_4}}$ in the third slot $\left( {{t_3}} \right)$, and in final slot  $\left( {{t_4}} \right)$ $U_N$ sends a new uplink signal $x_5$, meanwhile $R$ amplifies received signals $x_4$.

\item A CDRT system based on NOMA and PNC without jamming signal (`Ben2'):
There are three slots in each fading block. $S$ broadcasts superimposed signals ${\sqrt {{a_S}{P_S}} {x_1} + \sqrt {{{\bar a}_S}{P_S}} {x_2}}$ in the first slot $\left( {{t_1}} \right)$, $U_N$ and $U_F$ transmit ${\sqrt {{P_U}} {x_3}}$ and ${\sqrt {{P_U}} {x_4}}$ in the second slot $\left( {{t_2}} \right)$, and in the third slot $\left( {{t_3}} \right)$ $R$ amplifies and broadcasts the received signals in previous two time slots and $U_N$ transmits ${\sqrt {{P_U}} {x_5}}$ to $S$.
\end{enumerate}

Fig. \ref{fig2} shows simulation results for the exact ESR and numerical results for the lower bound of ESR versus $\rho$.
It is found that simulation and lower bound results of $x_1$ - $x_4$ match perfectly, which represent approximate the ESR of the considered system by using the lower bound is feasible.

\begin{figure}[t]
	\centering
	\subfigure[]{
		\label{fig31}
		\includegraphics[width = 0.319 \textwidth]{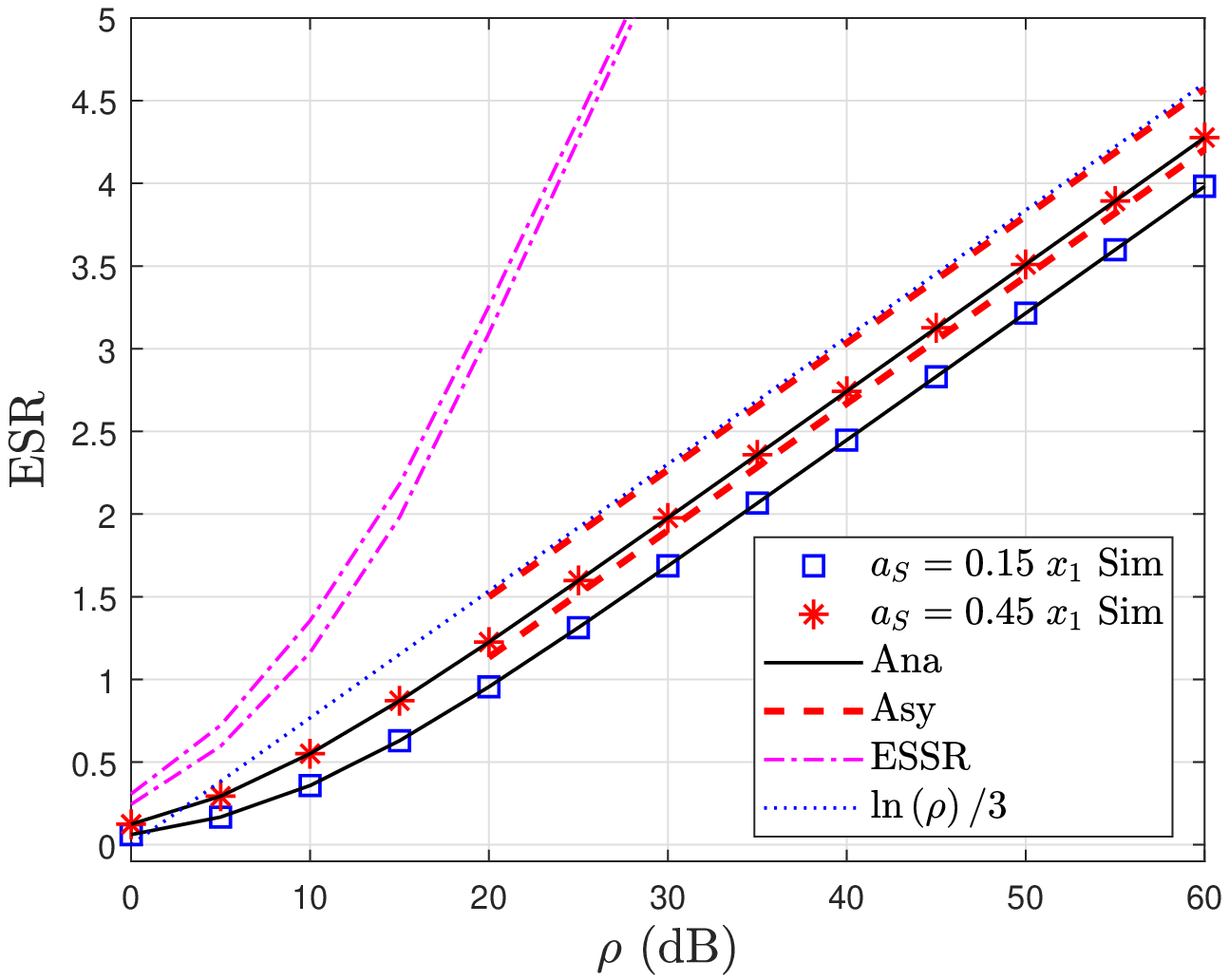}}
	\subfigure[]{
		\label{fig32}
		\includegraphics[width = 0.319 \textwidth]{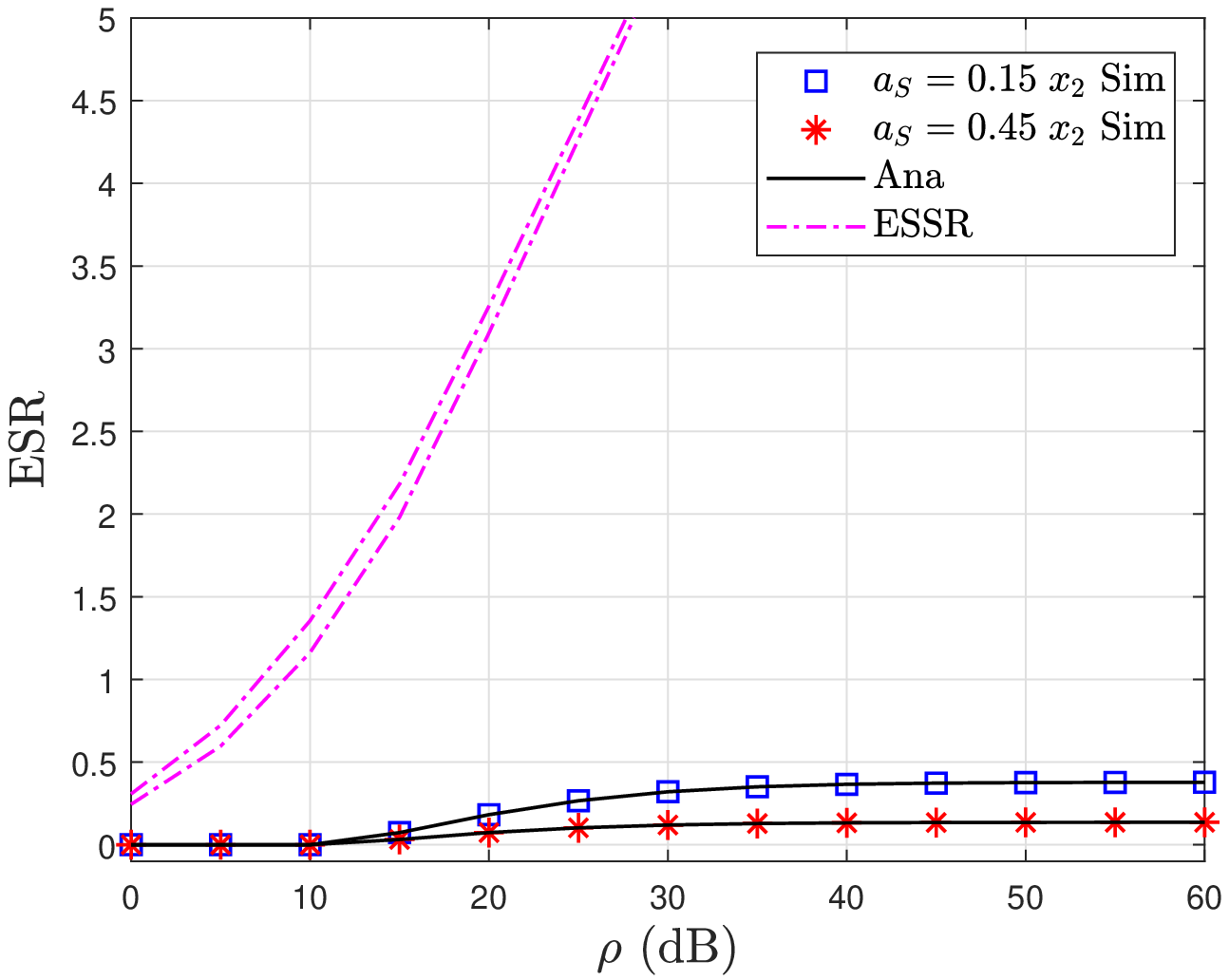}}
	\subfigure[]{
		\label{fig33}
		\includegraphics[width = 0.319 \textwidth]{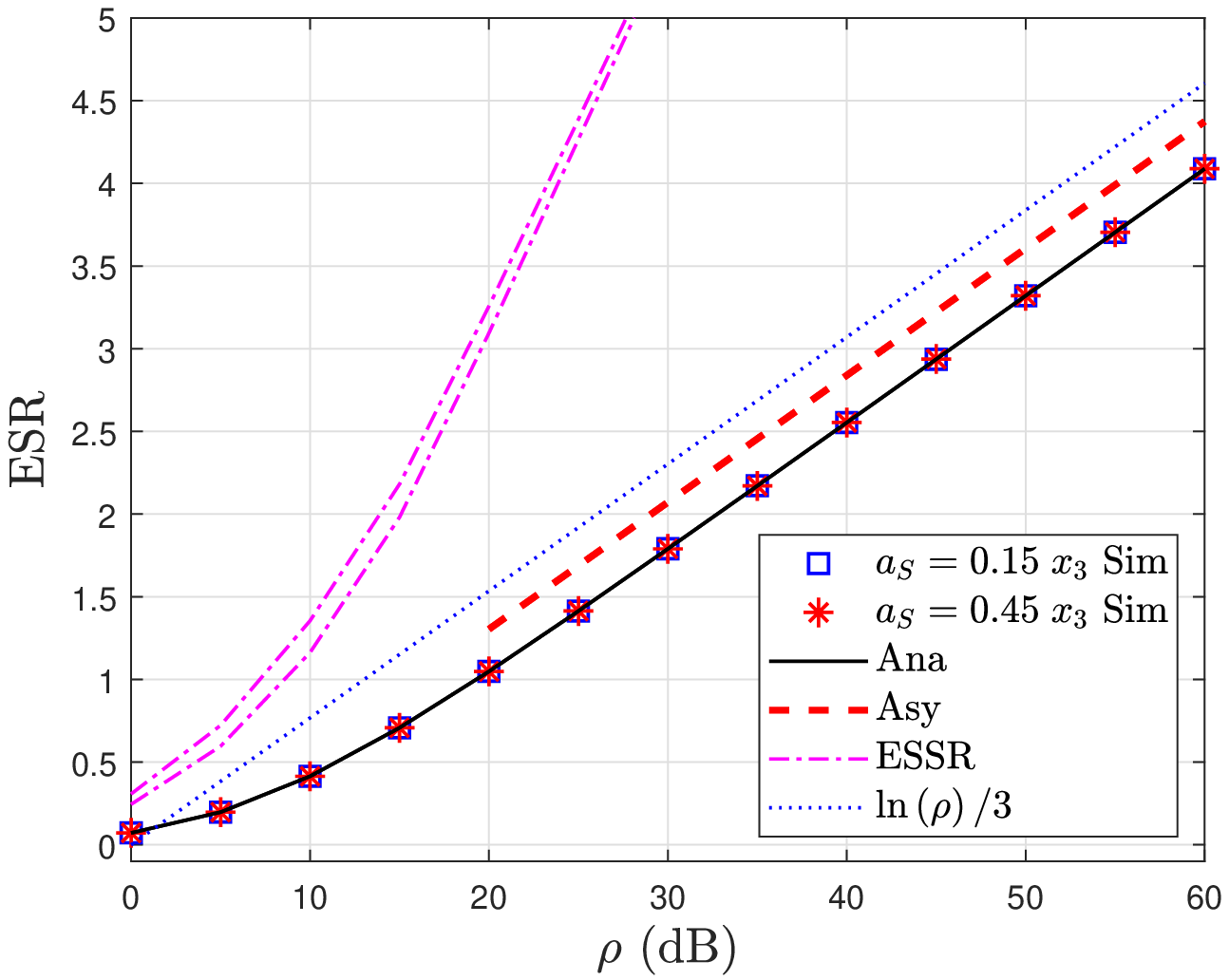}}
	\subfigure[]{
		\label{fig34}
		\includegraphics[width = 0.319 \textwidth]{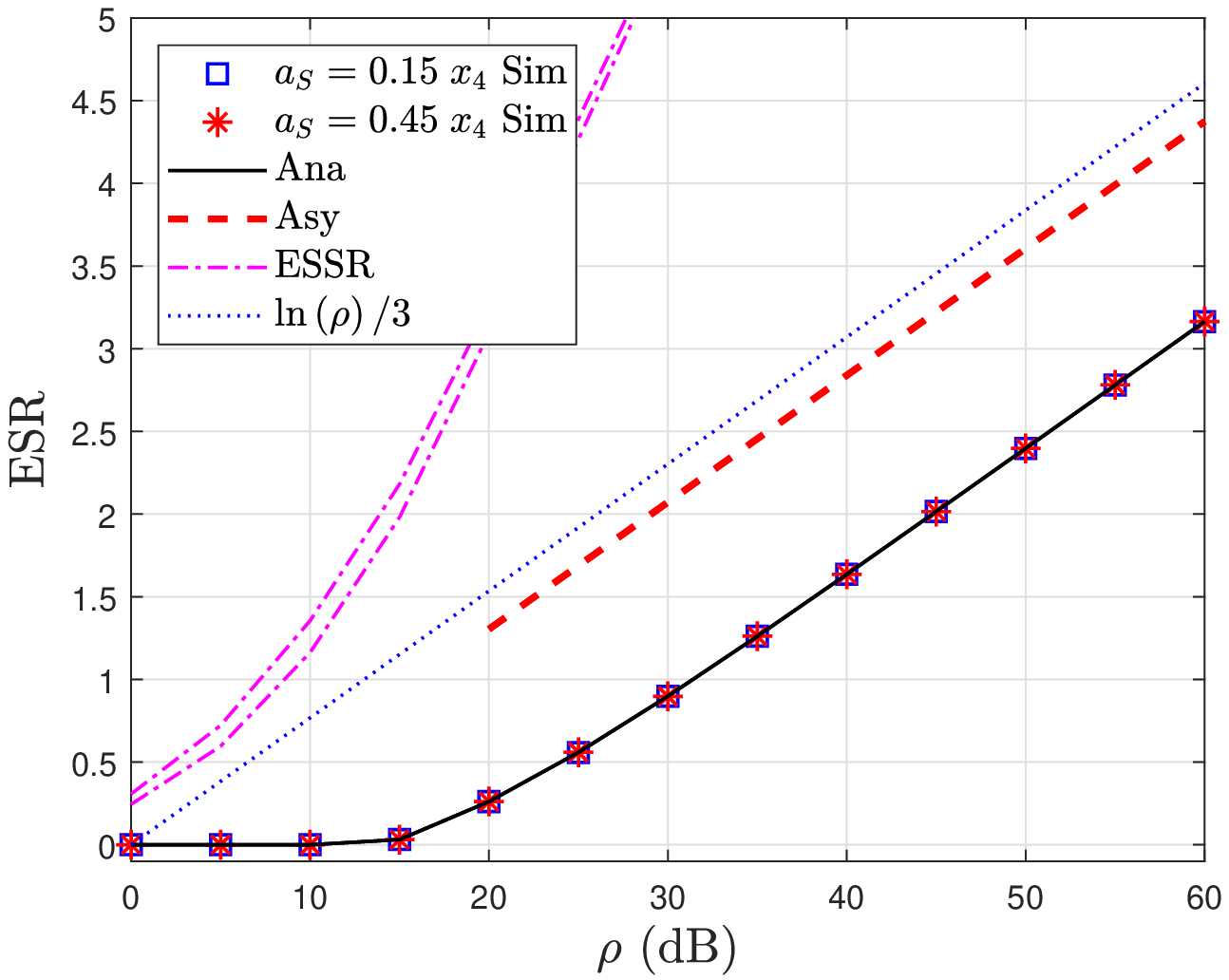}}
	\subfigure[]{
		\label{fig35}
		\includegraphics[width = 0.319 \textwidth]{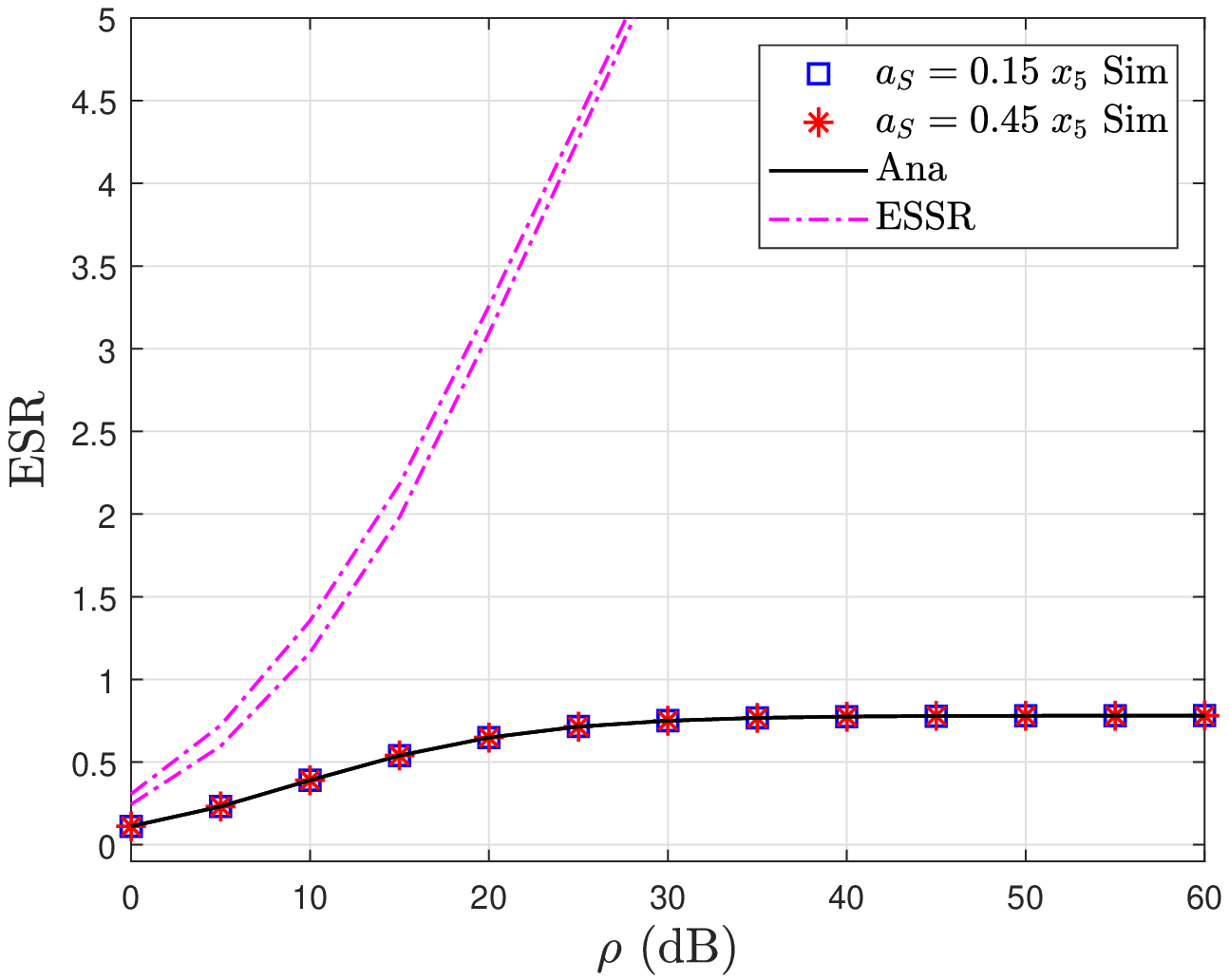}}
	\subfigure[]{
		\label{fig36}
		\includegraphics[width = 0.319 \textwidth]{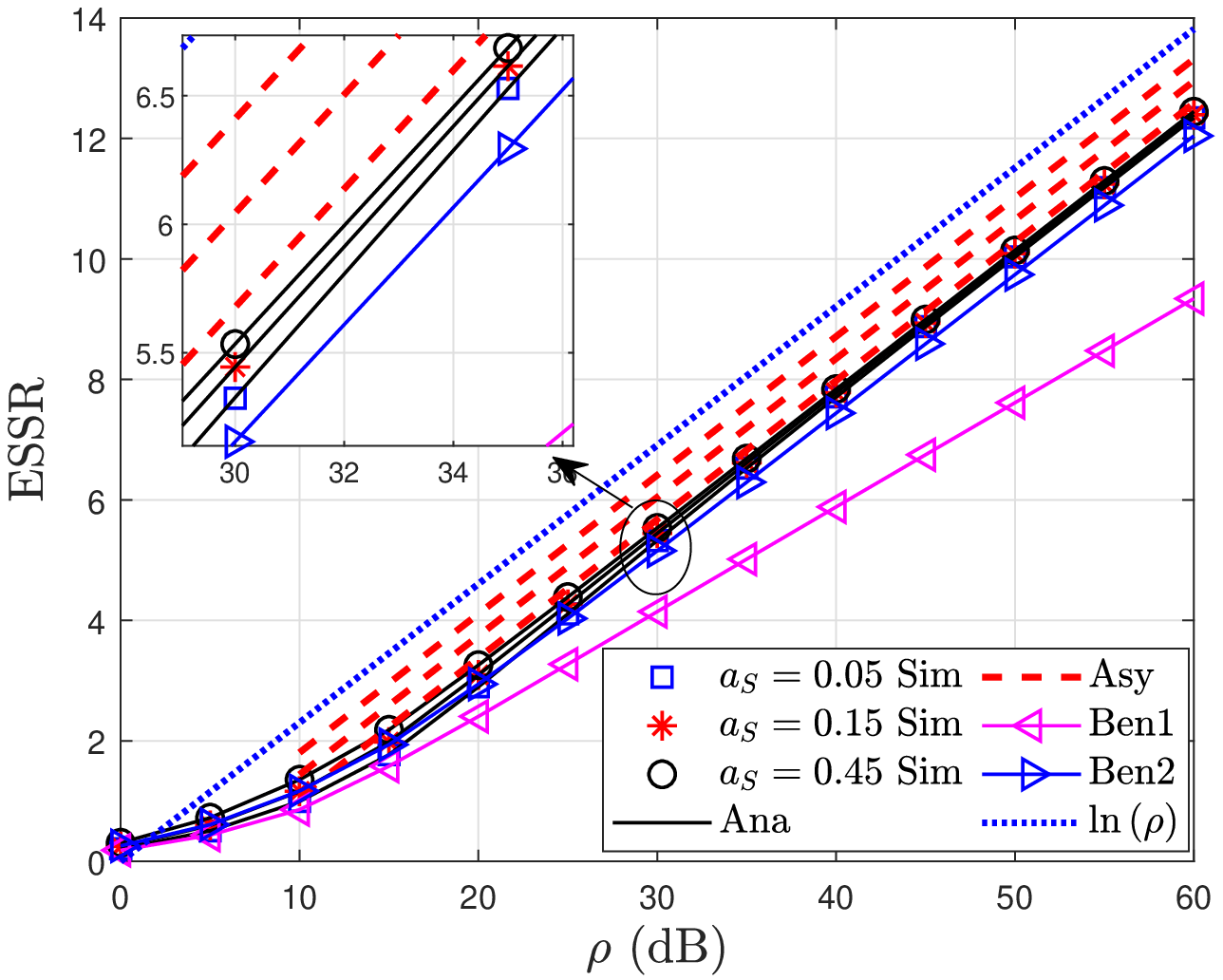}}
	\caption{ESRs and ESSR for varying ${\rho}$ and ${a_S}$.}
	\label{fig3}
\end{figure}
Fig. \ref{fig3} plots the effect of $\rho$ and $a_S$ on the simulation results for the lower bound of ESRs and ESSR. One can observe that simulation and numerical results match perfectly to verify the correctness of the analysis, and it is apparent that ESSR increases with increasing $\rho$ and $a_S$.
This is because the legitimate rates always increase and the eavesdropping rates initially increase and subsequently remain constant with increasing $\rho$. Different from $\rho$, $a_S$ only affects the ESR of $x_1$ and $x_2$.
The ESR of $x_1$ gets better and remain increasing with increasing $a_S$, correspondingly the ESR of $x_2$ gets worse with increasing $a_S$. Due to poor channel gain and the presence of inter-user interference, the slope of the decrease in the ESR of $x_2$ is less than the slope of the increase in the ESR of $x_1$, thus the ESSR increases with increasing $a_S$.

\begin{figure}[t]
	\centering
	\subfigure[]{
		\label{fig41}
		\includegraphics[width = 0.319 \textwidth]{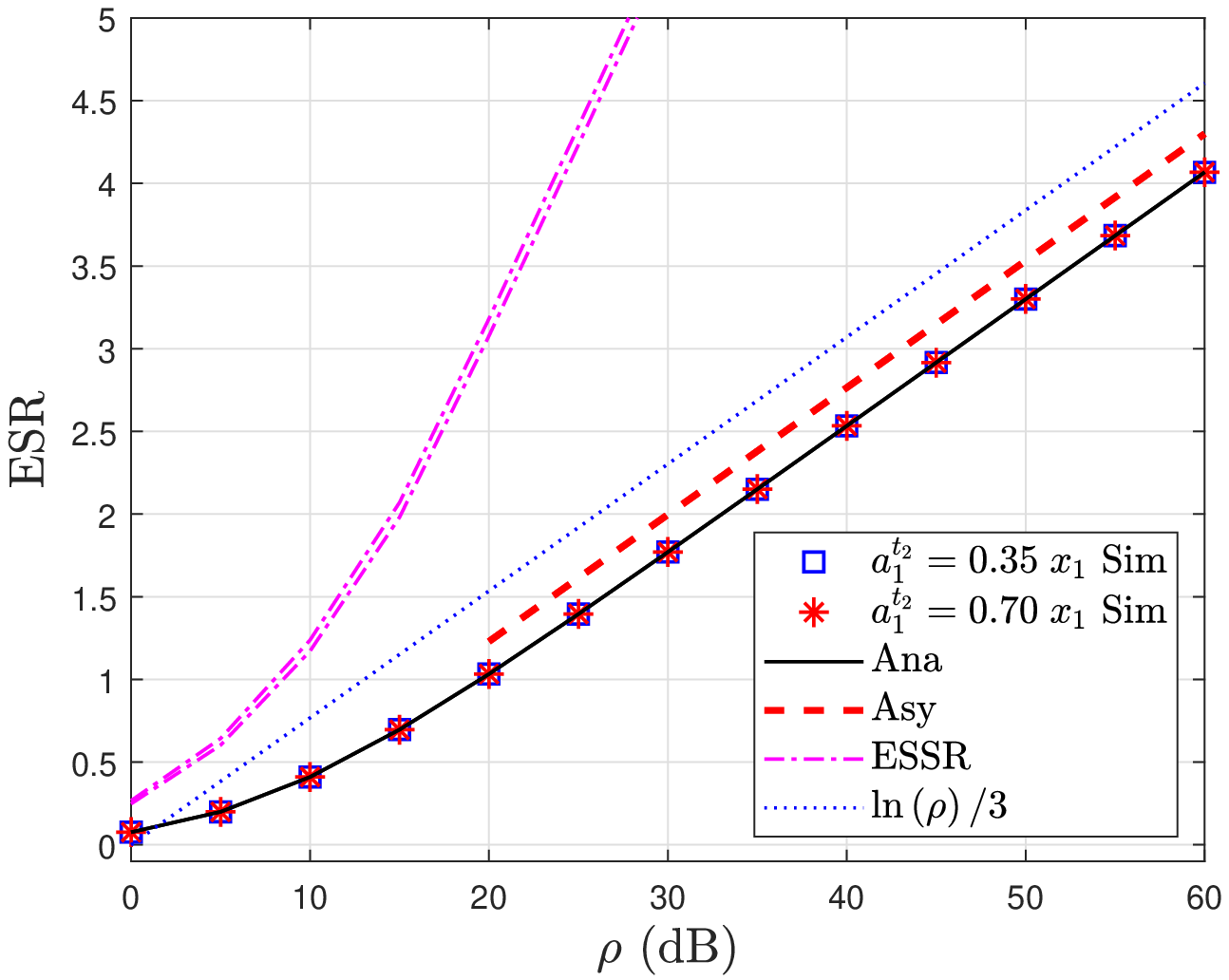}}
	\subfigure[]{
		\label{fig42}
		\includegraphics[width = 0.319 \textwidth]{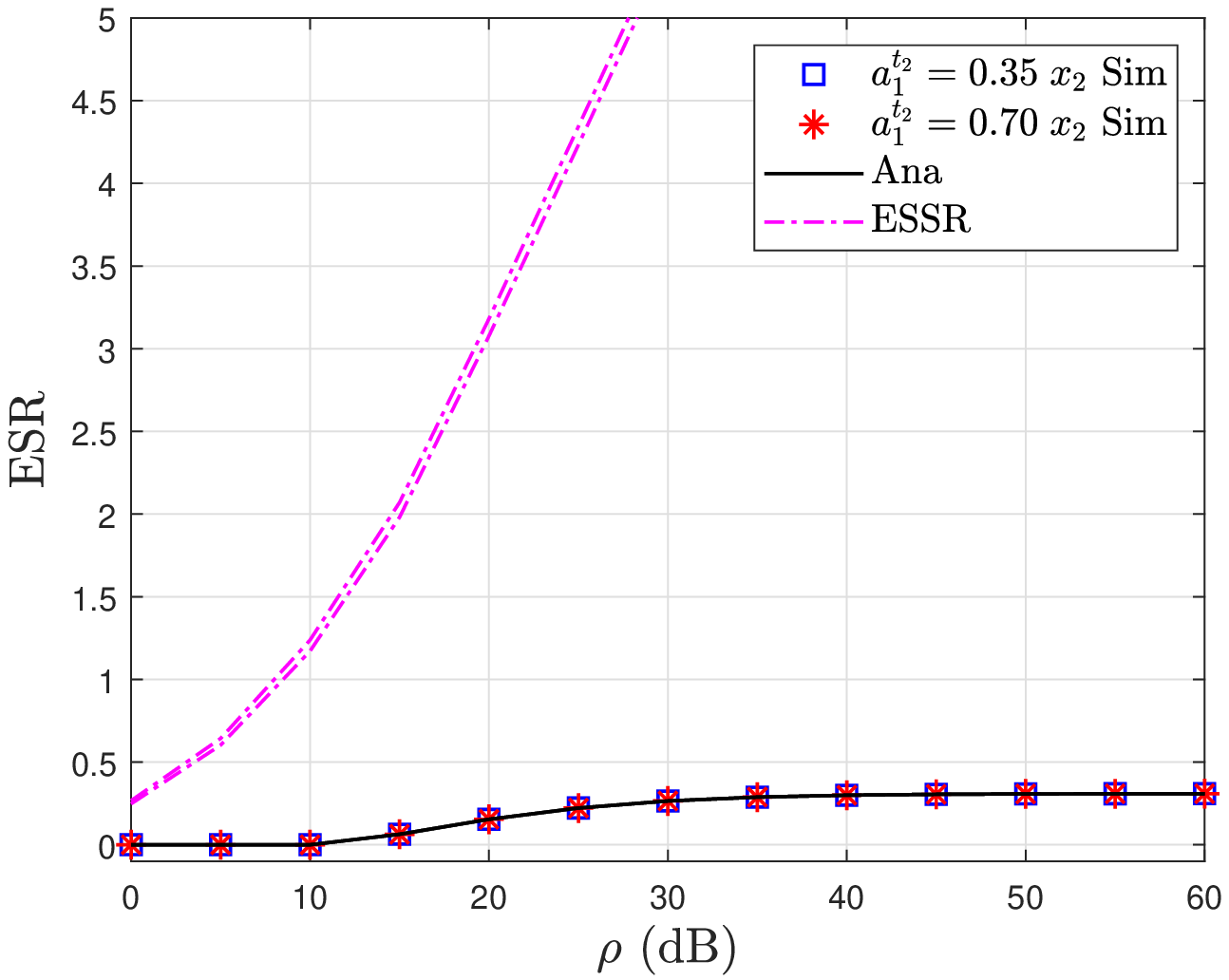}}
	\subfigure[]{
		\label{fig43}
		\includegraphics[width = 0.319 \textwidth]{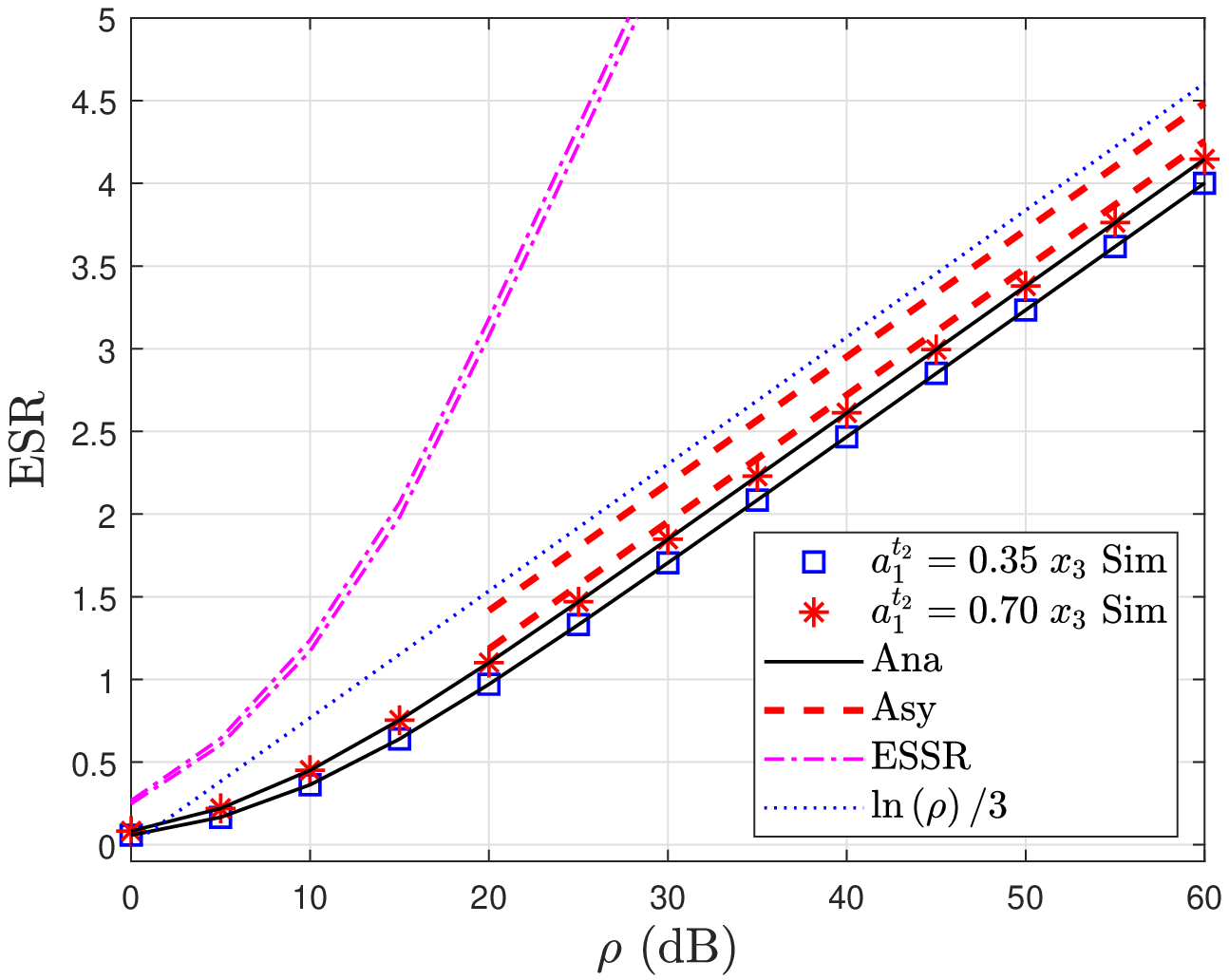}}
	\subfigure[]{
		\label{fig44}
		\includegraphics[width = 0.319 \textwidth]{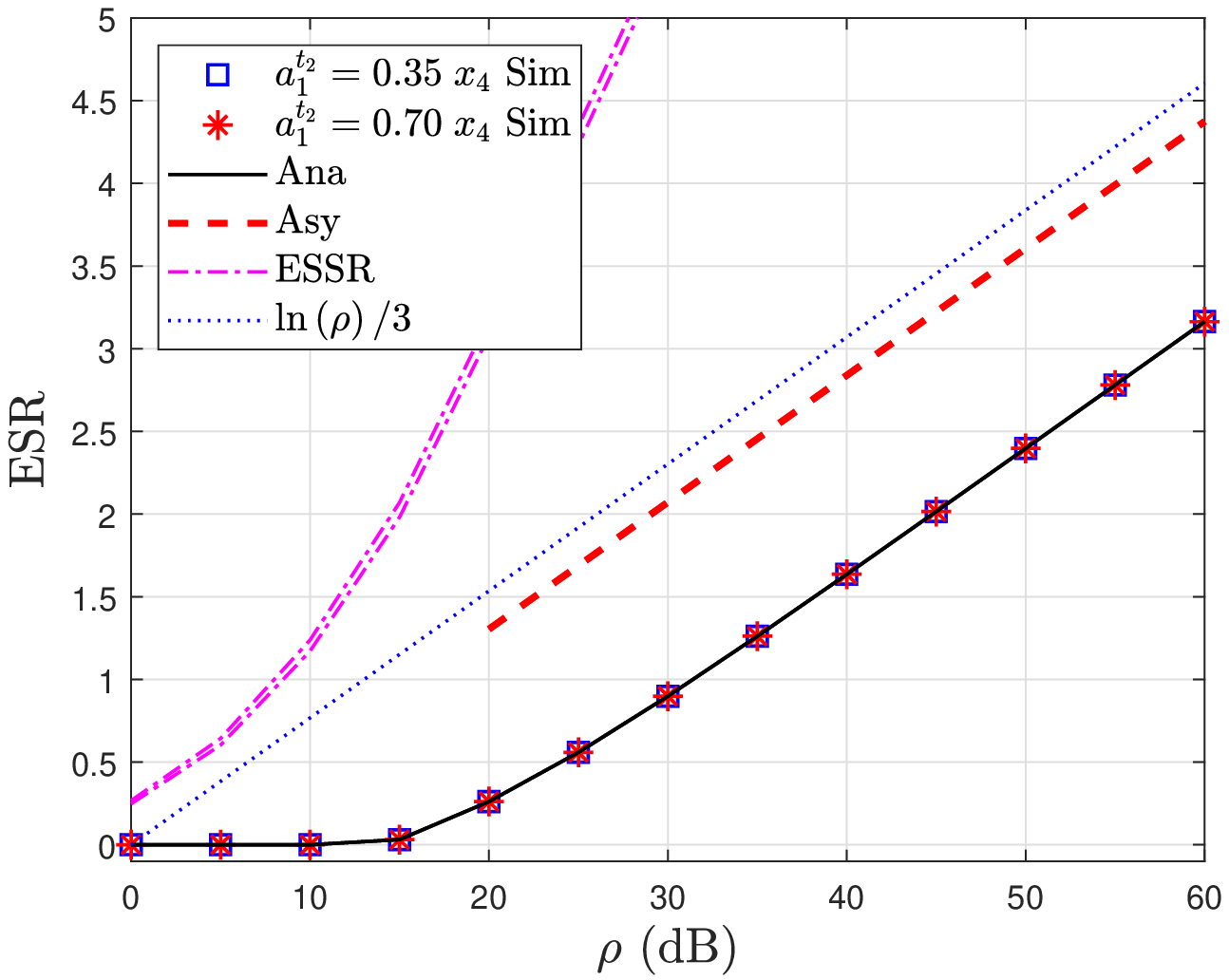}}
	\subfigure[]{
		\label{fig45}
		\includegraphics[width = 0.319 \textwidth]{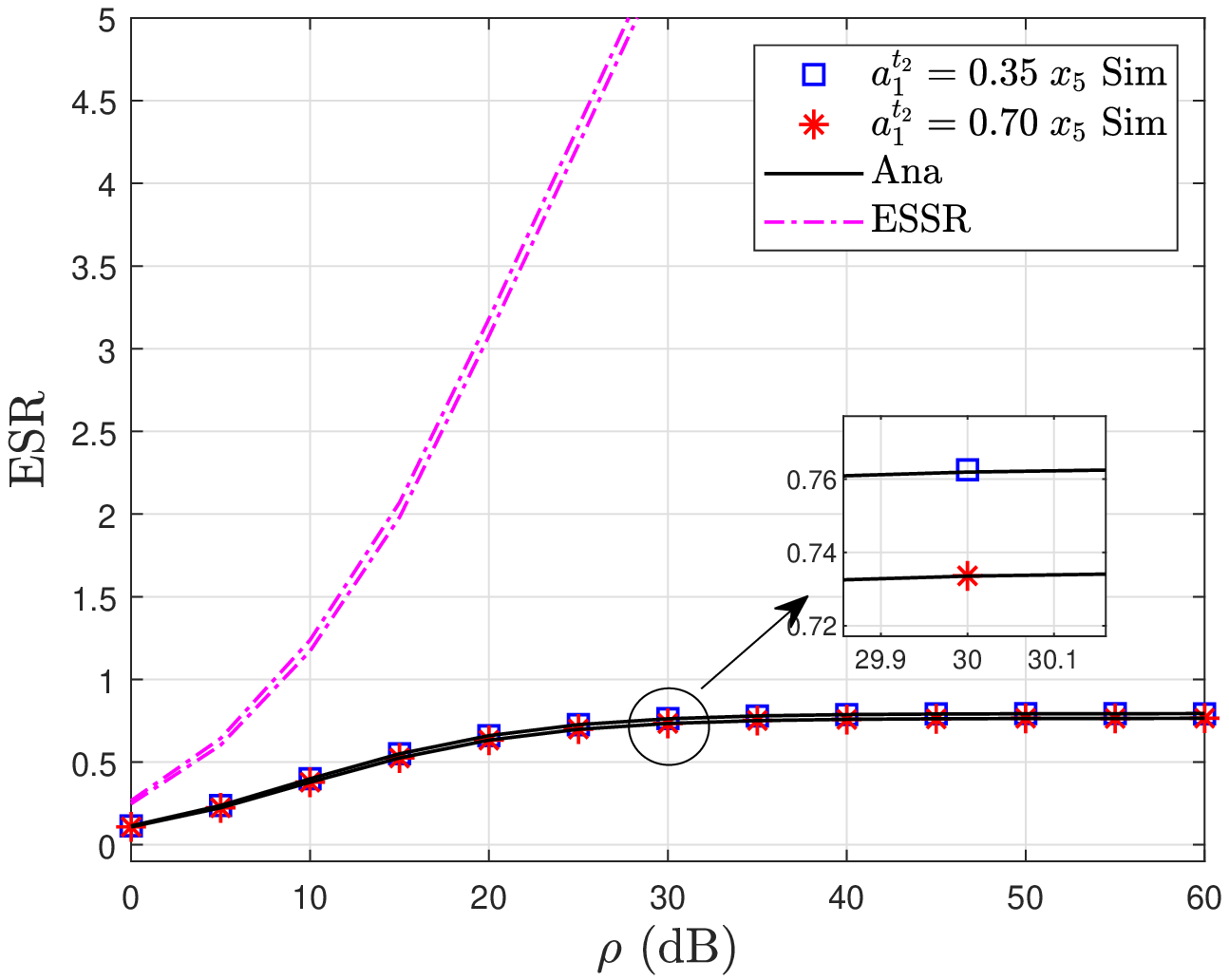}}
	\subfigure[]{
		\label{fig46}
		\includegraphics[width = 0.319 \textwidth]{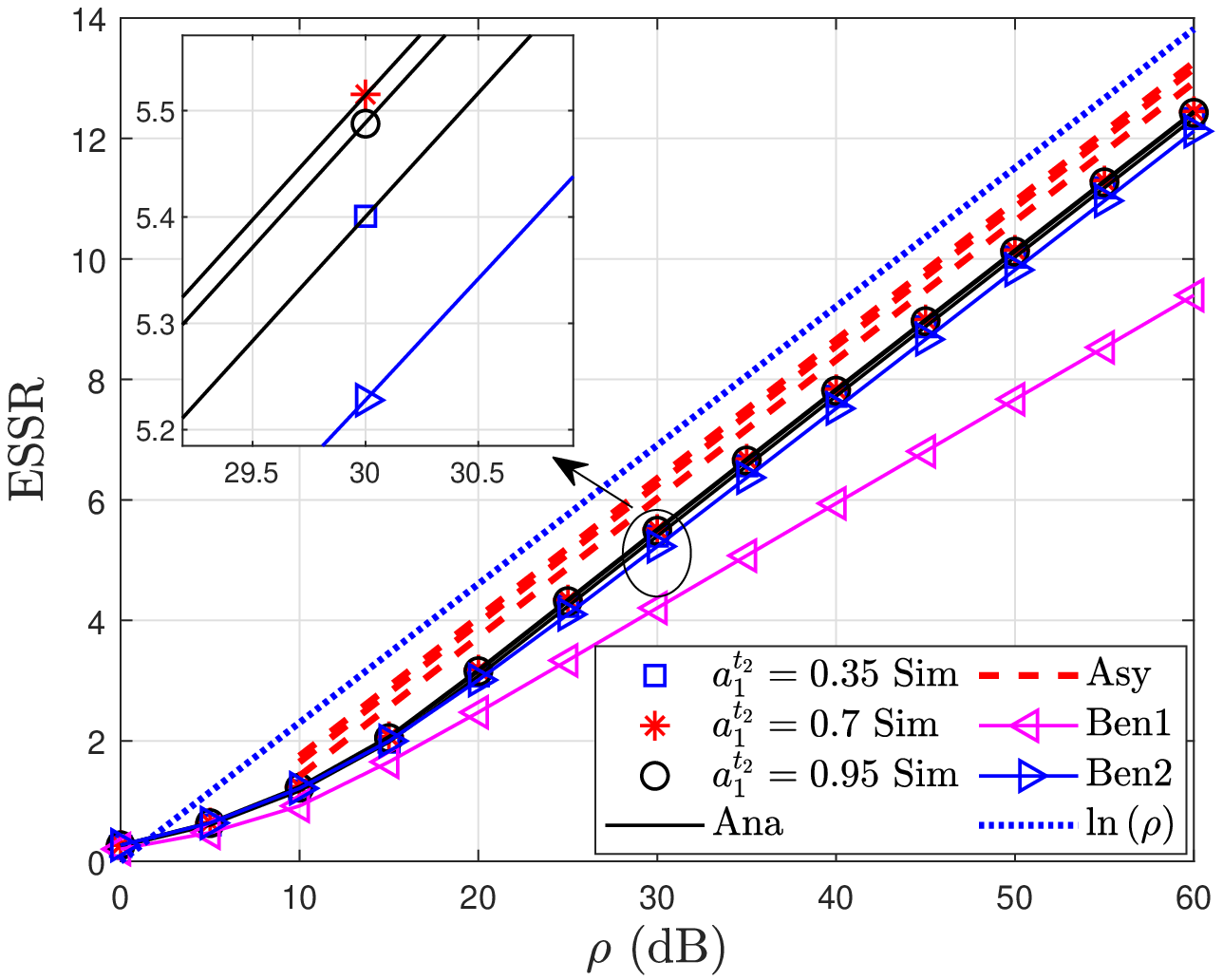}}
	\caption{ESRs and ESSR for varying ${\rho}$ and ${a_1^{{t_2}}}$.}
	\label{fig4}
\end{figure}
Fig. \ref{fig4} demonstrates the lower bound of ESRs and ESSR vs transmit SNR $\rho$ for varying ${a_1^{{t_2}}}$.
It can be observed that the ESSR of the proposed scheme increases with an increase in $\rho$. Different from $a_S$, one can observe the ESSR increases initially and subsequently decreases with increasing ${a_1^{{t_2}}}$. This verifies that there is  an optimal ${a_1^{{t_2}}}$ to achieve the optimal performance, which confirms the result in {\textit{Remark 4}}.

\begin{figure}[t]
	\centering
	\subfigure[]{
		\label{fig51}
		\includegraphics[width = 0.319 \textwidth]{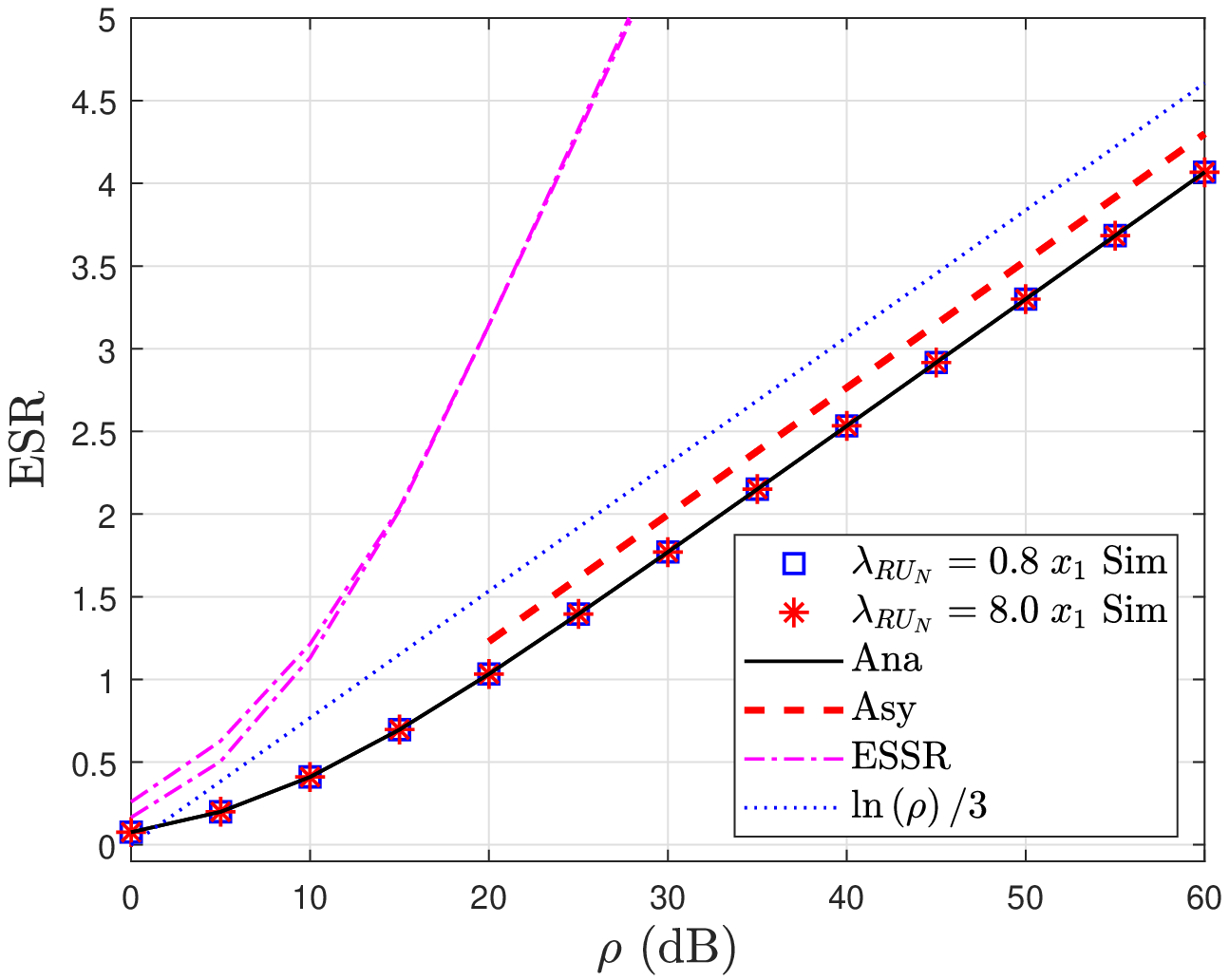}}
	\subfigure[]{
		\label{fig52}
		\includegraphics[width = 0.319 \textwidth]{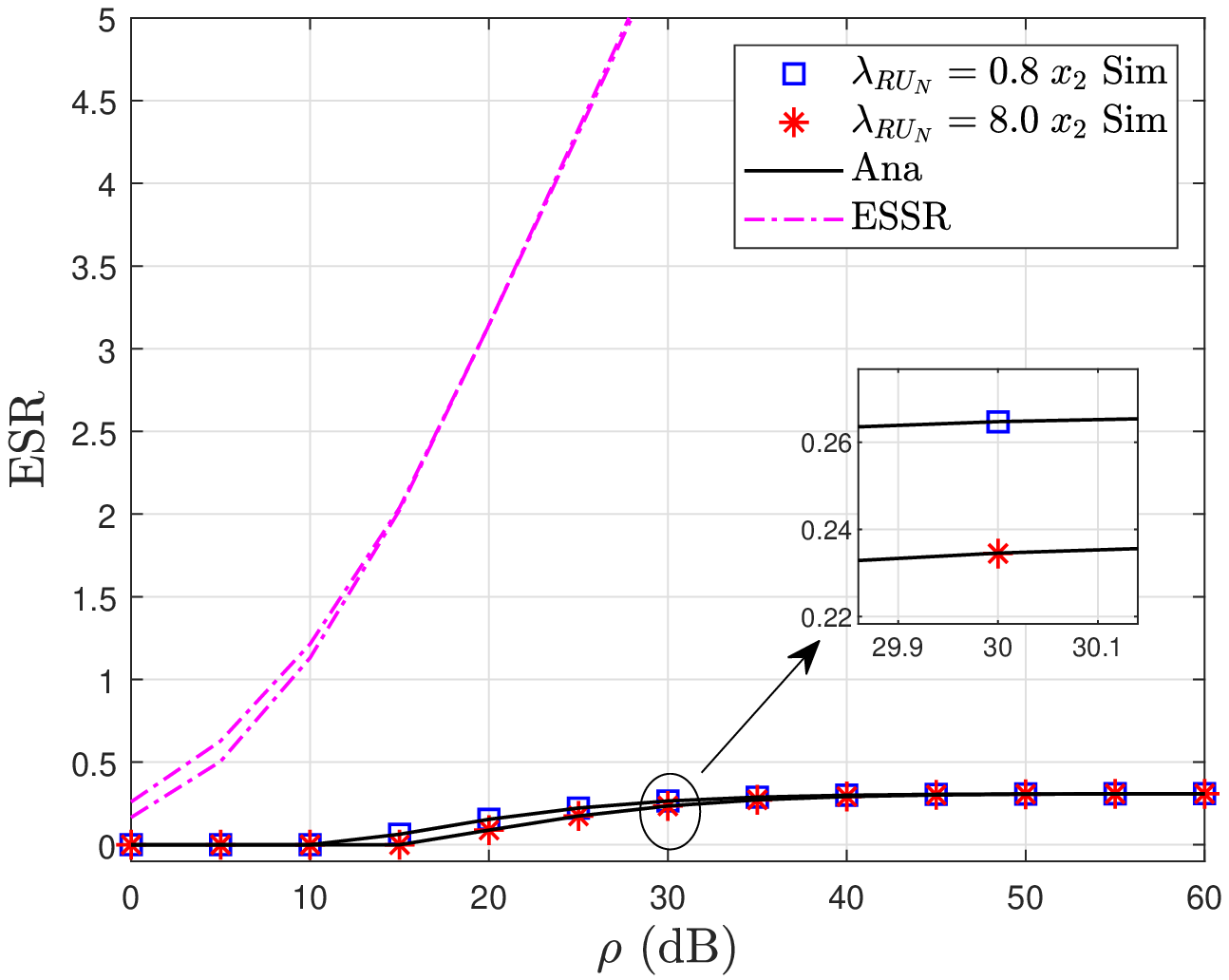}}
	\subfigure[]{
		\label{fig53}
		\includegraphics[width = 0.319 \textwidth]{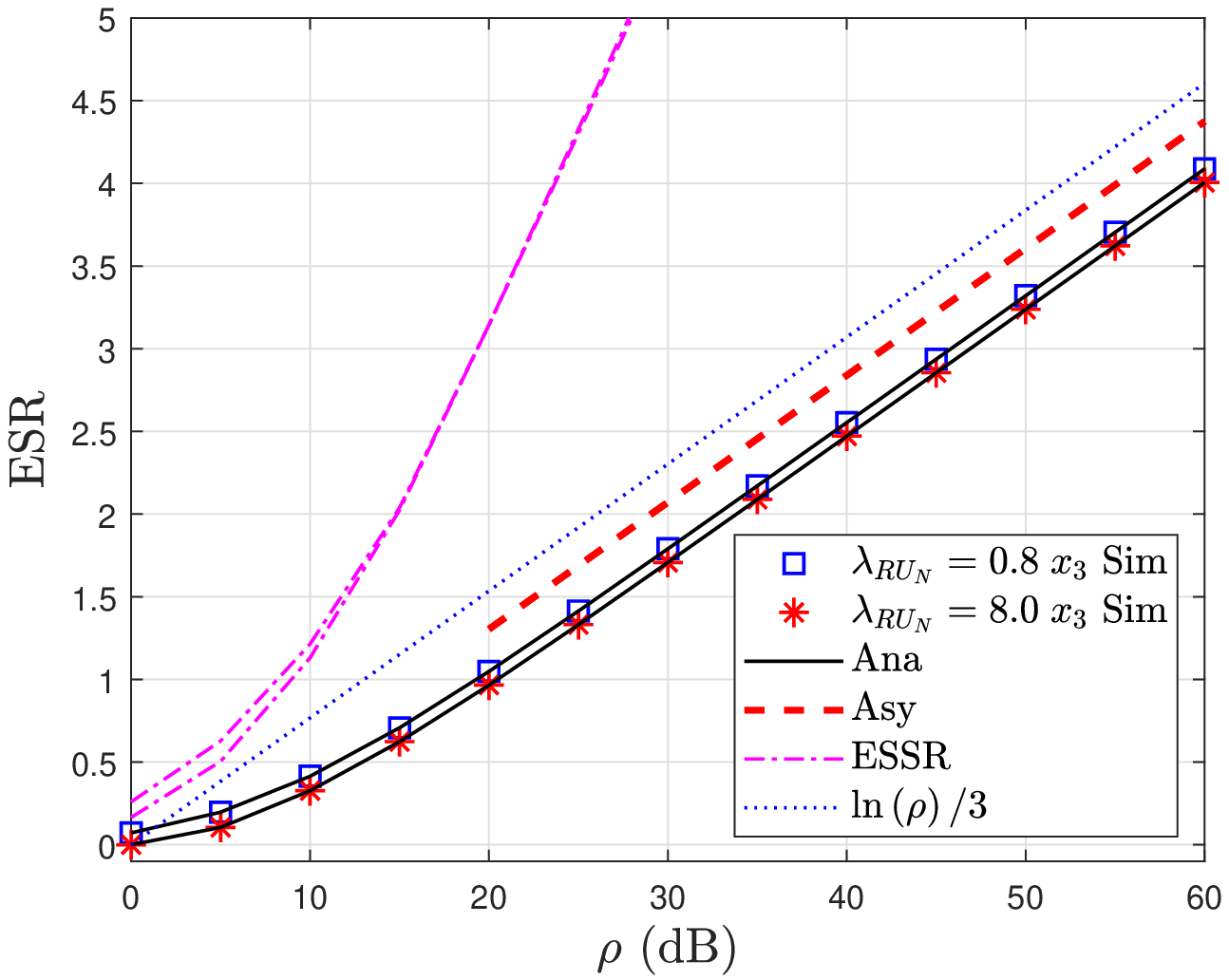}}
	\subfigure[]{
		\label{fig54}
		\includegraphics[width = 0.319 \textwidth]{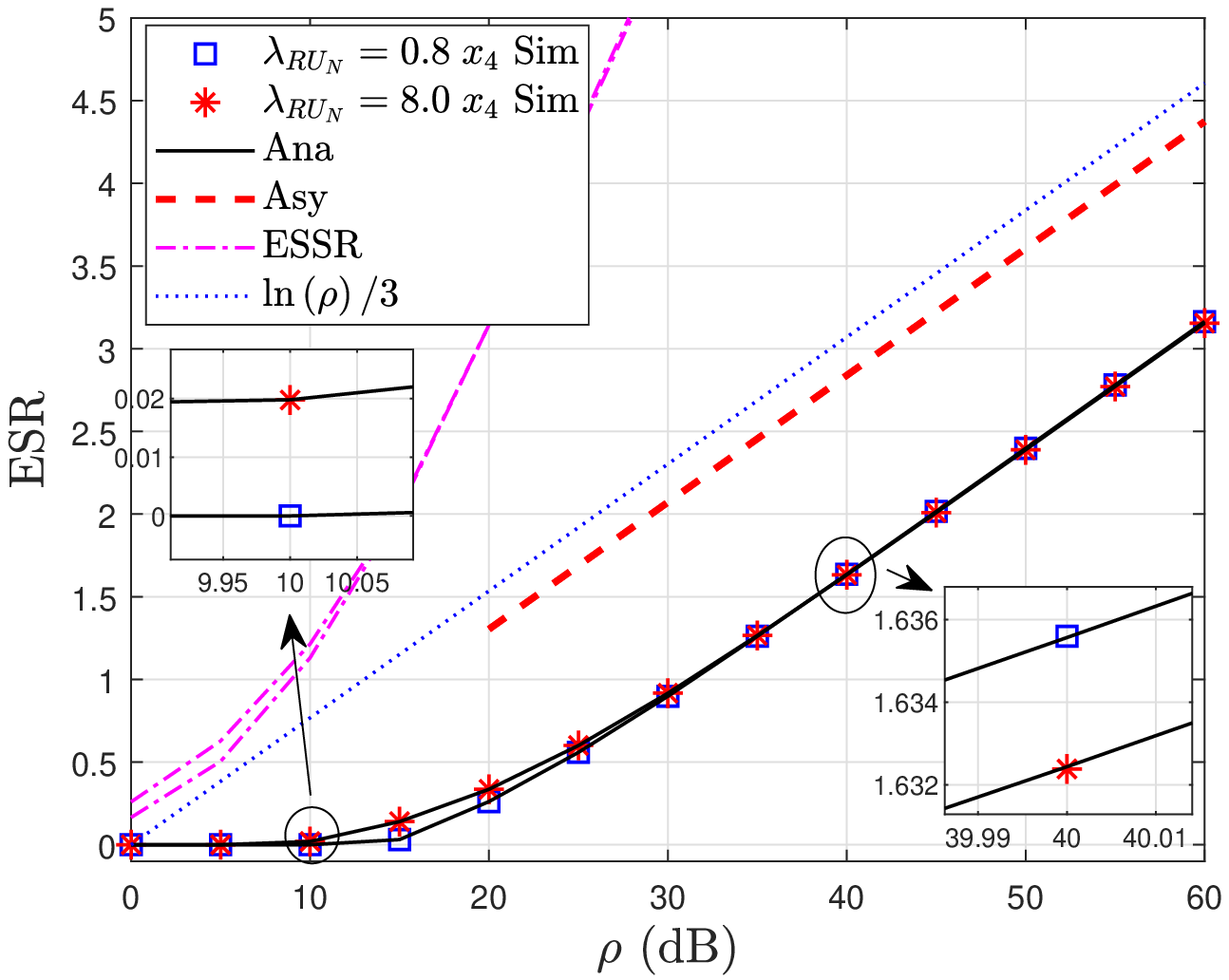}}
	\subfigure[]{
		\label{fig55}
		\includegraphics[width = 0.319 \textwidth]{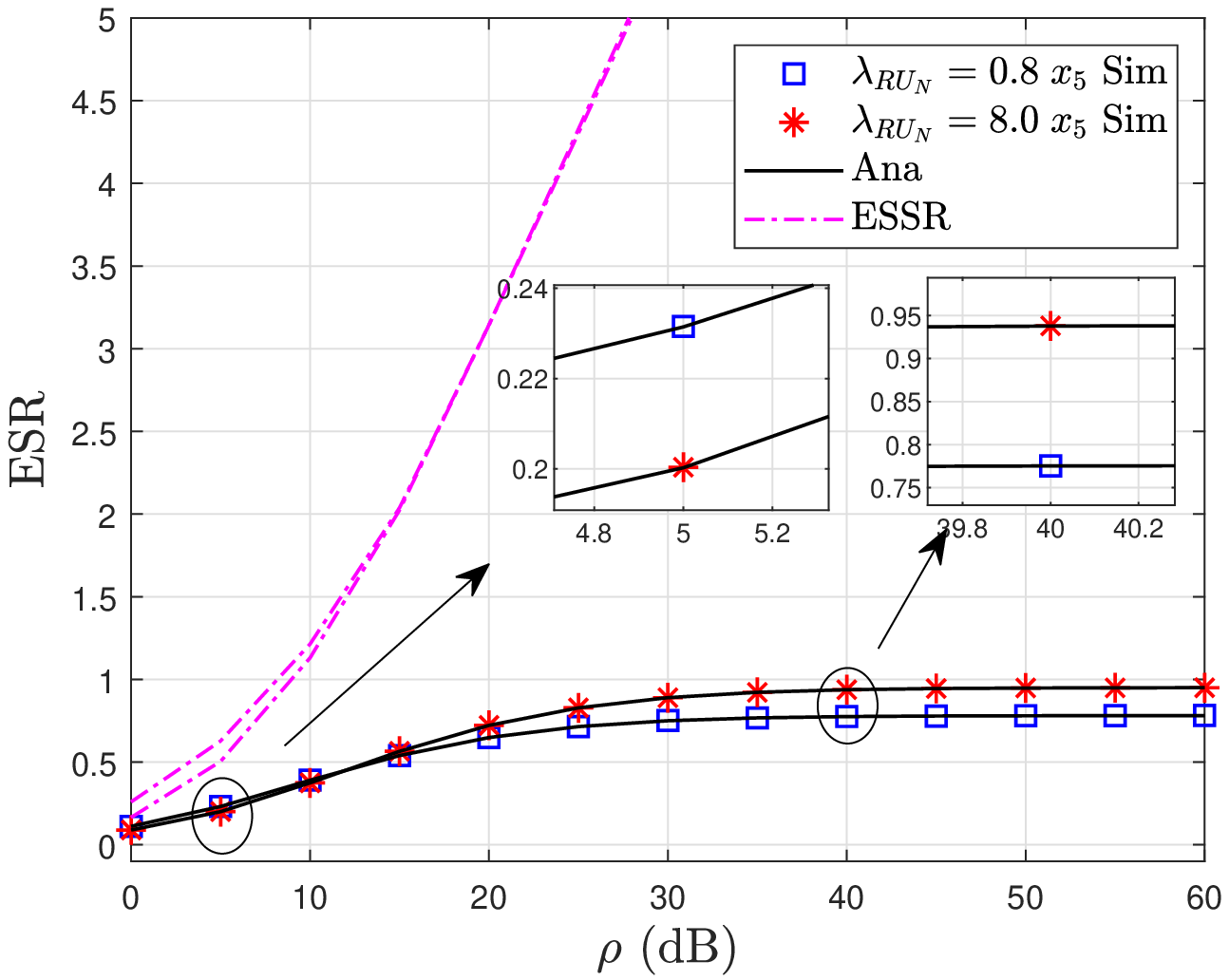}}
	\subfigure[]{
		\label{fig56}
		\includegraphics[width = 0.319 \textwidth]{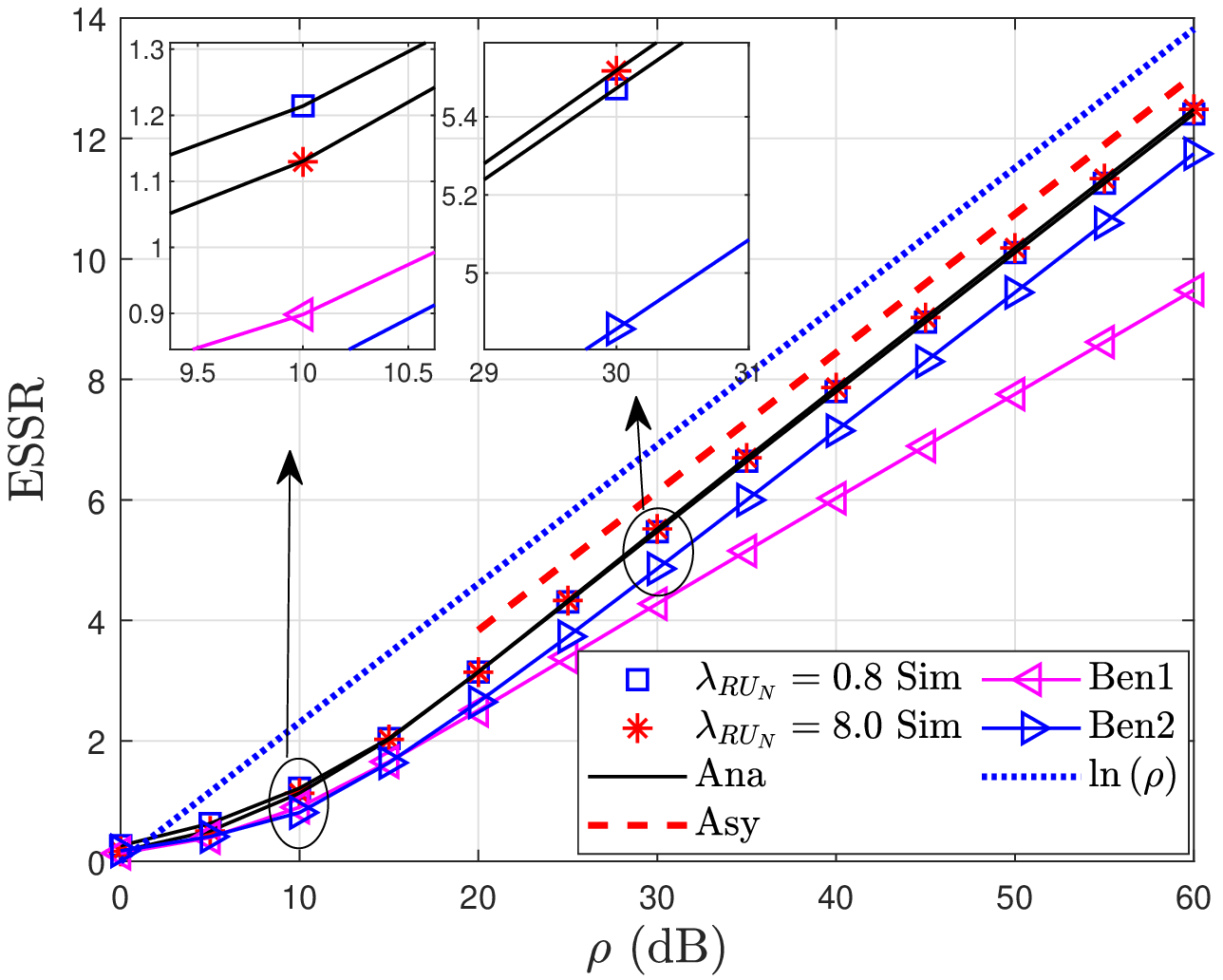}}
	\caption{ESRs and ESSR for varying ${\rho}$ and $\lambda _{R{U_N}}$.}
	\label{fig5}
\end{figure}
Fig. \ref{fig5} presents the impact of $\rho$ for varying $\lambda _{R{U_N}}$ on the lower bound of ESRs and ESSR.
We find that the effect of $\lambda _{R{U_N}}$ on the ESSR in the lower-$\rho$ region is different from that in the higher-$\rho$ region. The lower $\lambda _{R{U_N}}$, the larger ESSR in the lower-$\rho$ region while the lower $\lambda _{R{U_N}}$, the lower ESSR in the higher-$\rho$ region.
This is because the quality of $R$-$U_N$ link has a different influence on the ESR of $x_3$ and $x_4$.
Lower $\lambda _{R{U_N}}$ results in a poor eavesdropping rate of $x_3$, which leads to a higher ESR of $x_3$. However, lower $\lambda _{R{U_N}}$ results in the poor jamming quality of $z_2$, which leads to decreasing in ESR of $x_4$, even to zero.
In the higher-$\rho$ region, not only the jamming influence of $z_2$ is strong but also amplifying coefficient decreases, then both $\gamma ^{{x_4}}$ and $\gamma _R^{{x_4}}$ become worse. The ESR of $x_3$ is improved since the eavesdropping and jamming quality at $R$ are enhanced simultaneously.

\begin{figure}[t]
	\centering
	\subfigure[]{
		\label{fig61}
		\includegraphics[width = 0.319 \textwidth]{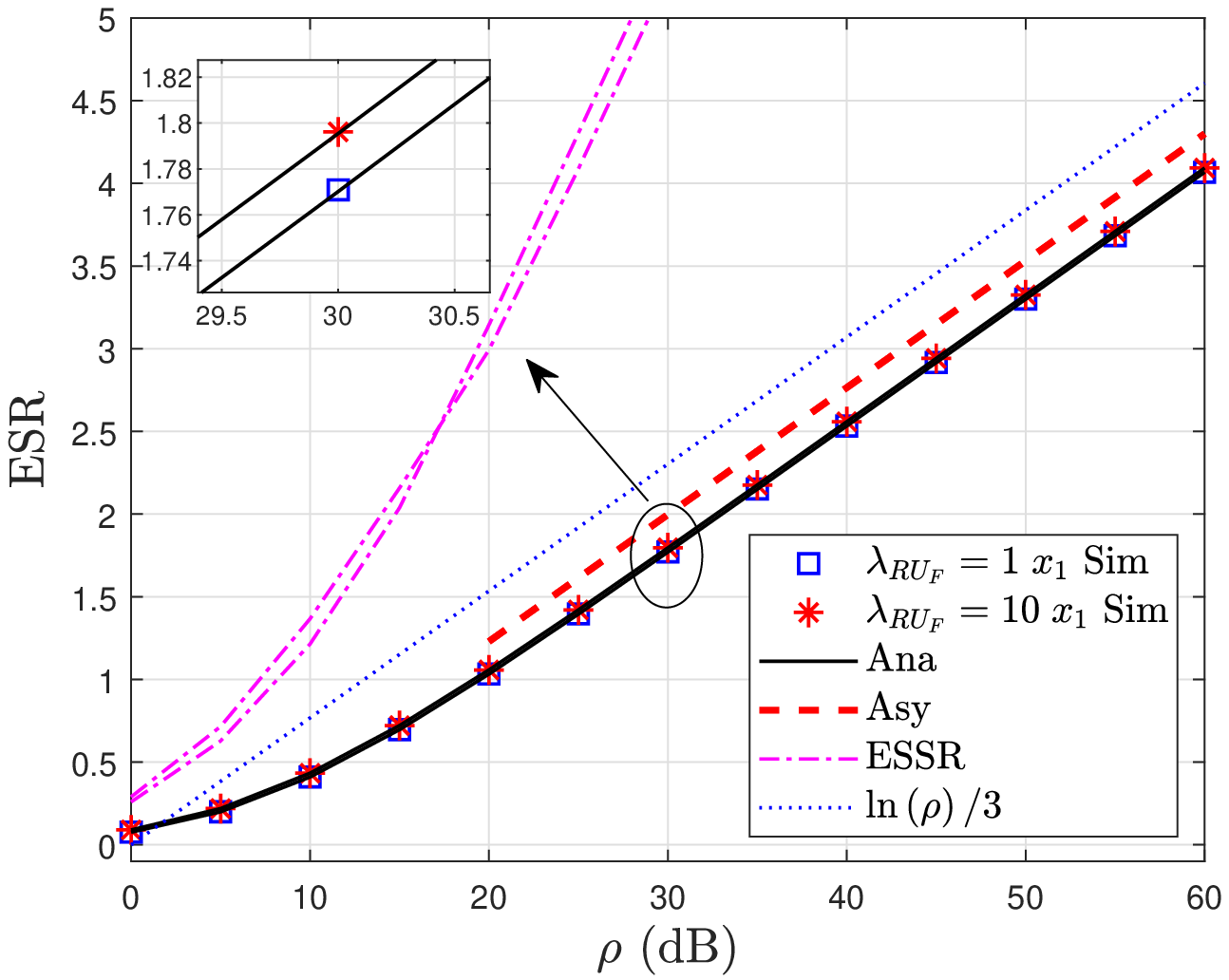}}
	\subfigure[]{
		\label{fig62}
		\includegraphics[width = 0.319 \textwidth]{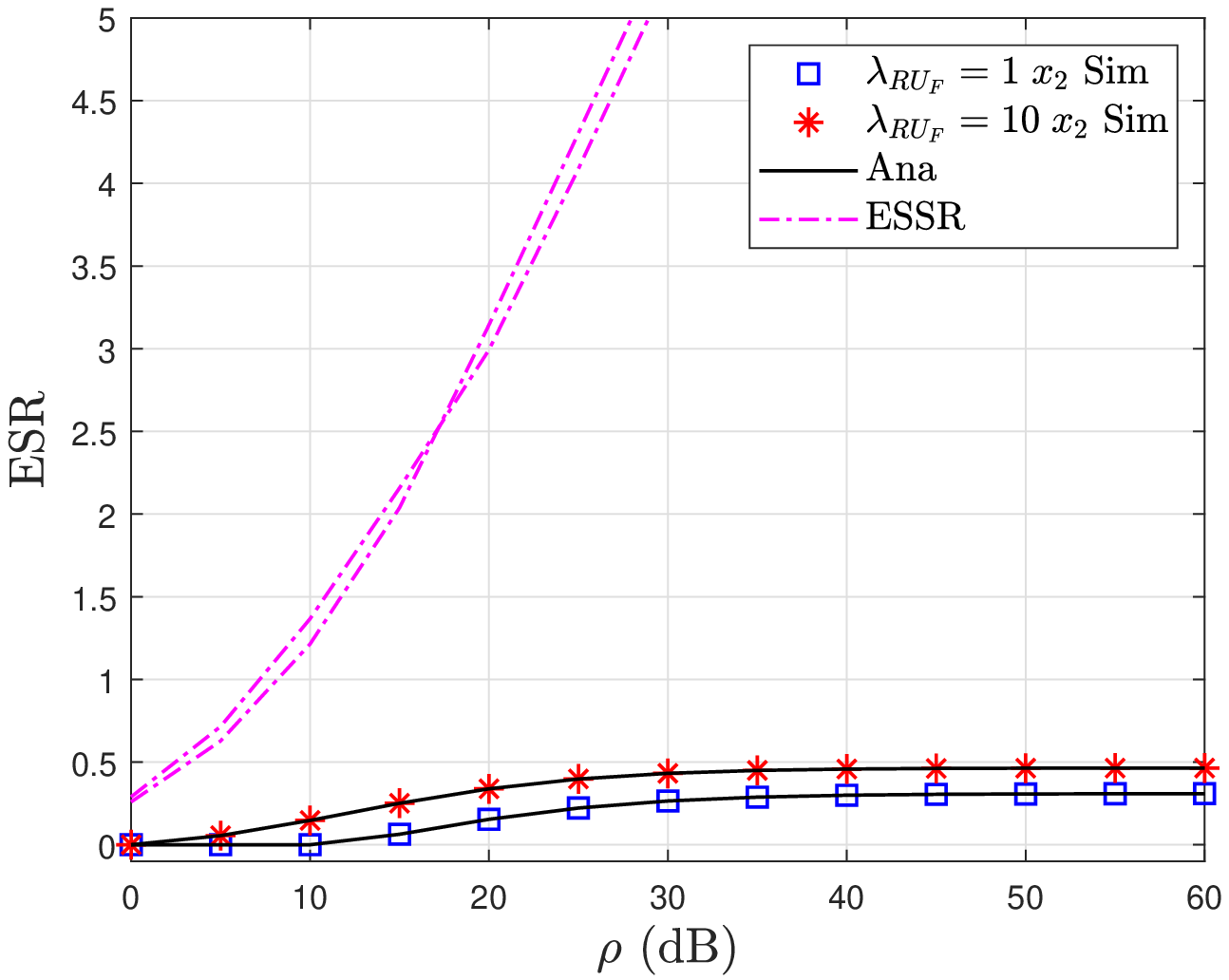}}
	\subfigure[]{
		\label{fig63}
		\includegraphics[width = 0.319 \textwidth]{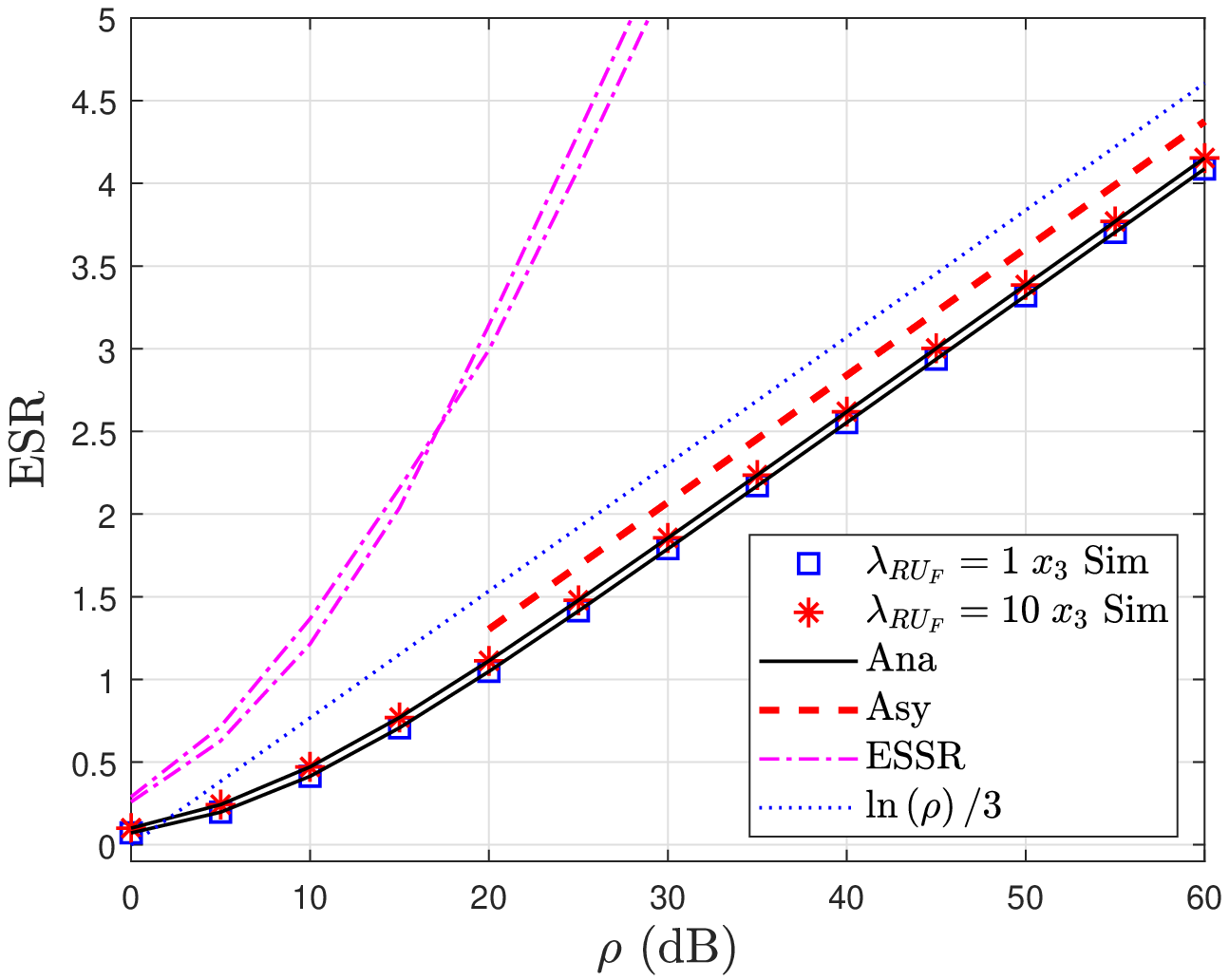}}
	\subfigure[]{
		\label{fig64}
		\includegraphics[width = 0.319 \textwidth]{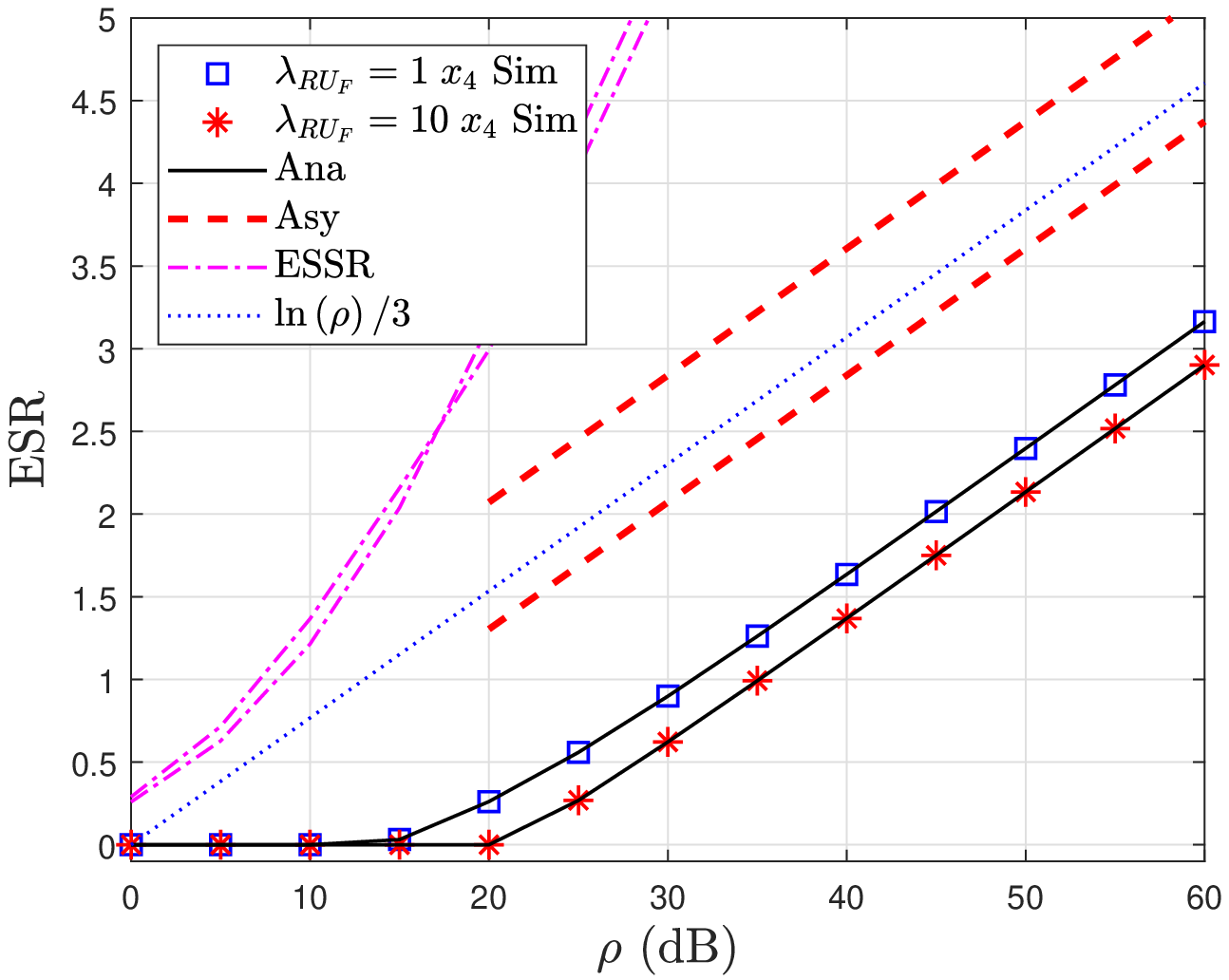}}
	\subfigure[]{
		\label{fig65}
		\includegraphics[width = 0.319 \textwidth]{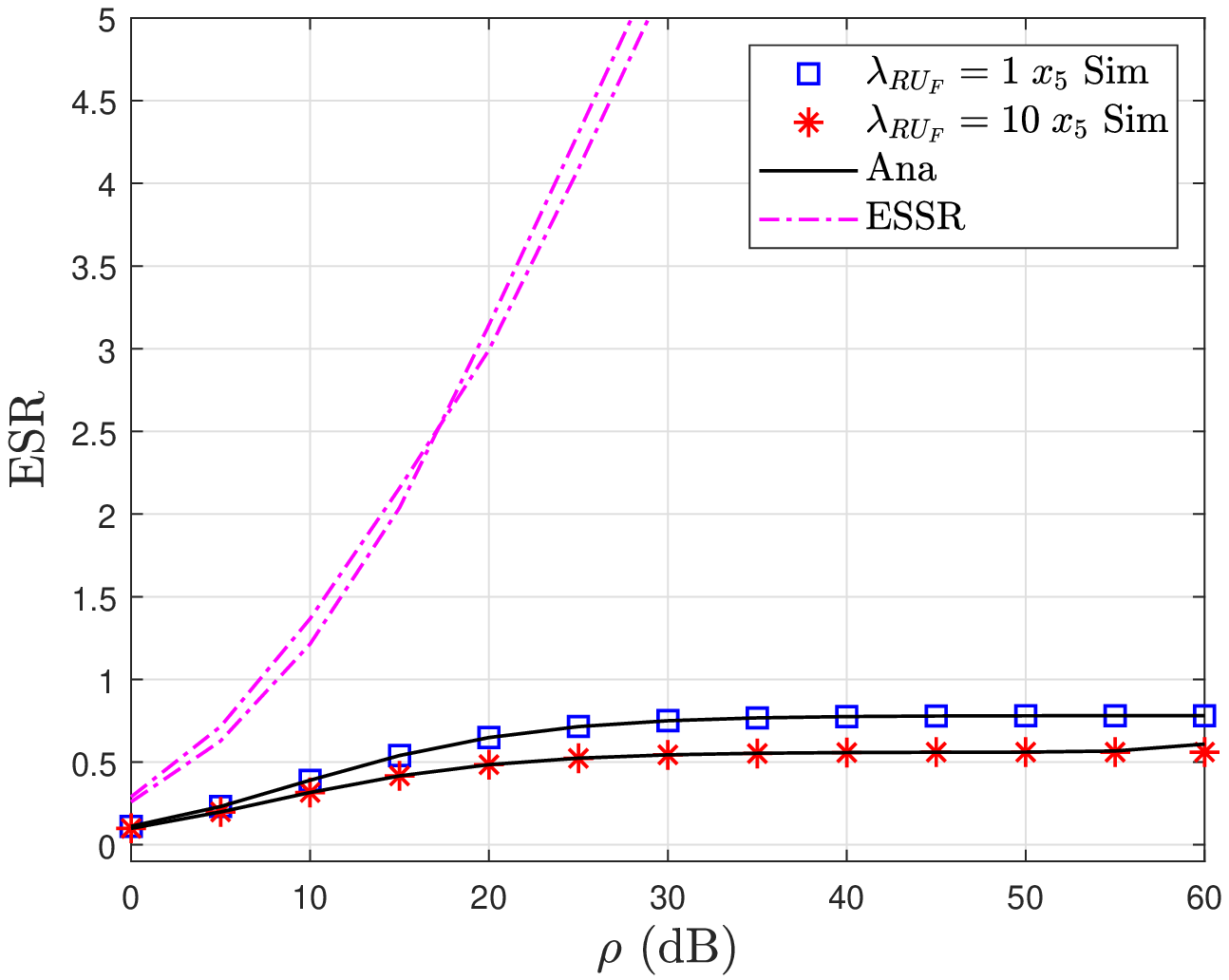}}
	\subfigure[]{
		\label{fig66}
		\includegraphics[width = 0.319 \textwidth]{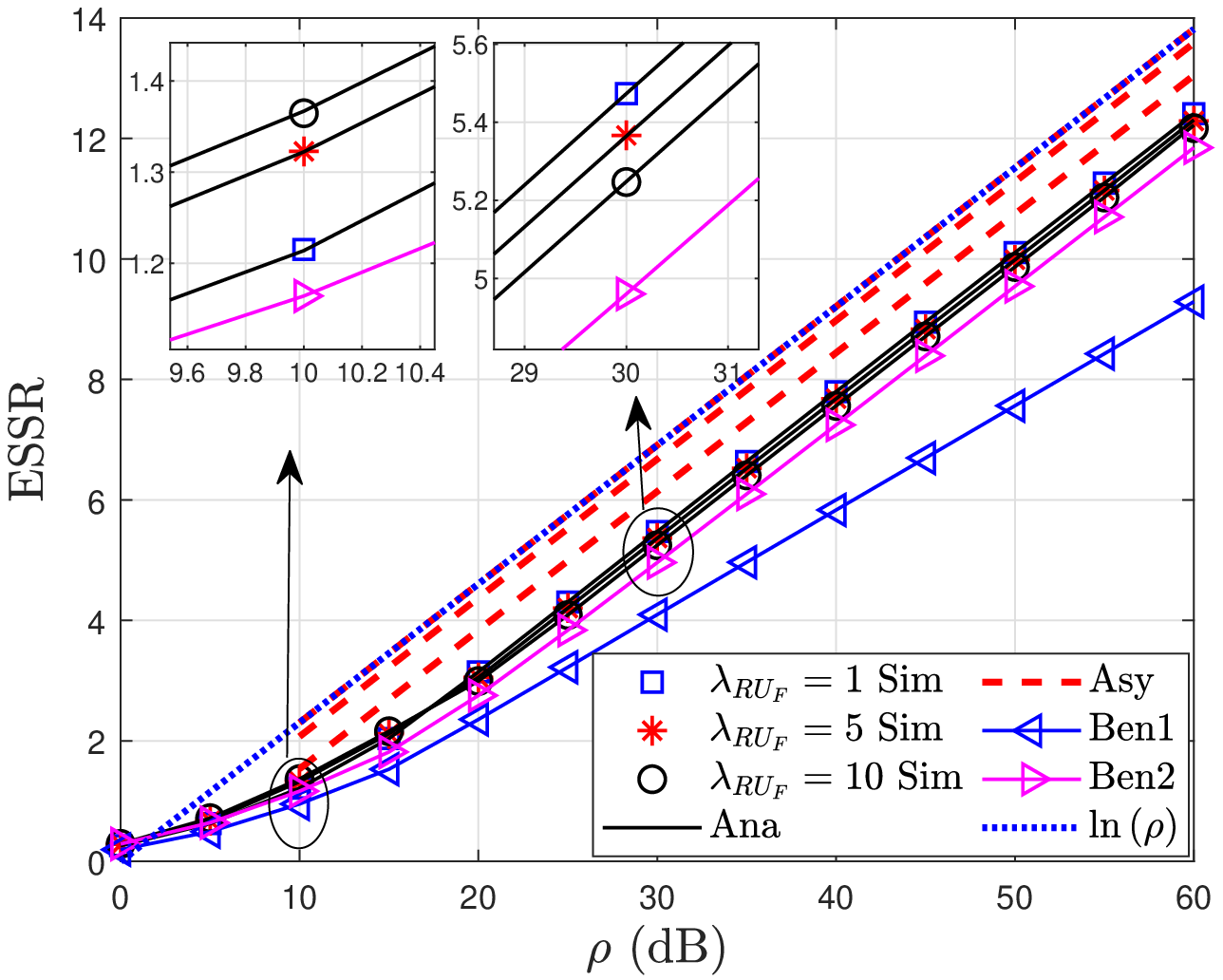}}
	\caption{ESRs and ESSR for varying ${\rho}$ and $\lambda _{R{U_F}}$.}
	\label{fig6}
\end{figure}
Fig. \ref{fig6} shows the lower bound of ESRs and ESSR with varying ${\rho}$ and $\lambda _{R{U_F}}$.
It is found that the ESSR with high $\lambda _{R{U_F}}$ in the lower-$\rho$ region outperforms that with lower $\lambda _{R{U_F}}$.
In the lower-$\rho$ region, the ESSR with low $\lambda _{R{U_F}}$ outperforms that with high $\lambda _{R{U_F}}$.
This is because ESRs of $x_1$ and $x_3$ play significant roles in the lower-$\rho$ region, while the variation of ESR of $x_4$ dominates in the higher-$\rho$ region.
ESRs of $x_1$ and $x_3$ increase and ESR of $x_4$ decreases as increases of $\lambda _{R{U_F}}$.

\begin{figure}[t]
	\centering
	\subfigure[]{
		\label{fig71}
		\includegraphics[width = 0.319 \textwidth]{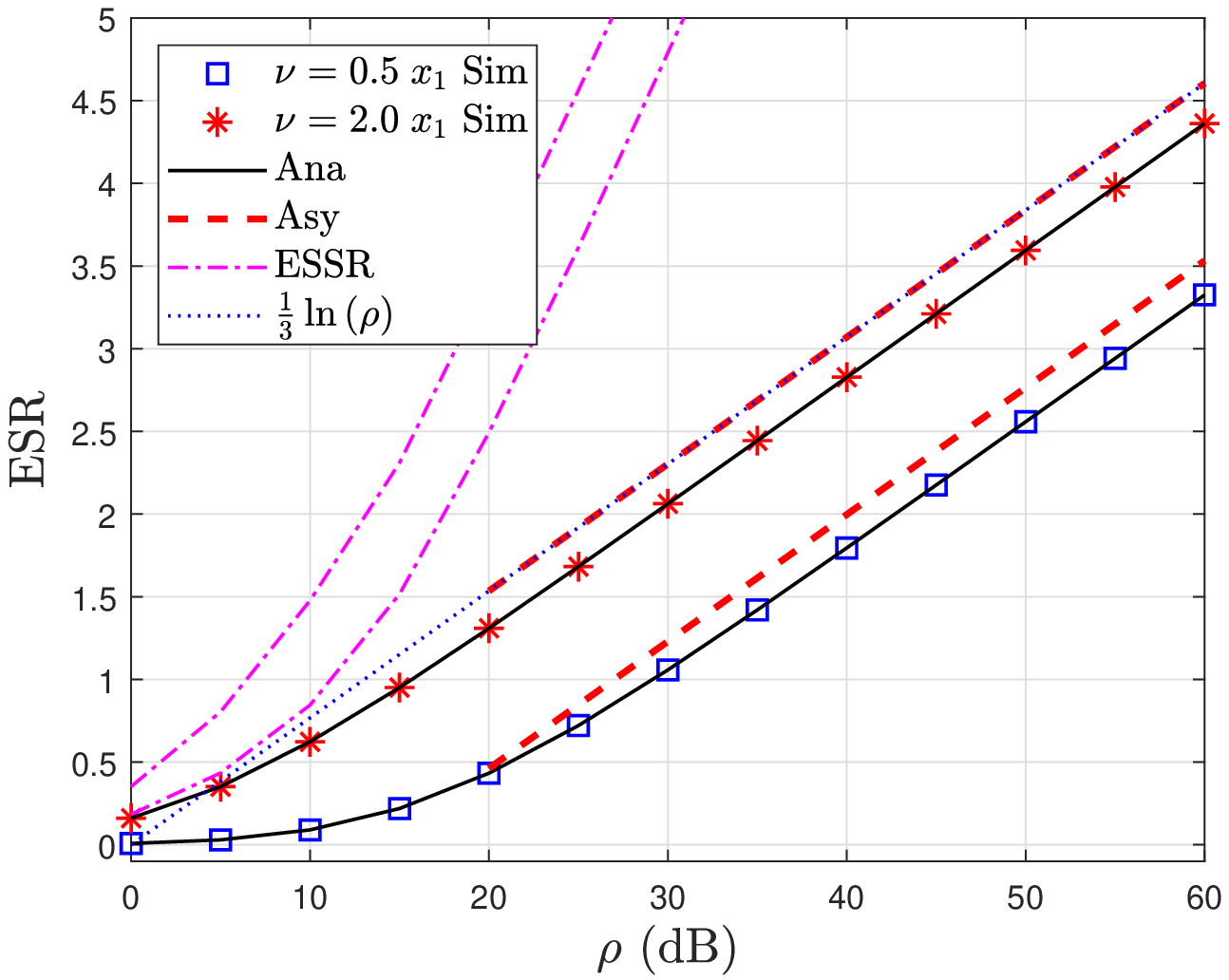}}
	\subfigure[]{
		\label{fig72}
		\includegraphics[width = 0.319 \textwidth]{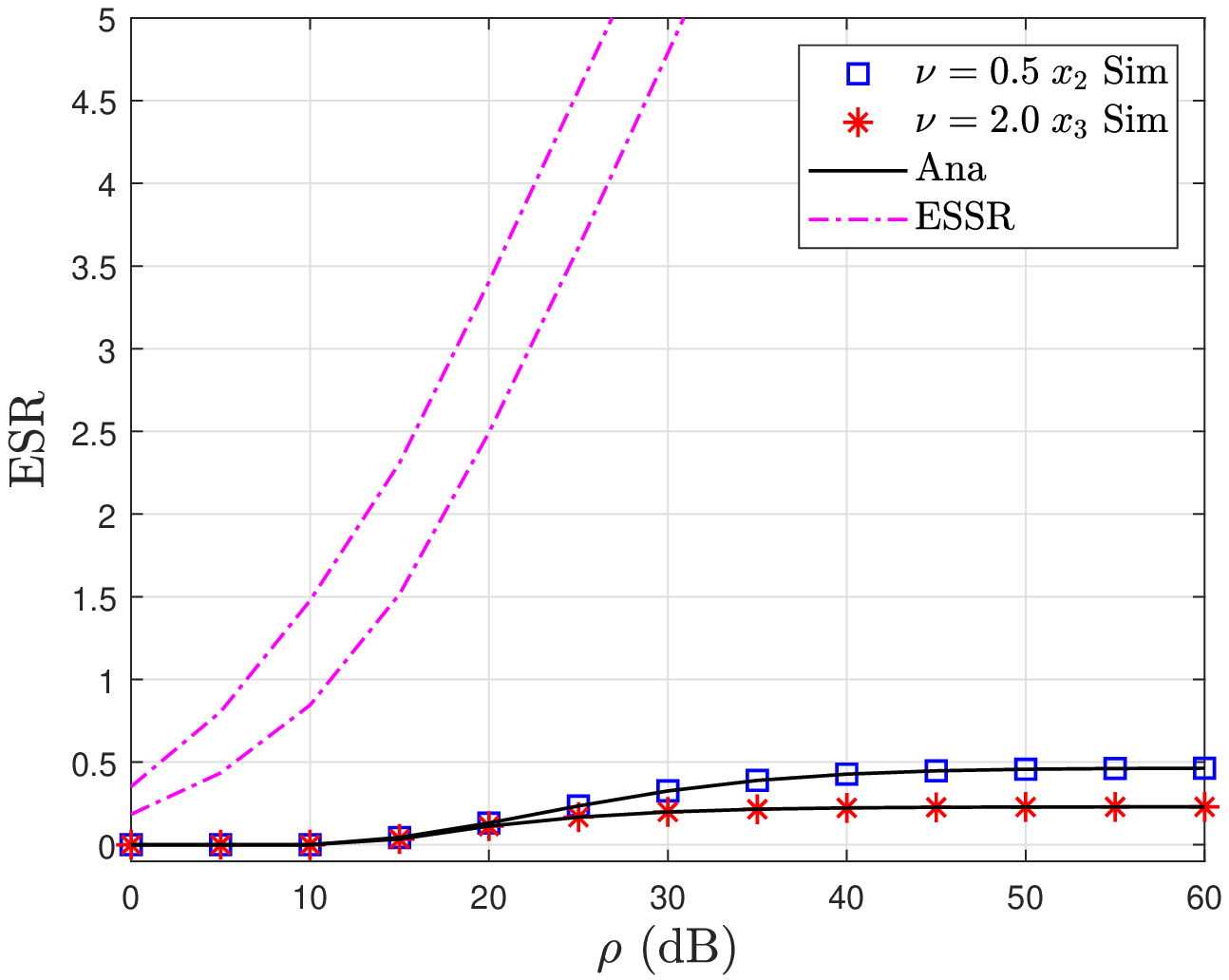}}
	\subfigure[]{
		\label{fig73}
		\includegraphics[width = 0.319 \textwidth]{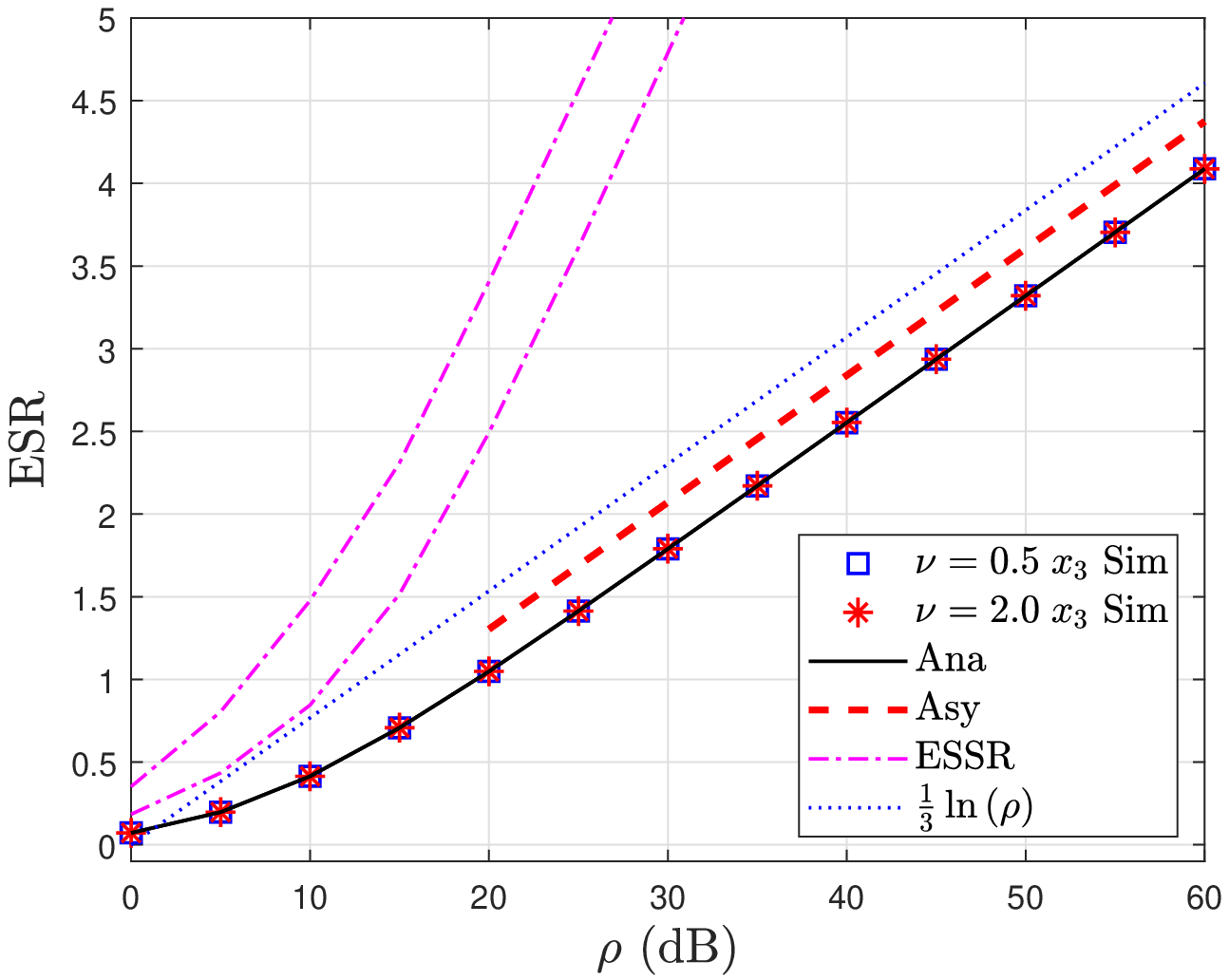}}
	\subfigure[]{
		\label{fig74}
		\includegraphics[width = 0.319 \textwidth]{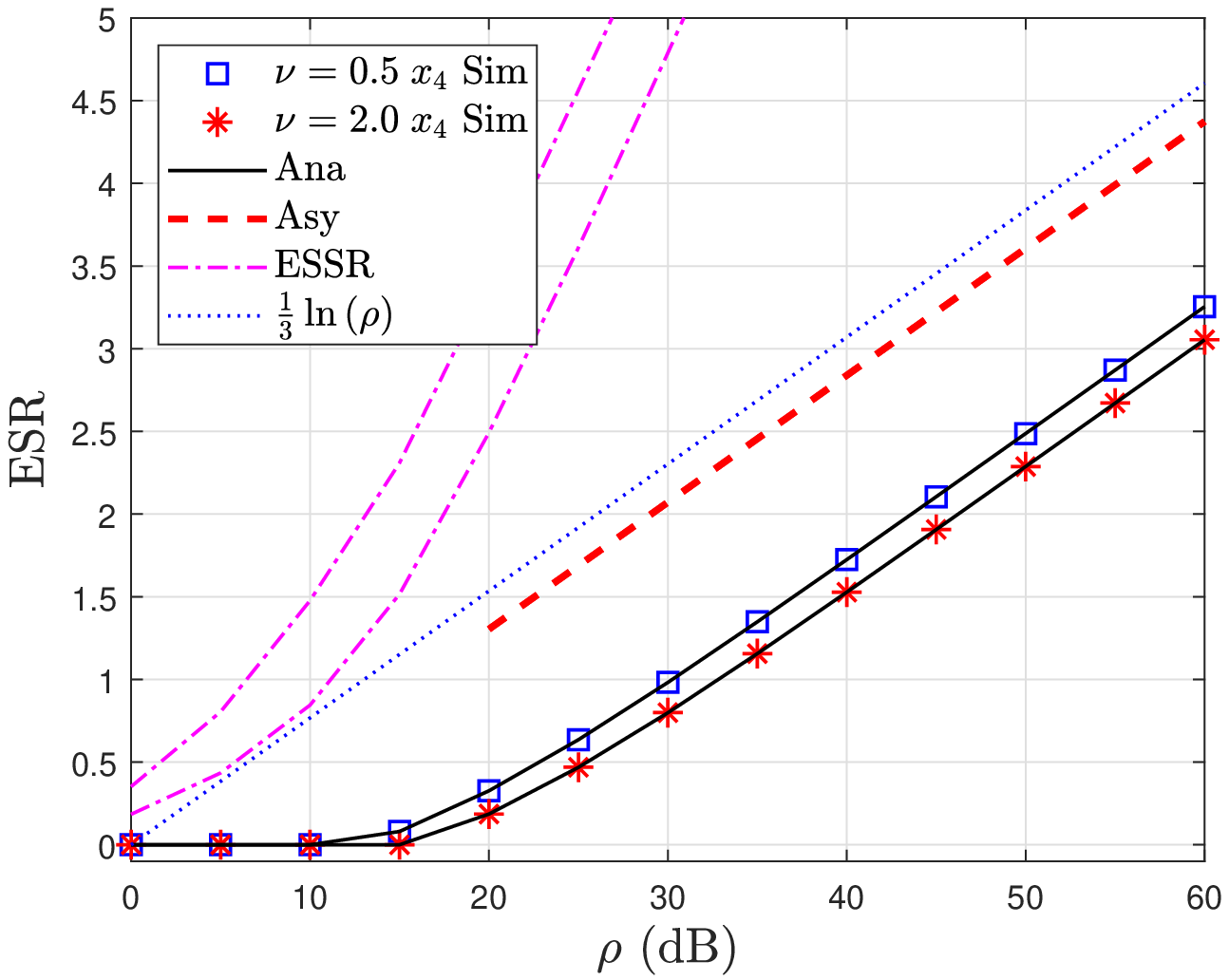}}
	\subfigure[]{
		\label{fig75}
		\includegraphics[width = 0.319 \textwidth]{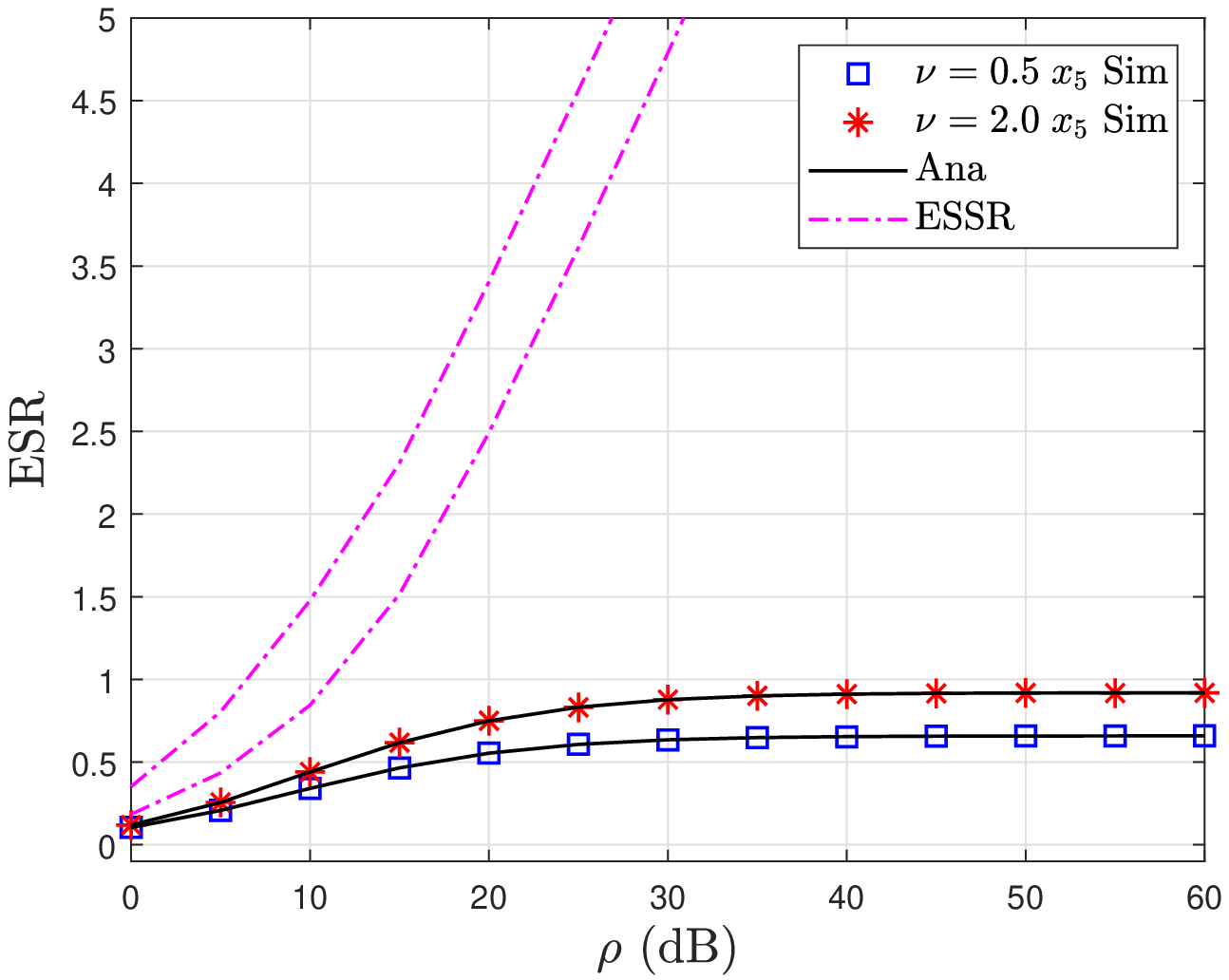}}
	\subfigure[]{
		\label{fig76}
		\includegraphics[width = 0.319 \textwidth]{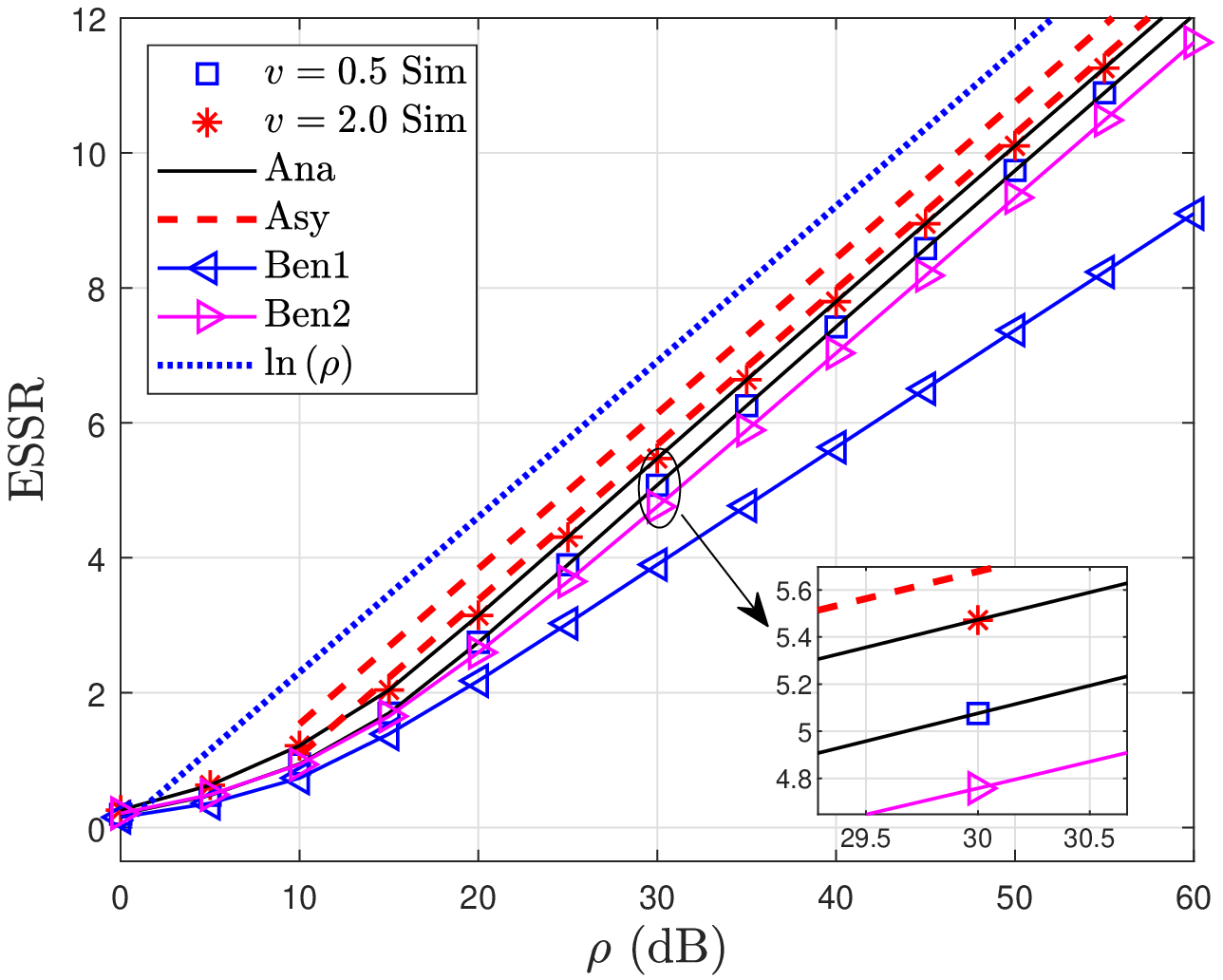}}
	\caption{ESRs and ESSR for varying ${\rho}$ and $\nu$.}
	\label{fig7}
\end{figure}
Fig. \ref{fig7} gives the comparison of the lower bound of ESRs and ESSR versus ${\rho}$ with varying $\nu$.
A trend is that ESSR increases with an increase in $\rho$ and $\nu$, which represent that the ESSR with higher $\nu$ (higher $\rho_S$) outperforms that with lower $\nu$, which is easy understand since increasing transmit SNR can enhance the secrecy performance.
The ESR of $x_1$ predominates in this scenario and ESR of $x_3$ almost is independent on $\nu$.
From Figs. \ref{fig3} - \ref{fig7}, it is easily observed that the scaling law of proposed scheme is obtained by asymptotic results. 

\begin{figure}[t]
	\centering
	\subfigure[]{
		\label{fig81}
		\includegraphics[width = 0.319 \textwidth]{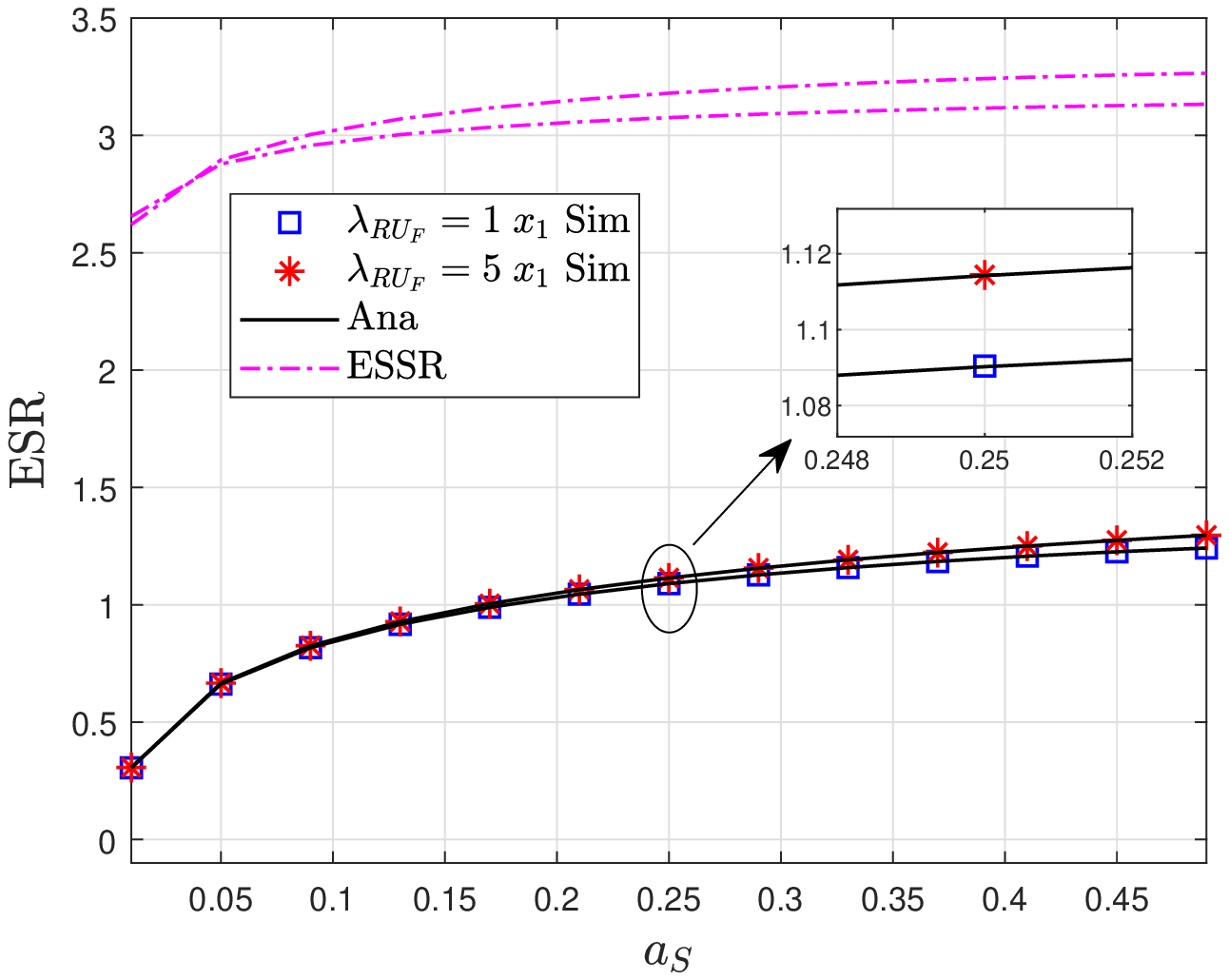}}
	\subfigure[]{
		\label{fig82}
		\includegraphics[width = 0.319 \textwidth]{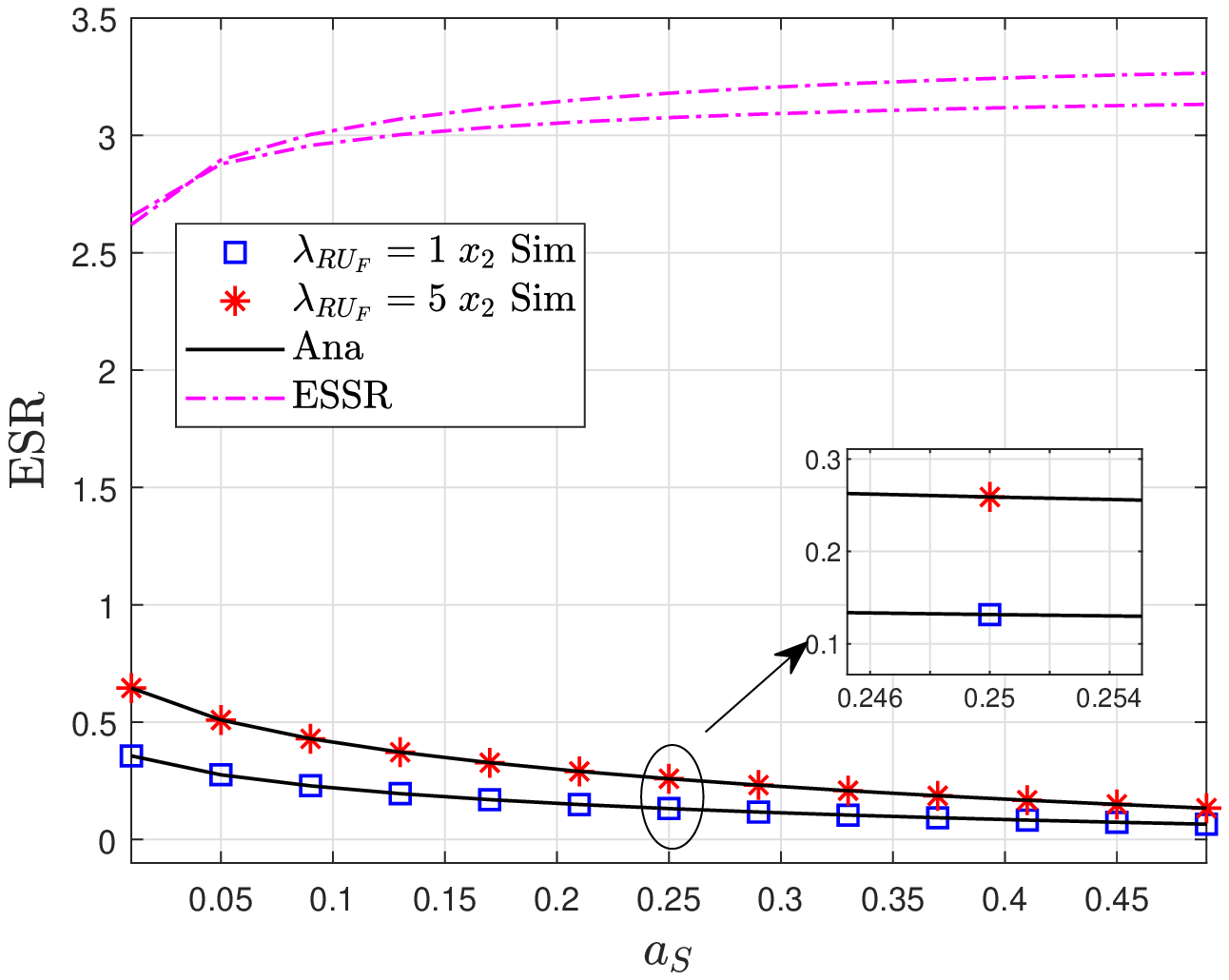}}
	\subfigure[]{
		\label{fig83}
		\includegraphics[width = 0.319 \textwidth]{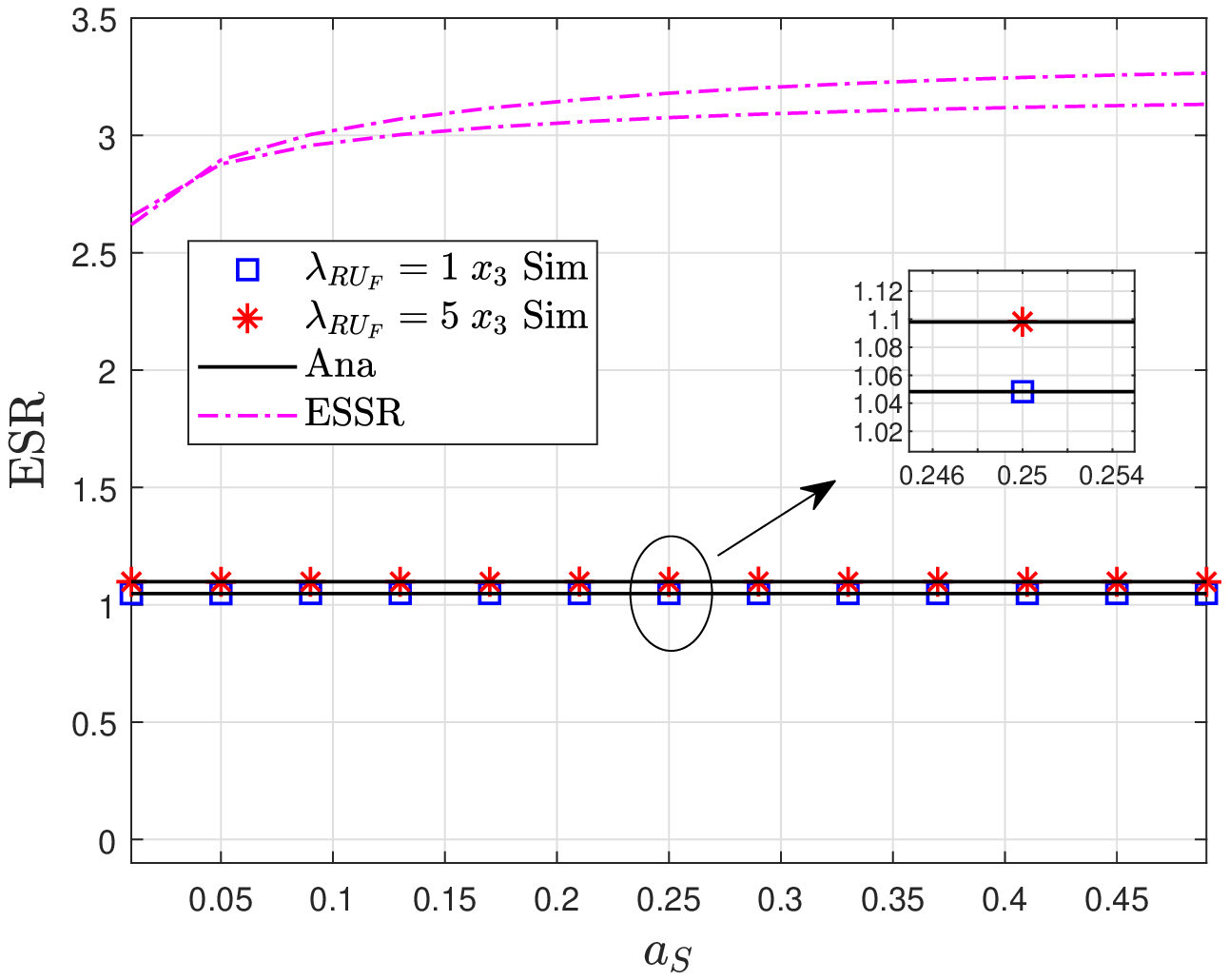}}
	\subfigure[]{
		\label{fig84}
		\includegraphics[width = 0.319 \textwidth]{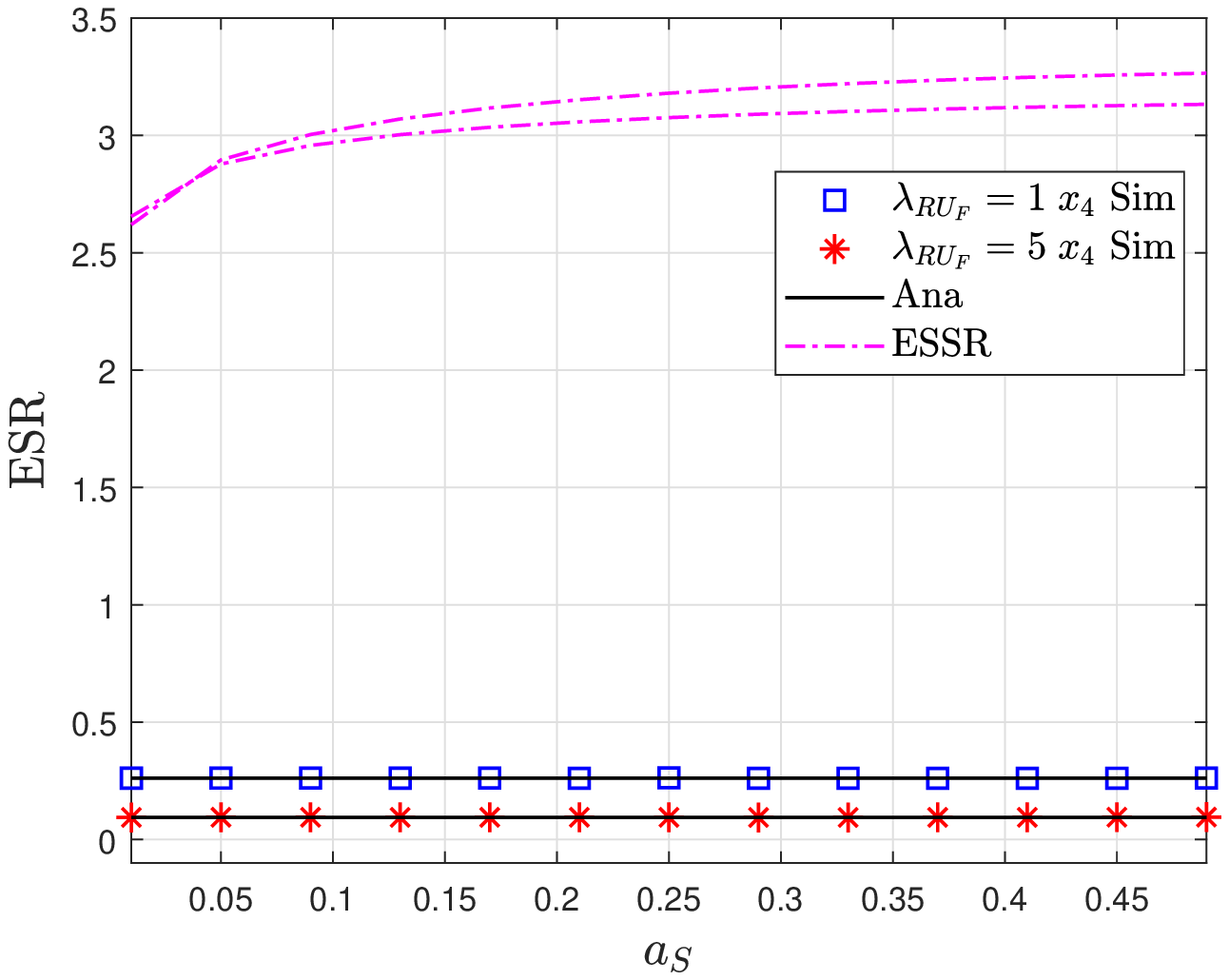}}
	\subfigure[]{
		\label{fig85}
		\includegraphics[width = 0.319 \textwidth]{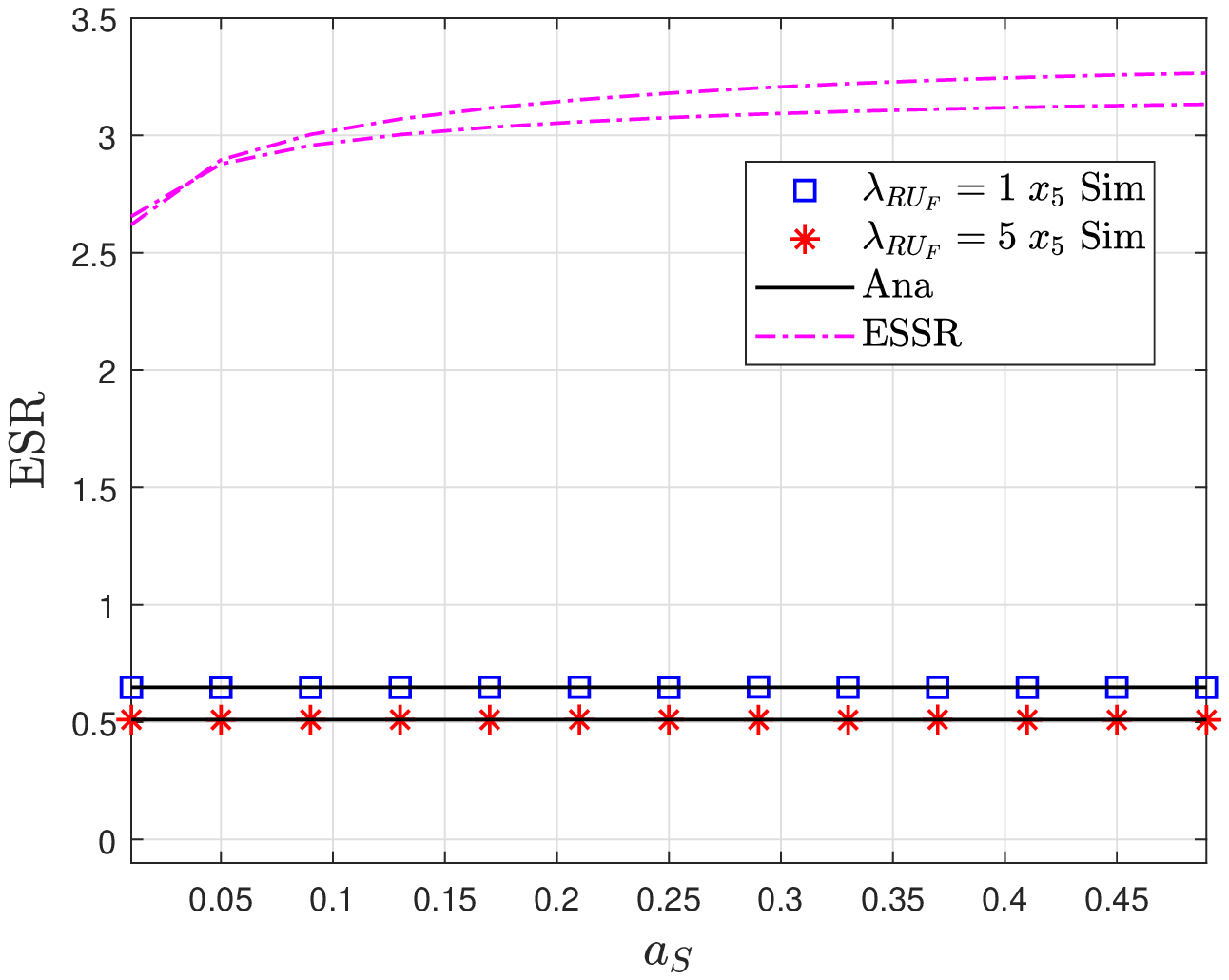}}
	\subfigure[]{
		\label{fig86}
		\includegraphics[width = 0.319 \textwidth]{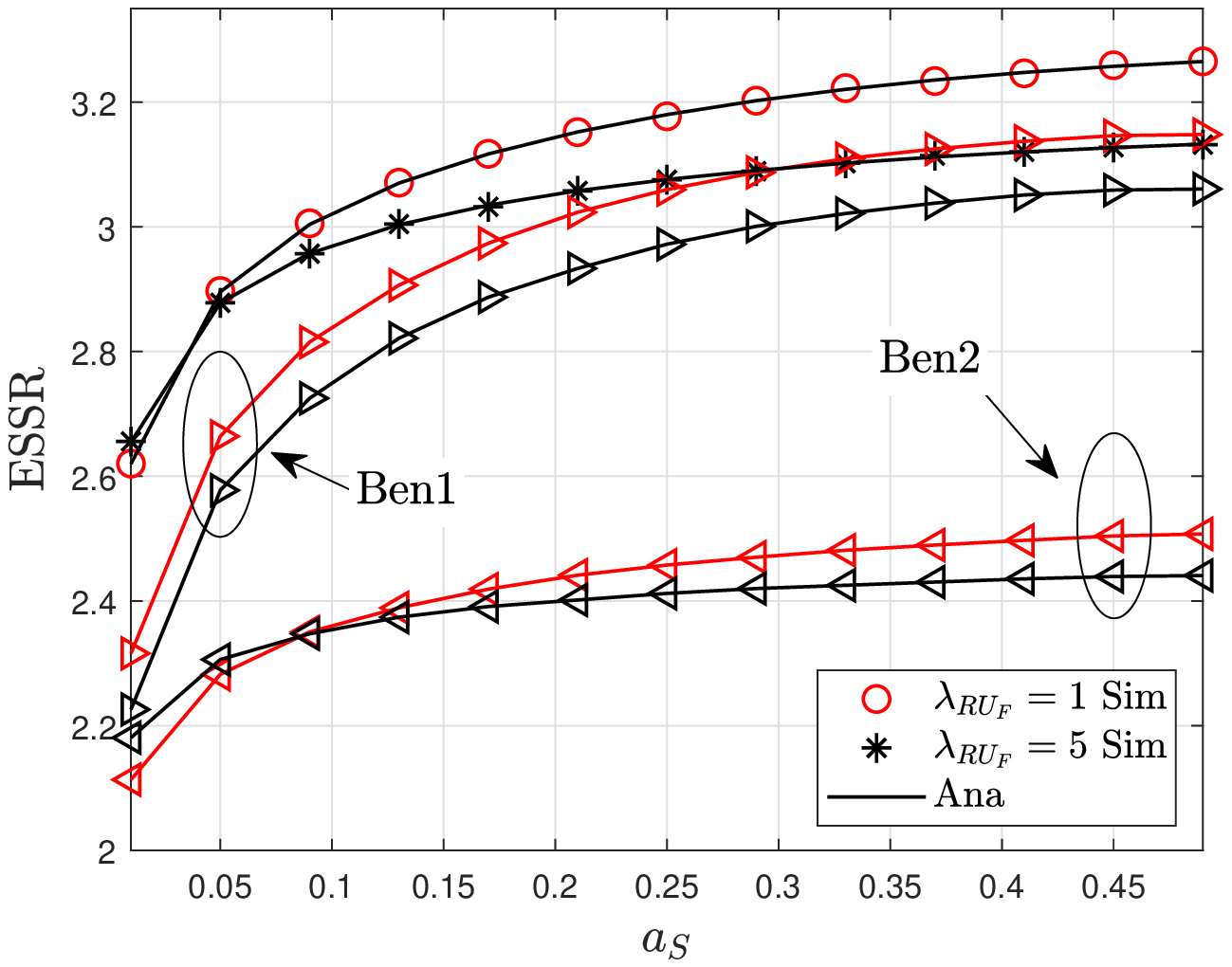}}
	\caption{ESRs and ESSR for varying ${\rho}$ and $\nu$.}
	\label{fig8}
\end{figure}
Fig. \ref{fig8} plots the simulation results for the lower bound of ESRs and ESSR for varying ${a_S}$ and $\lambda _{R{U_F}}$ with $\rho = 20$ dB.
It can be found that the ESSR increases with increasing ${a_S}$ while the rate of growth decreases.
This is because the ESR of $x_1$ increases with increasing ${a_S}$ and dominates.
Moreover, ESSR with lower $\lambda _{R{U_F}}$ outperforms that with higher $\lambda _{R{U_F}}$ in lower-$a_S$ region.
This is because ESRs of $x_1$, $x_2$, and $x_3$ increase and ESRs of $x_4$ and  $x_5$ decrease as increasing of $\lambda _{R{U_F}}$.
ESR of $x_4$ dominates in lower-$a_S$ region and ESR of $x_1$ dominates in higher-$a_S$ region.

\begin{figure}[h]
	\centering
	\subfigure[ESSR for varying ${\lambda _{R{U_F}}}$ with $\rho = 10$ dB.]{
		\label{fig91}
		\includegraphics[width = 0.483 \textwidth]{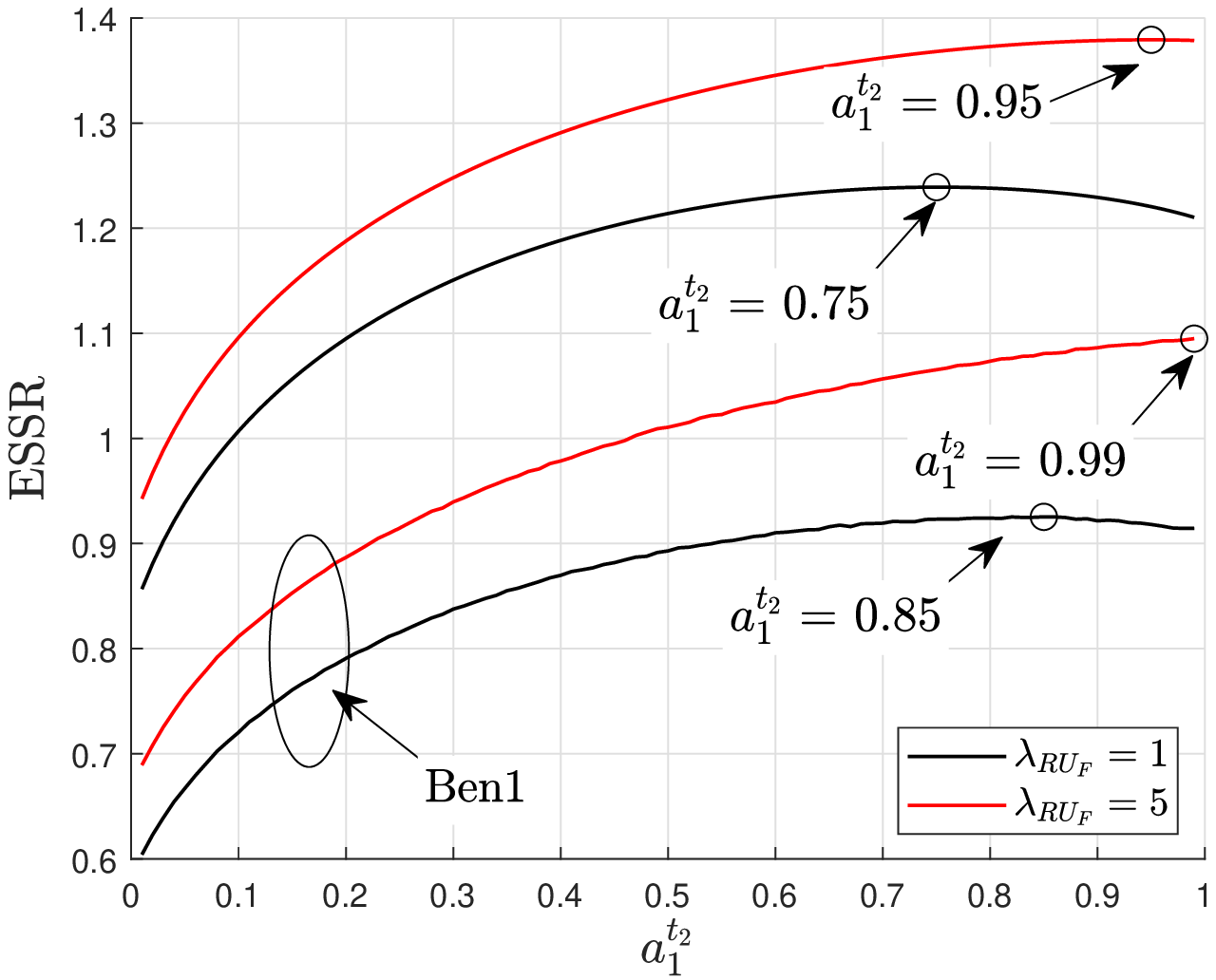}}
	\subfigure[ESSR for varying ${\lambda _{R{U_N}}}$ with $\rho = 20$ dB.]{
		\label{92}
		\includegraphics[width = 0.483 \textwidth]{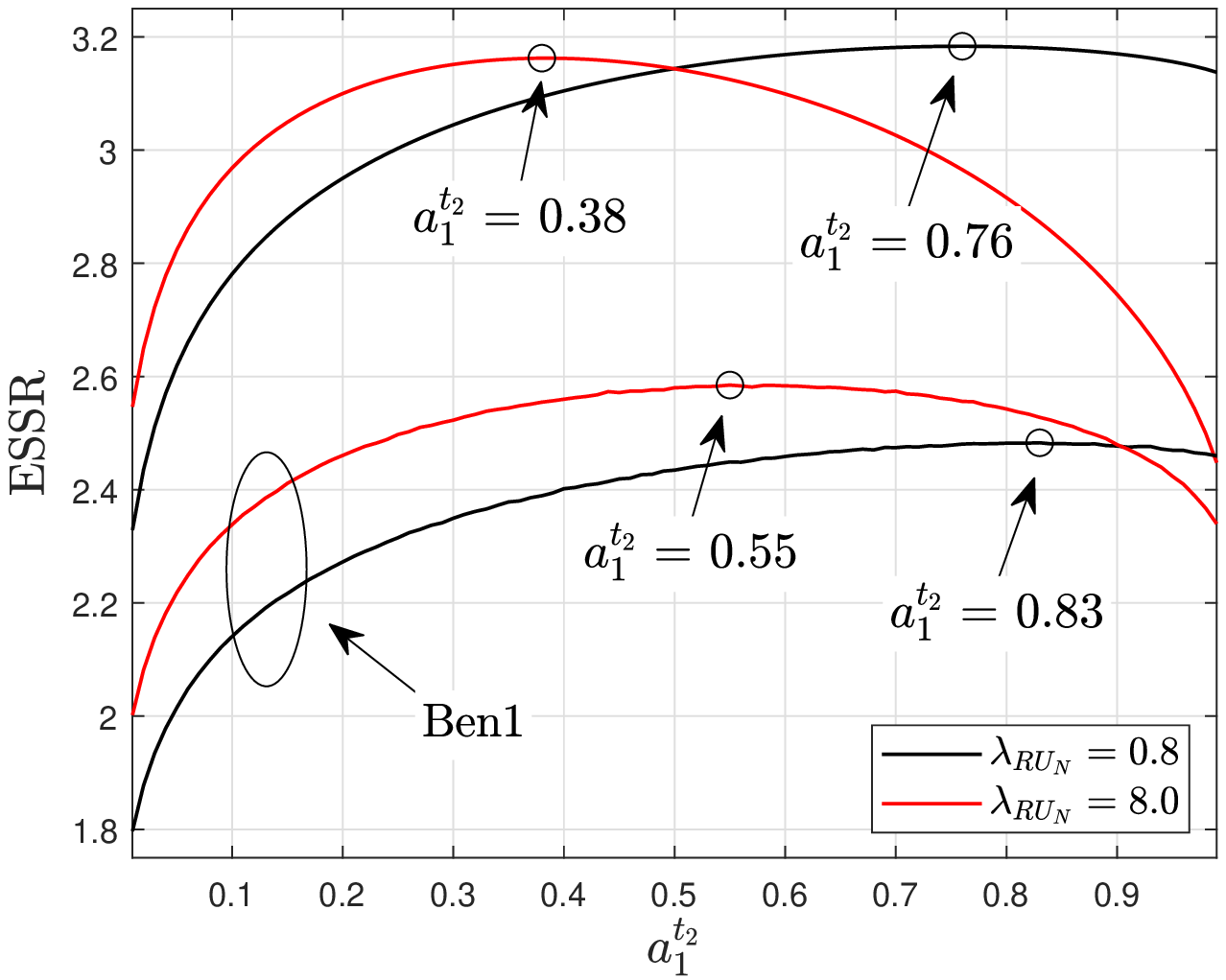}}
	\caption{ESRs and ESSR for varying ${{a_1^{{t_2}}}}$.}
	\label{fig9}
\end{figure}
Fig. \ref{fig9} shows the effect from ${{a_1^{{t_2}}}}$, ${\lambda _{R{U_F}}}$, and ${\lambda _{R{U_N}}}$ on the lower bound of ESRs and ESSR.
One can find there is an optimal ${{a_1^{{t_2}}}}$ to maximize the ESSR and the optimal ${{a_1^{{t_2}}}}$ depends on ${\lambda _{R{U_F}}}$ and ${\lambda _{R{U_N}}}$.
This is because ESR of $x_5$ is dominant in lower-${{a_1^{{t_2}}}}$ region and ESR of $x_3$ is dominant in higher-${{a_1^{{t_2}}}}$ region.
As ${{a_1^{{t_2}}}}$ increases, ESR of $x_3$ increases to the maximum then decreases.
From Figs. \ref{fig3} - \ref{fig9}, we observe that simulation and numerical results match perfectly to verify the correctness of our analysis.
Compared to the Benchmarks, the proposed scheme achieves higher secure performance.

\section{Conclusion}
\label{sec: Conclusion}

In this paper, we proposed a new scheme called NOMA-CDRT-PNC to provide reliable and secure communication for the CDRT system with uplink and downlink transmission. Cooperative jamming and inter-user interference prevented information leakage against the untrusted relay and NOMA and PNC schemes are utilized to enhance SE.
To characterize the security performance of the proposed schemes, we derived the expressions for lower bounds of the exact and asymptotic ESSR.
The effects of system parameters on the ESSR were analyzed in detail. Simulations were provided to verify the correctness of the analytical results.
The results showed that the proposed scheme obtained better security performance than conventional schemes.
An interesting work is to investigate the secrecy performance of CDRT systems with imperfect SIC and imperfect CSI, which will be part of our future work.
	
\begin{appendices}	
\section{Derivation of ${{\phi _5}\left( {a,b} \right)}$ }
\label{appendicesA}

Utilizing $\ln \left( y \right) = {\left. {\frac{{d{y^s}}}{{ds}}} \right|_{s = 0}}$, ${\phi _5}\left( {a,b} \right)$ is expressed as
\begin{equation}
 		\begin{aligned}
 		{\phi _5}\left( {a,b} \right) &= \int_0^\infty  {\frac{{{e^{ - by}}}}{{ay + 1}}} \ln \left( y \right)dy\\
 		&= \int_0^\infty  {\frac{{{e^{ - by}}}}{{ay + 1}}} {\left. {\frac{{d{y^s}}}{{ds}}} \right|_{s = 0}}dy\\
 		&= {\left. {\frac{d}{{ds}}\int_0^\infty  {\frac{{{y^s}{e^{ - by}}}}{{ay + 1}}} dy} \right|_{s = 0}}.
 		\label{psi5}
 	\end{aligned}
\end{equation}

Then utilizing \cite[(10), (11), (21)]{Adamchik1990} and \cite[(9.31.2), (9.31.5)]{Gradshteyn2007Book}, we obtain
\begin{equation}
	\begin{aligned}
      \int_0^\infty  {\frac{{{y^s}{e^{ - by}}}}{{ay + 1}}} dy &= \int_0^\infty  {{y^s}G_{1,1}^{1,1}\left[ {ay\left| {_0^0} \right.} \right]G_{0,1}^{1,0}\left[ {by\left| {_0^ - } \right.} \right]dy} \\
      &= {a^{ - s - 1}}G_{2,1}^{1,2}\left[ {_{1 + s}^{1,1 + s}\left| {\frac{a}{b}} \right.} \right].
	 \label{psi51}
	\end{aligned}
\end{equation}
Then utilizing residue theorem, we have
\begin{equation}
		\begin{aligned}
			{G_{2,1}^{1,2}\left( {_{1 + s}^{1,1 + s}\left| {\frac{a}{b}} \right.} \right)}
			&= \frac{1}{{2\pi i}}\int_L {\Gamma \left( p \right)\Gamma \left( { - s + p} \right)\Gamma \left( {s + 1 -
             p} \right){{\left( {\frac{a}{b}} \right)}^p}dp} \\
			&= \mathop {\lim }\limits_{p \to s} \left( {p - s} \right)\Gamma \left( p \right)\Gamma \left( { - s + p}
            \right)\Gamma \left( {s + 1 - p} \right){\left( {\frac{a}{b}} \right)^p}\\
			&+ \mathop {\lim }\limits_{p \to 0} p\Gamma \left( p \right)\Gamma \left( { - s + p} \right)\Gamma \left( {s
            + 1 - p} \right){\left( {\frac{a}{b}} \right)^p} + \mathcal{O}\left( {{a^s}} \right)\\
			&\simeq \mathop {\lim }\limits_{p \to s} \Gamma \left( p \right)\Gamma \left( { - s + p + 1} \right)\Gamma
            \left( {s + 1 - p} \right){\left( {\frac{a}{b}} \right)^p}\\
			&+ \mathop {\lim }\limits_{p \to 0} \Gamma \left( { - s + p} \right)\Gamma \left( {s + 1 - p} \right){\left( {\frac{a}{b}} \right)^p}\\
			&= \Gamma \left( s \right){\left( {\frac{a}{b}} \right)^s} + \Gamma \left( { - s} \right)\Gamma \left( {s + 1} \right),
			\label{psi53}
		\end{aligned}
\end{equation}
where $\mathcal{O}\left( a^s \right)$ denotes higher order terms.
Substituting (\ref{psi53}) into (\ref{psi5}), we have
\begin{equation}
	{\phi _5}\left( {a,b} \right) = {a^{ - 1}}{\left. {\frac{d{\Theta \left( s \right)}}{{ds}}} \right|_{s = 0}},
	\label{psi54}
\end{equation}
where
$\Theta \left( s \right) = {b^{ - s}}\Gamma \left( s \right) + {a^{ - s}}\Gamma \left( { - s} \right)\Gamma \left( {s + 1} \right)$.
By utilizing \cite[(8.322)]{Gradshteyn2007Book}, we have
\begin{equation}
	\Gamma \left( s \right) = \frac{{{e^{ - {\rm{C}}s}}}}{s}\prod\limits_{k = 1}^\infty  {{{\left( {1 + \frac{s}{k}} \right)}^{ - 1}}{e^{\frac{s}{k}}}}.
	\label{H1}
\end{equation}
By substituting ${{e^{ - {\rm{C}}s}}}$ with its taylor series, ${{e^{ - {\rm{C}}s}} = 1 - {\rm{C}}s + \frac{{{{\rm{C}}^2}{s^2}}}{2} + \mathcal{O}\left( {{s^2}} \right)}$, (\ref{H1}) is rewritten as
\begin{equation}
	\begin{aligned}
		\Gamma \left( s \right) &\mathop = \limits^{s \to 0} \frac{{1 - {\rm{C}}s + \frac{{{{\rm{C}}^2}{s^2}}}{2} + \mathcal{O}\left( {{s^2}} \right)}}{s}\\
		&= \frac{1}{s} - {\rm{C}} + \frac{{{{\rm{C}}^2}}}{2}s + \mathcal{O}\left( s \right)\\
		&= \frac{1}{s} - {\rm{C}} + {\omega _3},
		\label{H2}
	\end{aligned}
\end{equation}
 where ${{\omega _3} = \frac{{{{\rm{C}}^2}}}{2}s + \mathcal{O}\left( s \right)}$.
Similarly, we have
$\Gamma \left( { - s} \right) =  - \frac{1}{s} - {\rm{C}} - {\omega _3}$.
Then
$\Theta \left( s \right)$ is rewritten as
\begin{equation}
	\begin{aligned}
		\Theta \left( s \right) &= {b^{ - s}}\left( {\frac{1}{s} - {\rm{C}} + {\omega _3}} \right) + {a^{ - s}}\left( {\frac{1}{{ - s}} - {\rm{C}} - {\omega _3}} \right)\Gamma \left( {s + 1} \right)\\
		&= \underbrace {\frac{{{b^{ - s}}}}{s} - \frac{{{a^{ - s}}\Gamma \left( {s + 1} \right)}}{s}}_{{I_3}} + \underbrace {{\omega _3}{b^{ - s}} - {\rm{C}}{b^{ - s}} - {\rm{C}}{a^{ - s}}\Gamma \left( {s + 1} \right) - {\omega _3}{a^{ - s}}\Gamma \left( {s + 1} \right)}_{I_4}.
		\label{psi57}
	\end{aligned}
\end{equation}
Defining ${f_1}\left( s \right) = {b^{ - s}}$ and ${f_2}\left( s \right)={{a^{ - s}}\Gamma \left( {s + 1} \right)}$, then Taylor series expansions of ${f_1}\left( s \right)$ and ${f_2}\left( s \right)$ at $s = 0$  are  expressed as
\begin{equation}
	\begin{aligned}
		{f_1}\left( s \right) &= {b^{ - s}}\\
		&= {f_1}\left( 0 \right) + {f'_1}\left( 0 \right)\frac{s}{{1!}} + {f''_1}\left( 0 \right)\frac{{{s^2}}}{{2!}} + \mathcal{O}\left( {{s^2}} \right)\\
		&= 1 - \ln \left( b \right)s + \frac{{{{\ln }^2}\left( b \right){s^2}}}{2} + \mathcal{O}\left( {{s^2}} \right),
		\label{psi58}
	\end{aligned}
\end{equation}
and
\begin{equation}
	\begin{aligned}
		{f_2}\left( s \right) &= {a^{ - s}}\Gamma \left( {s + 1} \right)\\
		&= {f_2}\left( 0 \right) + {f'_2}\left( 0 \right)\frac{s}{{1!}} + {f''_2}\left( 0 \right)\frac{{{s^2}}}{{2!}} + \mathcal{O}\left( {{s^2}} \right)\\
		&= 1 + {f'_2}\left( 0 \right)s + {f''_2}\left( 0 \right){s^2} + \mathcal{O}\left( {{s^2}} \right),
		\label{psi59}
	\end{aligned}
\end{equation}
respectively, where
${f'_2\left( 0 \right)}$ and ${f''_2\left( 0 \right)}$ are given as
\begin{equation}
	\begin{aligned}
		{f'_2}\left( 0 \right) & =  \frac{d}{{ds}}\left( {{a^{ - s}}\Gamma \left( {s + 1} \right)} \right)\left| {_{s = 0}} \right.\\
		&=  - \ln \left( a \right){a^{ - s}}\Gamma \left( {s + 1} \right) + {a^{ - s}}\Gamma \left( {s + 1} \right){\varphi _0}\left( {s + 1} \right)\left| {_{s = 0}} \right.\\
		&= {\rm{C}} - \ln \left( a \right),
		\label{psi591}
	\end{aligned}
\end{equation}
and
\begin{equation}
	\begin{aligned}
		{f''_2}\left( 0 \right) &= \frac{{{d^2}}}{{d{s^2}}}\left( {{a^{ - s}}\Gamma \left( {s + 1} \right)} \right)\left| {_{s = 0}} \right.\\
		&= {\left( {{a^{ - s}}} \right)^{\prime \prime }}\Gamma \left( {s + 1} \right) + 2{\left( {{a^{ - s}}} \right)^\prime }{\left( {\Gamma \left( {s + 1} \right)} \right)^\prime } + {a^{ - s}}{\left( {\Gamma \left( {s + 1} \right)} \right)^{\prime \prime }}\left| {_{s = 0}} \right.\\
		&= {\left( {\ln \left( a \right)} \right)^2}\Gamma \left( 1 \right) - 2\ln \left( a \right){\Gamma ^\prime }\left( 1 \right) + {\Gamma ^{\prime \prime }}\left( 1 \right)\\
		&= {\ln ^2}\left( a \right) + 2{\rm{C}}\ln \left( a \right) + {\psi ^{\left( 1 \right)}}\left( 1 \right) + {{\rm{C}}^2},
		\label{psi61}
	\end{aligned}
\end{equation}
respectively,
where ${\Gamma ^\prime }\left( 1 \right) = \Gamma \left( 1 \right){\psi ^{\left( 0 \right)}}\left( 1 \right) =  - {\rm{C}}$,
${\Gamma ^{\prime \prime }}\left( 1 \right) = {\psi ^{\left( 1 \right)}}\left( 1 \right) + {\left( {{\Gamma ^\prime }\left( 1 \right)} \right)^2} = {\psi ^{\left( 1 \right)}}\left( 1 \right) + {{\rm{C}}^2}$, and ${{\psi ^{\left( k \right)}}\left( x \right)}$ is the  polygamma  function as defined by \cite[(8.363.8)]{Gradshteyn2007Book}.

Then $I_3$ is obtained as
\begin{equation}
	\begin{aligned}
		{I_3} &= \frac{{1 - \ln \left( b \right)s + \frac{{{{\ln }^2}\left( b \right)}}{{2!}}{s^2}}}{s} - \frac{{1 + {f'_2}\left( 0 \right)s + \frac{{{f''_2}\left( 0 \right)}}{2}{s^2}}}{s}\\
		&=  - \ln \left( b \right) + \frac{{s{{\ln }^2}\left( b \right)}}{{2}} - {f'_2}\left( 0 \right) - {f''_2}\left( 0 \right)\frac{{{s}}}{{2}}.
		\label{psi60}
	\end{aligned}
\end{equation}
Substituting (\ref{psi591}) and (\ref{psi61}) into (\ref{psi60}), ${I_3}^\prime$ is obtained as
\begin{equation}
	\begin{aligned}
		{I_3}^\prime  &= \frac{{{{\ln }^2}\left( b \right)}}{{2!}} - \frac{{{{f''}_2}\left( 0 \right)}}{2}\\
		&= \frac{{{{\ln }^2}\left( b \right)}}{2} - \frac{{{{\ln }^2}\left( a \right) + 2{\rm{C}}\ln \left( a \right) + {\psi ^{\left( 1 \right)}}\left( 1 \right) + {{\rm{C}}^2}}}{2}.
		\label{dI3}
	\end{aligned}
\end{equation}

Similar as (\ref{dI3}), we obtain
\begin{equation}
	\begin{aligned}
		{I_4}^\prime \left| {_{s = 0}} \right. &= \left( { - {\rm{C}}{b^{ - s}} + {\omega _3}{b^{ - s}} - {\rm{C}}{a^{ - s}}\Gamma \left( {s + 1} \right)} \right.{\left. { - {\omega _3}{a^{ - s}}\Gamma \left( {s + 1} \right)} \right)^\prime }\\
		&= \left( {{\rm{C}} - {\omega _3}} \right){b^{ - s}}\ln \left( b \right) + \frac{{{{\rm{C}}^2}}}{2}{b^{ - s}} + \left( {\ln \left( a \right) - {{\varphi ^{\left( 0 \right)}}}\left( {s + 1} \right)} \right){\rm{C}}{a^{ - s}}\Gamma \left( {s + 1} \right)\\
		&-\frac{{{{\rm{C}}^2}{a^{ - s}}\Gamma \left( {s + 1} \right)}}{2}\left( {1 + \left( {{{\varphi ^{\left( 0 \right)}}}\left( {s + 1} \right) - \ln \left( a \right)} \right)s} \right)\\
		&= {\rm{C}}\ln \left( {ab} \right) + {{\rm{C}}^2}.
		\label{dI4}
	\end{aligned}
\end{equation}
Substituting (\ref{dI3}) and (\ref{dI4}) into (\ref{psi54}), we obtain
\begin{equation}
	\begin{aligned}
		{\phi _5}\left( {a,b} \right) &= \frac{1}{a}\left( {{I_3}^\prime  + {I_4}^\prime \left| {_{s = 0}} \right.} \right)\\
		&= \frac{{{{\ln }^2}\left( b \right)}}{{2a}} - \frac{{{{\ln }^2}\left( a \right) + 2{\rm{C}}\ln \left( a \right) + {\psi ^{\left( 1 \right)}}\left( 1 \right) + {{\rm{C}}^2}}}{{2a}} + \frac{{\rm{C}}}{a}\ln \left( {ab} \right) + \frac{{{{\rm{C}}^2}}}{a}\\
		&= \frac{1}{{2a}}\ln \left( {\frac{b}{a}} \right)\ln \left( {ab} \right) - \frac{1}{{2a}}{\psi ^{\left( 1 \right)}}\left( 1 \right) + \frac{{\rm{C}}}{a}\ln \left( b \right) + \frac{{{{\rm{C}}^2}}}{{2a}}.
		\label{psi63}
	\end{aligned}
\end{equation}

\end{appendices}	


\end{document}